\documentclass[pdftex,twocolumn,epjc3_preprint,runningheads]{svjour3}

\usepackage[T1]{fontenc}
\usepackage{lmodern}
\usepackage{calc}
\usepackage{graphicx}
\usepackage{booktabs}
\usepackage{textcomp}
\usepackage{xspace}
\usepackage{relsize}
\usepackage{amssymb}
\usepackage{amsmath}
\usepackage{listings}
\usepackage{microtype}
\usepackage{multirow}
\usepackage{tabularx}
\usepackage{array}
\usepackage{placeins}
\usepackage{cuted}
\usepackage{soul} 
\usepackage{fixltx2e}
\usepackage{slashed}
\usepackage{bm}
\usepackage[numbers,sort&compress]{natbib}
\usepackage[labelfont=bf,font=small]{caption}
\usepackage[skip=-2pt]{subcaption}
\usepackage[colorlinks,citecolor=blue,urlcolor=blue,linkcolor=blue,breaklinks=breakall]{hyperref}
\usepackage{breakurl}
\usepackage[dvipsnames]{xcolor}
\usepackage[clockwise,figuresright]{rotating}
\usepackage{siunitx}
\usepackage{tikz}
\usepackage[normalem]{ulem}
\usepackage[utf8]{inputenc}

\usepackage{etoolbox}
\AfterEndEnvironment{strip}{\leavevmode}

\allowdisplaybreaks

\newcolumntype{L}{>{\raggedright\let\newline\\\arraybackslash\hspace{0pt}}X}
\newcolumntype{R}{>{\raggedleft\let\newline\\\arraybackslash\hspace{0pt}}X}
\newcolumntype{C}{>{\centering\let\newline\\\arraybackslash\hspace{0pt}}X}

\newcommand{\gambitinstitute}[1]{\expandafter\csname #1\endcsname \label{#1}}

\newcommand{\aachen}{Institute for Theoretical Particle Physics and Cosmology (TTK), RWTH Aachen University, D-52056 Aachen, Germany}

\newcommand{\imperial}{Department of Physics, Imperial College London, Blackett Laboratory, Prince Consort Road, London SW7 2AZ, UK}

\newcommand{\monash}{School of Physics and Astronomy, Monash University, Melbourne, VIC 3800, Australia}

\newcommand{\mcgill}{Department of Physics, McGill University, 3600 rue University, Montr\'eal, Qu\'ebec H3A 2T8, Canada}

\newcommand{\desy}{DESY, Notkestra\ss e 85, D-22607 Hamburg, Germany}

\makeatletter

\newcommand{\preprintnumber}[1]{\gdef\@preprintnumber{\begin{flushright}{#1}\end{flushright}}}

\g@addto@macro\bfseries{\boldmath}
\makeatother

\newcommand{\subparagraph}{} 
\usepackage{titlesec}
\titleformat*{\paragraph}{\bfseries}

\journalname{Eur. Phys. J. C}
\smartqed
\let\underscore\_
\renewcommand{\_}{\discretionary{\underscore}{}{\underscore}}

\makeatletter
\let\orgdescriptionlabel\descriptionlabel
\renewcommand*{\descriptionlabel}[1]{%
  \let\orglabel\label
  \let\label\@gobble
  \phantomsection
  \protected@edef\@currentlabel{#1}%
  \let\label\orglabel
  \orgdescriptionlabel{#1}%
}
\makeatother

\newcommand\postnewlinemarker{\hbox{\ensuremath{\hookrightarrow}}}
\newcommand\cpp[1]{{\lstinline!#1!}}  

\newcommand\yaml[1]{{\lstset{style=yaml}\lstinline!#1!\lstset{style=cpp}}}

\newcommand\term[1]{{\lstset{style=terminal}\lstinline!#1!\lstset{style=cpp}}}
\newcommand\fortran[1]{{\lstset{style=fortran}\lstinline!#1!\lstset{style=cpp}}}
\newcommand\py[1]{{\lstset{style=python}\lstinline!#1!\lstset{style=cpp}}}
\newcommand\customtilde{{\raisebox{0.2ex}{\scalebox{0.6}{\boldmath$\sim$}}}}
\newcommand\mathematica[1]{{\lstset{style=Mathematica}\lstinline!#1!\lstset{style=cpp}}}

\lstnewenvironment{lstlistingyaml}{\lstset{style=yaml}}{\lstset{style=cpp}}
\lstnewenvironment{lstlistingterm}{\lstset{style=terminal}}{\lstset{style=cpp}}
\lstnewenvironment{lstlistingfortran}{\lstset{style=fortran}}{\lstset{style=cpp}}
\lstnewenvironment{lstcpp}{\lstset{style=cpp}}{\lstset{style=cpp}}
\lstnewenvironment{lstcppalt}{\lstset{style=cppalt}}{\lstset{style=cpp}}
\lstnewenvironment{lstcppnum}{\lstset{style=cppnum}}{\lstset{style=cpp}}
\lstnewenvironment{lstyaml}{\lstset{style=yaml}}{\lstset{style=cpp}}
\lstnewenvironment{lstterm}{\lstset{style=terminal}}{\lstset{style=cpp}}
\lstnewenvironment{lsttermalt}{\lstset{style=terminalalt}}{\lstset{style=cpp}}
\lstnewenvironment{lsttext}{\lstset{style=text}}{\lstset{style=cpp}}
\lstnewenvironment{lstfortran}{\lstset{style=fortran}}{\lstset{style=cpp}}
\lstnewenvironment{lstpy}{\lstset{style=python}}{\lstset{style=cpp}}
\lstnewenvironment{lstmathematica}{\lstset{style=mathematica}}{\lstset{style=cpp}}

\newcommand{\tmpname}{}
\newcommand{\tmplistingname}{}
\makeatletter
\newif\ifATOlabelname
\lst@Key{labelname}{Listing}{\def\ATOlabelname{#1}\global\ATOlabelnametrue}
\makeatother
\lstnewenvironment{lstcpplabel}[1][]{
  \lstset{style=cpp,#1} 
  \ifATOlabelname
    \renewcommand{\tmpname}{\lstlistingname}
    \renewcommand{\tmplistingname}{\lstlistlistingname}
    \renewcommand{\lstlistingname}{\ATOlabelname}
    \renewcommand{\lstlistlistingname}{List of \lstlistingname s}
  \fi
}{
  \renewcommand{\lstlistingname}{\tmpname}
  \renewcommand{\lstlistlistingname}{\tmplistingname}
  \lstset{style=cpp}
}
\definecolor{solarized@base03}{HTML}{002B36}
\definecolor{solarized@base02}{HTML}{073642}
\definecolor{solarized@base01}{HTML}{586e75}
\definecolor{solarized@base00}{HTML}{657b83}
\definecolor{solarized@base0}{HTML}{839496}
\definecolor{solarized@base1}{HTML}{93a1a1}
\definecolor{solarized@base2}{HTML}{EEE8D5}
\definecolor{solarized@base3}{HTML}{FDF6E3}
\definecolor{solarized@yellow}{HTML}{B58900}
\definecolor{solarized@orange}{HTML}{CB4B16}
\definecolor{solarized@red}{HTML}{DC322F}
\definecolor{solarized@magenta}{HTML}{D33682}
\definecolor{solarized@violet}{HTML}{6C71C4}
\definecolor{solarized@blue}{HTML}{268BD2}
\definecolor{solarized@cyan}{HTML}{2AA198}
\definecolor{solarized@green}{HTML}{859900}
\definecolor{darkred}{HTML}{550003}
\definecolor{darkgreen}{HTML}{00AA00}

\newcommand\YAMLstringstyle{\footnotesize\color{solarized@green}\mdseries}
\newcommand\YAMLkeystyle{\footnotesize\color{solarized@blue}\ttfamily}
\newcommand\YAMLvaluestyle{\footnotesize\color{blue}\mdseries}
\newcommand\ProcessThreeDashes{\llap{\color{cyan}\mdseries-{-}-}}

\newcommand\CPPcommentstyle{\color{solarized@violet}\footnotesize\ttfamily}
\newcommand\CPPdirectivestyle{\color{solarized@magenta}\footnotesize\ttfamily}
\newcommand\termplainstyle{\footnotesize\ttfamily}

\newcommand\processLongMacroDelimiter
{%
\CPPdirectivestyle%
\#define%
}

\lstdefinestyle{cpp}
{
  language=C++,
  basicstyle=\footnotesize\ttfamily,
  basewidth={0.53em,0.44em}, 
  numbers=none,
  tabsize=2,
  breaklines=true,
  escapeinside={@}{@},
  showstringspaces=false,
  numberstyle=\tiny\color{solarized@base01},
  keywordstyle=\color{solarized@orange},
  stringstyle=\color{solarized@red}\ttfamily,
  identifierstyle=\color{solarized@blue},
  commentstyle=\CPPcommentstyle,
  directivestyle=\CPPdirectivestyle,
  emphstyle=\color{solarized@green},
  frame=single,
  rulecolor=\color{solarized@base2},
  rulesepcolor=\color{solarized@base2},
  literate={~} {\customtilde}1,
  moredelim=*[directive]\ \ \#,
  moredelim=*[directive]\ \ \ \ \#
}

\lstdefinestyle{cppalt}
{
  language=C++,
  basicstyle=\footnotesize\ttfamily,
  basewidth={0.53em,0.44em}, 
  numbers=none,
  tabsize=2,
  breaklines=true,
  escapeinside={*@}{@*},
  showstringspaces=false,
  numberstyle=\tiny\color{solarized@base01},
  keywordstyle=\color{solarized@orange},
  stringstyle=\color{solarized@red}\ttfamily,
  identifierstyle=\color{solarized@blue},
  commentstyle=\CPPcommentstyle,
  directivestyle=\CPPdirectivestyle,
  emphstyle=\color{solarized@green},
  frame=single,
  rulecolor=\color{solarized@base2},
  rulesepcolor=\color{solarized@base2},
  literate={~}{\customtilde}1,
  moredelim=**[is][\processLongMacroDelimiter]{BeginLongMacro}{EndLongMacro} 
}

\lstdefinestyle{cppnum}
{
  language=C++,
  basicstyle=\footnotesize\ttfamily,
  basewidth={0.53em,0.44em}, 
  numbers=none,
  tabsize=2,
  breaklines=true,
  escapeinside={@}{@},
  numberstyle=\tiny\color{solarized@base01},
  showstringspaces=false,
  numberstyle=\tiny\color{solarized@base01},
  keywordstyle=\color{solarized@orange},
  stringstyle=\color{solarized@red}\ttfamily,
  identifierstyle=\color{solarized@blue},
  commentstyle=\CPPcommentstyle,
  directivestyle=\CPPdirectivestyle,
  emphstyle=\color{solarized@green},
  frame=single,
  rulecolor=\color{solarized@base2},
  rulesepcolor=\color{solarized@base2},
  literate={~} {\customtilde}1,
  moredelim=*[directive]\ \ \#,
  moredelim=*[directive]\ \ \ \ \#
}

\lstdefinestyle{python}
{
  language=Python,
  basicstyle=\footnotesize\ttfamily,
  basewidth={0.53em,0.44em},
  numbers=none,
  tabsize=2,
  breaklines=true,
  escapeinside={@}{@},
  showstringspaces=false,
  numberstyle=\tiny\color{solarized@base01},
  keywordstyle=\color{blue},
  stringstyle=\color{orange}\ttfamily,
  identifierstyle=\color{darkred},
  commentstyle=\color{purple},
  emphstyle=\color{green},
  frame=single,
  rulecolor=\color{solarized@base2},
  rulesepcolor=\color{solarized@base2},
  literate = {~}{\customtilde}1
             {\ as\ }{{\color{blue}\ as\ \color{black}}}3
}

\lstdefinestyle{fortran}
{
  language=Fortran,
  basicstyle=\footnotesize\ttfamily,
  basewidth={0.53em,0.44em},
  numbers=none,
  tabsize=2,
  breaklines=true,
  escapeinside={@}{@},
  showstringspaces=false,
  numberstyle=\tiny\color{solarized@base01},
  keywordstyle=\color{blue},
  stringstyle=\color{orange}\ttfamily,
  identifierstyle=\color{Periwinkle},
  commentstyle=\color{purple},
  emphstyle=\color{green},
  morekeywords={and, or, true, false},
  frame=single,
  rulecolor=\color{solarized@base2},
  rulesepcolor=\color{solarized@base2},
  literate={~}{\customtilde}1
}

\lstdefinestyle{terminal}
{
  language=bash,
  basicstyle=\termplainstyle,
  numbers=none,
  tabsize=2,
  breaklines=true,
  escapeinside={@}{@},
  frame=single,
  showstringspaces=false,
  numberstyle=\tiny\color{solarized@base01},
  keywordstyle=\color{solarized@orange},
  stringstyle=\color{solarized@red}\ttfamily,
  identifierstyle=\color{black},
  commentstyle=\color{solarized@violet},
  emphstyle=\color{solarized@green},
  frame=single,
  rulecolor=\color{solarized@base2},
  rulesepcolor=\color{solarized@base2},
  morekeywords={gambit, cmake, make, mkdir},
  deletekeywords={test},
  literate = {\ gambit}{{\ }{\color{black}}gambit}7
             {/gambit}{{/}{\color{black}}gambit}6
             {gambit/}{{\color{black}}gambit{/}}6
             {/include}{{/}{\color{black}}include}8
             {cmake/}{{\color{black}}cmake/}6
             {.cmake}{{.}{\color{black}}cmake}6
             {~}{\customtilde}1
}

\lstdefinestyle{terminalalt}
{
  language=bash,
  basicstyle=\footnotesize\ttfamily,
  numbers=none,
  tabsize=2,
  breaklines=true,
  escapeinside={*@}{@*},
  frame=single,
  showstringspaces=false,
  numberstyle=\tiny\color{solarized@base01},
  keywordstyle=\color{solarized@orange},
  stringstyle=\color{solarized@red}\ttfamily,
  identifierstyle=\color{black},
  commentstyle=\color{solarized@violet},
  emphstyle=\color{solarized@green},
  frame=single,
  rulecolor=\color{solarized@base2},
  rulesepcolor=\color{solarized@base2},
  morekeywords={gambit, cmake, make, mkdir},
  deletekeywords={test},
  literate = {\ gambit}{{\ }{\color{black}}gambit}7
             {/gambit}{{/}{\color{black}}gambit}6
             {gambit/}{{\color{black}}gambit{/}}6
             {/include}{{/}{\color{black}}include}8
             {cmake/}{{\color{black}}cmake/}6
             {.cmake}{{.}{\color{black}}cmake}6
             {~}{\customtilde}1
}

\lstdefinestyle{text}
{
  language={},
  basicstyle=\footnotesize\ttfamily,
  identifierstyle=\color{black},
  numbers=none,
  tabsize=2,
  breaklines=true,
  escapeinside={*@}{@*},
  showstringspaces=false,
  frame=single,
  rulecolor=\color{solarized@base2},
  rulesepcolor=\color{solarized@base2},
  literate={~}{\customtilde}1
}

\lstdefinestyle{yaml}
{
  language=bash,
  escapeinside={@}{@},
  keywords={true,false,null},
  otherkeywords={},
  keywordstyle=\color{solarized@base0}\bfseries,
  basicstyle=\footnotesize\color{black}\ttfamily,
  identifierstyle=\YAMLkeystyle,
  sensitive=false,
  commentstyle=\color{solarized@orange}\ttfamily,
  morecomment=[l]{\#},
  morecomment=[s]{/*}{*/},
  stringstyle=\YAMLstringstyle\ttfamily,
  moredelim=**[s][\YAMLkeystyle]{,}{:},   
  moredelim=**[l][\YAMLvaluestyle]{:},    
  morestring=[b]',
  morestring=[b]",
  literate =    {---}{{\ProcessThreeDashes}}3
                {>}{{\textcolor{solarized@red}\textgreater}}1
                {|}{{\textcolor{solarized@red}\textbar}}1
                {\ -\ }{{\mdseries\color{black}\ -\ \negmedspace}}3
                {\}}{{{\color{black} \}}}}1
                {\{}{{{\color{black} \{}}}1
                {[}{{{\color{black} [}}}1
                {]}{{{\color{black} ]}}}1
                {~}{\customtilde}1,
  breakindent=0pt,
  breakatwhitespace,
  columns=fullflexible
}

\lstdefinestyle{mathematica}
{
  language={Mathematica},
  basicstyle=\footnotesize\ttfamily,
  basewidth={0.53em,0.44em},
  numbers=none,
  tabsize=2,
  breaklines=true,
  escapeinside={@}{@},
  numberstyle=\tiny\color{black},
  showstringspaces=false,
  numberstyle=\tiny\color{solarized@base01},
  keywordstyle=\color{solarized@orange},
  stringstyle=\color{solarized@red}\ttfamily,
  identifierstyle=\color{solarized@orange}\ttfamily,
  commentstyle=\color{solarized@gray}\ttfamily,
  directivestyle=\color{solarized@orange}\ttfamily,
  emphstyle=\color{solarized@green},
  frame=single,
  rulecolor=\color{solarized@base2},
  rulesepcolor=\color{solarized@base2},
  literate={~} {\customtilde}1,
  moredelim=*[directive]\ \ \#,
  moredelim=*[directive]\ \ \ \ \#,
  mathescape=true
}

\newcommand{\doublecross}[2]{\hyperref[#2]{\textbf{#1}}}
\newcommand{\doublecrosssf}[2]{\hyperref[#2]{\textbf{\textsf{#1}}}}

\newcommand{\startglossary}{\section{Glossary}\label{glossary}Here we explain some terms that have specific technical definitions in \GB.\begin{description}}
\newcommand{\finishglossary}{\end{description}}

\newcommand{\bcode}{\begin{lstlisting}}
\newcommand{\ecode}{\end{lstlisting}}

\newcommand{\sss}{\scriptscriptstyle}
\newcommand{\ms}{m_{\sss S}}
\newcommand{\msms}{\overline{m}_{\sss S}}
\newcommand{\lhs}{\lambda_{h\sss S}}
\newcommand{\ls}{\lambda_{\sss S}}
\newcommand{\lh}{\lambda_{h}}
\newcommand{\mh}{m_h}
\newcommand{\msmh}{\overline{m}_h}

\newcommand{\MSbar}{$\MSBar$\xspace}
\newcommand{\MSBar}{\overline{MS}}

\newcommand{\ztwo}{$\mathbb{Z}_2$\xspace}
\newcommand{\zthree}{$\mathbb{Z}_3$\xspace}

\newcommand{\gambit}{\textsf{GAMBIT}\xspace}

\newcommand{\darkbit}{\textsf{DarkBit}\xspace}

\newcommand{\specbit}{\textsf{SpecBit}\xspace}
\newcommand{\decaybit}{\textsf{DecayBit}\xspace}
\newcommand{\precisionbit}{\textsf{PrecisionBit}\xspace}
\newcommand{\scannerbit}{\textsf{ScannerBit}\xspace}

\newcommand{\GB}{\gambit}

\newcommand{\ds}{\textsf{DarkSUSY}\xspace}
\newcommand{\darksusy}{\ds}

\newcommand{\micromegas}{\textsf{micrOMEGAs}\xspace}

\newcommand\flexiblesusy{\FlexibleSUSY}
\newcommand\FlexibleSUSY{\textsf{FlexibleSUSY}\xspace}
\newcommand\FlexibleEFTHiggs{\textsf{FlexibleEFTHiggs}\xspace}

\newcommand\SARAH{\textsf{SARAH}\xspace}

\newcommand\pippi{\textsf{pippi}\xspace}

\newcommand\twalk{\textsf{T-Walk}\xspace}
\newcommand\diver{\textsf{Diver}\xspace}
\newcommand\ddcalc{\textsf{DDCalc}\xspace}

\newcommand\beq{\begin{equation}}
\newcommand\eeq{\end{equation}}

\renewcommand{\url}[1]{\href{#1}{#1}}

\newcommand{\sarah}{\textsf{SARAH} }

\newcommand{\softsusy}{\textsf{SOFTSUSY} }

\usepackage{dsfont}

\graphicspath{ {Figures/} }

\begin{document}

\preprintnumber{TTK-18-21, DESY 18-102, CoEPP-MN-18-3}

\title{Impact of vacuum stability, perturbativity and XENON1T on global fits of $\mathbb{Z}_2$ and $\mathbb{Z}_3$ scalar singlet dark matter}
\author
{
Peter Athron\thanksref{inst:a} \and
Jonathan M. Cornell\thanksref{inst:b} \and
Felix Kahlhoefer\thanksref{inst:c} \and
James McKay\thanksref{inst:d,e2} \and
Pat Scott\thanksref{inst:d,e3} \and
Sebastian Wild\thanksref{inst:e}
}

\institute
{ \monash\label{inst:a}\and
 \mcgill\label{inst:b}\and
 \aachen\label{inst:c}\and
 \imperial\label{inst:d}\and
 \desy\label{inst:e}
}

\thankstext{e2}{j.mckay14@imperial.ac.uk}
\thankstext{e3}{p.scott@imperial.ac.uk}

\titlerunning{Vacuum stability, perturbativity and XENON1T vs $\mathbb{Z}_2$ and $\mathbb{Z}_3$ scalar singlet dark matter}
\authorrunning{Athron et.\ al.}

\date{Received: date / Accepted: date}

\maketitle

\begin{abstract}

Scalar singlet dark matter is one of the simplest and most predictive realisations of the WIMP (weakly-interacting massive particle) idea. Although the model is constrained from all directions by the latest experimental data, it still has viable regions of parameter space. Another compelling aspect of scalar singlets is their ability to stabilise the electroweak vacuum. Indeed, models of scalar dark matter are not low-energy effective theories, but can be valid all the way to the Planck scale. Using the GAMBIT framework, we present the first global fit to include both the low-energy experimental constraints and the theoretical constraints from UV physics, considering models with a scalar singlet charged under either a $\mathbb{Z}_2$ or a $\mathbb{Z}_3$ symmetry. We show that if the model is to satisfy all experimental constraints, completely stabilise the electroweak vacuum up to high scales, and also remain perturbative to those scales, one is driven to a relatively small region of parameter space.  This region has a Higgs-portal coupling slightly less than 1, a dark matter mass of 1--2\,TeV and a spin-independent nuclear scattering cross-section around $10^{-45}$\,cm$^2$.

\end{abstract}


\section{Introduction}
\label{intro}

The discovery of the Higgs boson at the LHC makes a strong case for the existence of fundamental scalar particles. This observation immediately raises the question of whether there are other fundamental scalars that may address some of the open problems of particle physics. For example, the Standard Model (SM) of particle physics can be extended by a gauge-singlet scalar field with a stabilising symmetry in order to obtain a dark matter  (DM) candidate \cite{SilveiraZee,McDonald94,Burgess01}. Such a scalar singlet naturally interacts with the SM by coupling to the Higgs field and thus obtains a thermal relic abundance via the freeze-out mechanism. In spite of its simplicity, the model possesses a number of viable parameter regions that are consistent with all experimental constraints. Scalar singlets are therefore arguably the simplest realisation of the idea of weakly-interacting massive particles (WIMPs).

A remarkable feature of the scalar singlet DM model is that, just like the SM, it remains valid up to very high energies --- potentially up to the Planck scale $M_\mathrm{Pl} \sim \mathcal{O}(10^{19})$\,GeV. This is in sharp contrast to many alternative DM models, which are conceived only as effective low-energy theories. In fact, scalar singlets can even resolve a potential problem of the SM at high energies: for the measured values of the Higgs boson and top quark masses, the electroweak vacuum is found to be metastable, because the Higgs quartic coupling becomes negative on scales $\gtrsim \mathcal{O}(10^{15})$\,GeV.  Even though the expected lifetime of the electroweak vacuum state far exceeds the age of Universe, it is an appealing feature of scalar singlet models that the additional coupling between the Higgs and the scalar singlet affects the running of the Higgs quartic coupling at high scales and can prevent it from becoming negative~\cite{Gonderinger2010,Drozd2011,2012JHEP...05..061K,EliasMiro:2012ay,Belanger2013a,2014PhRvD..89a5017G,Khan2014,Alanne2014,Kahlhoefer15,Han2015a,Kanemura2015}.

In this work we present the most comprehensive study of scalar singlet DM to date by combining the information from low-energy observables, such as the relic abundance of scalar singlets and experimental constraints, with a study of the properties of the model at high energies, in particular perturbativity and vacuum stability. For this purpose we use the \GB global fitting package \cite{gambit}, which enables the user to incorporate existing software via a backend system.  Specifically, we use \flexiblesusy \textsf{2.0.1} \cite{Athron:2014yba,Athron:2017fvs} and \SARAH \textsf{4.12.2} \cite{Staub:2009bi,Staub:2010jh,Staub:2012pb,Staub:2013tta} for the renormalisation group evolution needed to study vacuum stability, \ddcalc \textsf{2.0.0} \cite{DarkBit} for DM direct detection, \textsf{gamLike} \textsf{1.0.0} \cite{DarkBit} and \darksusy \textsf{5.1.3} \cite{darksusy,darksusy4} for DM indirect detection with gamma-rays, \micromegas \textsf{3.6.9.2} \cite{Belanger:2013oya} for the relic density calculation, and \diver \textsf{1.0.4} and \twalk \textsf{1.0.1} \cite{ScannerBit} for efficient sampling of the parameter space. This approach makes it possible to study the parameter space relevant for scalar singlets together with a number of nuisance parameters reflecting uncertainties in SM couplings and masses, the DM halo distribution, and nuclear matrix elements important for calculating nuclear scattering cross-sections. Moreover, it is easily possible to extract additional information, such as the scale at which perturbativity is violated and the expected age of the Universe in a metastable scenario.

Most studies in the past have focused on the case where a $\mathbb{Z}_2$ symmetry stabilises the singlet, and makes it a viable DM candidate~\cite{Goudelis09,Yaguna09,Profumo2010a,Arina11,Mambrini11,Djouadi:2011aa,Cheung:2012xb,Cline:2013gha,Endo:2014cca,Craig:2014lda,Urbano:2014hda,Feng15,Duerr15,Duerr16,He:2016mls,Escudero:2016gzx,Han:2016gyy,Ko:2016xwd,Cuoco:2016jqt,Beniwal,SSDM,Ghorbani:2018yfr}.  Perturbativity, vacuum stability, direct detection and the relic density of scalar singlets with a $\mathbb{Z}_3$ symmetry have also been investigated~\cite{Belanger2013a,2017JHEP...10..088B}. The latter case introduces new phenomenology due to an additional cubic $S$ coupling, leading to semi-annihilations.  This annihilation channel can open up regions of parameter space that would otherwise be ruled out by direct detection~\cite{Belanger2013a}, and impact indirect detection by modifying the injection spectra of light particles~\cite{D'Eramo:2010ep}. However, vacuum stability considerations limit the magnitude of the responsible coupling, leading to an interesting interplay of different constraints.  Variants with a residual local $\mathbb{Z}_2$ or $\mathbb{Z}_3$ symmetry, arising from the breaking of a new $U(1)$ symmetry and including an associated $Z'$ boson, have also been studied \cite{2014JCAP...05..047K,2015JCAP...01..023K,2015PhLB..747..255B}.

In this paper, we present an updated global analysis of the model with a global \ztwo symmetry, and carry out the first global fit of the model with a global \zthree symmetry. In particular, we improve on our earlier analysis of the $\mathbb{Z}_2$ model~\cite{SSDM} by treating the singlet self-coupling $\ls$ as a free parameter, including full RGE running, considering vacuum stability and perturbativity, and incorporating the latest results from XENON1T \cite{Aprile:2017iyp,Aprile:2018dbl} and PandaX \cite{Cui:2017nnn}. As we will see, these new direct detection results are particularly relevant, as XENON1T has sufficient sensitivity to probe the most interesting regions of parameter space. Intriguingly, rather than ruling out the model, XENON1T observes an upward fluctuation in their data, which can be interpreted as slight preference for the model that we consider.

We give details of the models in Sec.~\ref{sec:singletdm:model}, of our input parameters and scanning procedure in Sec.~\ref{sec:singletdm:input}, and of our observable calculations and likelihood functions in Sec.~\ref{sec:singletdm:physics}.  The results for the $\mathbb{Z}_2$ and $\mathbb{Z}_3$ models appear in Secs.~\ref{sec:singletdm:z2uv} and \ref{sec:singletdm:z3uv}, respectively. We summarise our findings in Sec.~\ref{sec:singletdm:conc}.

\GB software can be downloaded from \href{https://gambit.hepforge.org}{gambit.hepforge.org}, and all samples, input files and best-fit points from this paper are available from Zenodo \cite{the_gambit_collaboration_2018_1298566}.

\section{Model}
\label{sec:singletdm:model}

\subsection{$\mathbb{Z}_2$-symmetric model}

Let us first consider the case where a real scalar singlet $S$ is stabilised by making the Lagrangian invariant under the $\mathbb{Z}_2$ transformation $S\rightarrow -S$. The most general renormalisable scalar potential permitted by the $\mathbb{Z}_2$, Lorentz and gauge symmetries is then~\cite{Davoudiasl2005}
\begin{align}
\mathrm{V}_{\mathbb{Z}_2} =& \mu_{\sss H}^2 |H|^2 + \frac12 \lh |H|^4 + \frac12\lhs S^2|H|^2 + \frac12 \mu_{\sss S}^2 S^2 \nonumber \\
& + \frac14\ls S^4.
\label{eqn:SS_pot}
\end{align}
The terms proportional to $\mu_{\sss H}^2$ and $\mu_{\sss S}^2$ are the Higgs and singlet bare masses, the terms proportional to $\lh$ and $\ls$ are their quartic self-couplings, and the $S^2|H|^2$ term is the portal coupling that connects the two bosons. As $S$ never obtains a vacuum expectation value (VEV), the singlet extension is fully specified by the three parameters $\mu_{\sss S}^2$, $\lhs$ and $\ls$. After electroweak symmetry breaking we can replace $H \rightarrow \left[0, (v_0+h)/\sqrt{2}\right]^\text{T}$, with $h$ being the SM Higgs field and $v_0 = 246$\,GeV  the VEV of the electroweak vacuum. Then, the portal term proportional to $\lhs$ induces couplings of $h$ to the scalar singlet $S$ via the terms $h^2S^2$ and $v_0hS^2$. Moreover, after symmetry breaking the \MSbar singlet mass is given by
\begin{equation}
\msms = \sqrt{\mu_{\sss S}^2 + \frac12{\lhs v_0^2}}\label{Eq:m_S_MSBar}\ ,
\end{equation}
where $v_0$ is the $\MSBar$ Higgs VEV. The singlet pole mass, $\ms$, can
be obtained from this using
\begin{equation}
\ms^2 = \msms^2 + \mathrm{\Sigma_S},
\label{Eq:m_S_pole}
\end{equation}
where $\mathrm{\Sigma_S}$ represents loop corrections that shift the $\MSBar$ mass to the pole.

The only renormalisable interaction of $S$ with the SM is through the ``Higgs portal'' $S^2H^2$ term.  It is this term that makes it possible to have thermal production of DM in the early Universe.  This portal coupling also provides potential annihilation signals \cite{Yaguna09,Profumo2010a,Arina11}, direct detection and $h\to SS$ decays \cite{Mambrini11}. Notice that for scalar masses less than a few TeV, the couplings $\ls$ and $\lhs$ necessary to explain the DM relic density remain sufficiently small to preserve perturbativity. The scalar field in this model can also feature in theories of inflation \cite{Lerner09,Lebedev:2011aq,Herranen15,Kahlhoefer15} and baryogenesis \cite{Profumo07,Barger09,JimKimmo}.

Several viable parts of the parameter space of the scalar singlet model have yet to be probed, with the DM phenomenology essentially given  by $\ms$ and $\lhs$. Specifically, the parameters that have been identified to be compatible with current experimental data exist in a number of regions~\cite{Cline:2013gha,Beniwal,SSDM}:
\begin{enumerate}
\item A resonance region around $\ms\sim\mh/2$, where in spite of very small couplings ($\lhs \lesssim 10^{-2}$) the singlet can nevertheless account for the entire observed relic abundancce of DM.
\item The resonant ``neck'' region at $\ms=\mh/2$, which can escape detection by the combination of large couplings and an extremely small relic $S$ density.
\item A high-mass region with $\lhs$ of order one.
\end{enumerate}
Eventually, direct detection is expected to probe much of this remaining parameter space, leaving only large values of $\lhs$ at which the theory begins to become non-perturbative \cite{Cline:2013gha} and a small part of the resonance region at $\ms\sim\mh/2$ untested.

Although the relic density and searches at (in)direct detection and collider experiments only probe the mass $\ms$ and the portal coupling $\lhs$, the quartic self-coupling $\ls$ of the scalar singlet does become relevant for the stability of the electroweak vacuum.\footnote{The quartic self-coupling also induces DM self-interactions, which can in princple be constrained by astrophyical observations (e.g.~\cite{Tenkanen16}). However, for the range of singlet masses that we consider, the self-interaction cross section is too small to be observable even for very large values of $\lambda_s$.} The possibility to further enlarge the expected lifetime of the electroweak vacuum or to even render it absolutely stable is an appealing feature of scalar extensions of the SM, and one of the prime motivations for our study of the scalar singlet DM model.

\subsection{$\mathbb{Z}_3$-symmetric model}

The symmetry group that stabilises $S$ is not necessarily $\mathbb{Z}_2$.  We will also consider a complex scalar singlet charged under a $\mathbb{Z}_3$ symmetry, with $S$ transforming as $S\rightarrow e^{2\pi i/3}S$.  This is particularly interesting because, due to the cubic $S^3$ term allowed by this symmetry, it is the simplest DM theory involving semi-annihilations \cite{2010PhLB..683...39H,2009JHEP...01..028H,D'Eramo:2010ep}, i.e.~processes where two DM particles annihilate to an SM particle and another DM particle.\footnote{It is also possible to have an $S^3$ term if the Lagrangian is not symmetric under any $\mathbb{Z}_n$ symmetry.  However, such a model also requires quite some tuning to keep the DM sufficiently metastable so that its lifetime is long compared to the age of the Universe.} The most general scalar potential respecting the $\mathbb{Z}_3$ and SM symmetries is given by
\begin{align}
\mathrm{V}_{\mathbb{Z}_3} =&\mu_{\sss H}^2 |H|^2 + \frac12 \lh |H|^4 + \lhs S^\dagger S|H|^2 +\mu_{\sss S}^2 S^\dagger S\nonumber\\& + \ls (S^\dagger S)^2 + \frac{\mu_3}{2}(S^3+S^{\dagger 3}),
\end{align}
where $S^{\dagger}$ denotes the Hermitian conjugate of $S$.  Unlike the $\mathbb{Z}_2$ model, the scalar is no longer a self-adjoint field.  Instead, we have both $S^*$ and $S$ particles, both of which contribute to the relic abundance.

This model has received significantly less attention than the $\mathbb{Z}_2$-symmetric theory, but has been studied in the context of neutrino masses \cite{2008PhLB..662...49M}, baryogenesis \cite{2018JHEP...02..115K} and in terms of DM phenomenology \cite{Belanger2013a}.  The latter included constraints from vacuum stability and perturbativity along with the relic density, direct detection and invisible Higgs decays.  Singlet masses below $\sim$$53\,$GeV were ruled out by invisible Higgs decays, and the semi-annihilation process was shown to allow the model to avoid direct detection constraints in parts of parameter space where the $\mathbb{Z}_2$ model is excluded.  However, as we will discuss in Section~\ref{sec:singletdm:vs}, vacuum stability sets a limit on $\mu_3$ and thus on the strength of semi-annihilations, so eventually this model also comes within reach of tonne-scale direct detection experiments.

\section{Input parameters and sampling}\label{sec:singletdm:input}

\subsection{Parameters and nuisances}\label{sec:singletdm:params}

In Ref.~\cite{SSDM}, we studied the direct phenomenological implications of the $\mathbb{Z}_2$ symmetric scalar singlet model defined at a low energy scale, without considering renormalisation of the theory, running couplings nor vacuum stability.  In this sense, Ref.~\cite{SSDM} treated the scalar singlet as an effective field theory at the scale of the scalar mass.  In this study, we will go on to examine the implications of considering the scalar singlet as a UV-complete theory.  The input parameters and their required ranges are necessarily different for each of these studies.

The parameters and ranges that we scan over in our fits, along with those that we hold fixed, are presented in Tables~\ref{tab:param}--\ref{tab:fixed_params}.

\begin{table}
\begin{center}
\caption{\label{tab:param} Model parameters that we vary in our fits, as well as the ranges over which we vary them, and the types of priors that we apply to the sampling. The mixed prior for the parameter $\mu_3$ consists of two separate scans.  One scan employs a flat prior between 0 and 1\,GeV and a logarithmic prior from 1\,GeV to 4\,TeV, whereas the other scan employs a flat prior for the full range.}
\begin{tabular}{lccccc}
\hline
Parameter & Minimum & Maximum & Prior \\
\hline
$\lhs$                   & $10^{-4}$ & $\sqrt{4\pi}=3.54$      & log \\
$\ls$                   & $10^{-4}$ & $\sqrt{4\pi}=3.54$      & log  \\
$\ms$ (full-range scan)  & 45\,GeV   & 10\,TeV & log  \\
$\ms$ (low-mass scan)    & 45\,GeV   & 70\,GeV & flat \\
$\mu_3$ ($\mathbb{Z}_3$ model only)    & 0\,GeV   & 4\,TeV & mixed\\
\hline
\end{tabular}
\end{center}
\end{table}

Table \ref{tab:param} gives the parameters of the scalar singlet models and the priors on them that we adopt in our scans.  We carry out two main types of scans: the first considering masses across the entire parameter space, from 45\,GeV to 10\,TeV, and a second focussed on masses at and below the Higgs resonance $\ms\sim \mh/2$, in order to obtain better sampling of this region. Notice that we do not scan over DM masses below 45\,GeV, as this part of parameter space is robustly excluded by the combination of direct detection searches, constraints on the invisible decay width of the Higgs, and the singlet relic density. Note also that although the \MSbar mass $\msms$ (at scale $\msms$) is the actual input parameter in our scans, our effective prior range is defined in terms of the $S$ pole mass, as we scan over a larger range of $\msms$ but apply a cut on the $S$ pole mass after spectrum generation.

To study the phenomenology of a given model, one must be able to compute perturbative expressions, such as pole masses and loop-corrected scattering cross-sections.  We thus demand perturbativity as part of the likelihood analysis, by invalidating points in parameter space where any of the dimensionless coupling parameters exceed $\sqrt{4\pi}$.  The choice of this value is similar to that in other studies \cite{Lerner09,Krauss:2017xpj}.

\begin{table}[tp]
\centering
\caption{Names and ranges of SM, nuclear and halo nuisance parameters that we vary simultaneously with scalar singlet parameters in our fits.  We sample all these parameters using flat priors.} \label{tab:SMparams}
\begin{tabular}{l@{\ }c@{\ }r}
\hline
Parameter & & Value($\pm$Range)\\
\hline
Local DM density & \phantom{$^{\MSBar}$}$\rho_0$\phantom{$^{\MSBar}$} &  0.2--0.8\,GeV\,cm$^{-3}$\\
Mean DM speed &$v_\textrm{mean}$ &240(24) km\,s$^{-1}$\\
Galactic escape speed &$v_\mathrm{esc}$ &533(96) km\,s$^{-1}$\\
Nuclear matrix el. (strange)  & \phantom{$^{\MSBar}$}$\sigma_s$\phantom{$^{\MSBar}$} & $43(24)$\,MeV\\
Nuclear matrix el. (up + down) & \phantom{$^{\MSBar}$}$\sigma_l$\phantom{$^{\MSBar}$} & $50(45)$\,MeV\\
Strong coupling & $\alpha_s^{\MSBar}(m_Z)$      & $0.1181(33)$\\
Higgs $\MSBar$ mass & $\msmh(\msms)$ & 130(50)\,GeV\\
Top pole mass  & \phantom{$^{\MSBar}$}$m_t$\phantom{$^{\MSBar}$}  &  $173.34(2.28)$\,GeV\\
\hline\end{tabular}
\end{table}

In addition to the scalar singlet parameters, we also vary a number of nuclear, SM and astrophysical parameters within their allowed experimental or observational uncertainties.   Table~\ref{tab:SMparams} gives the full ranges of all the nuisance parameters that we consider, along with the central values that we adopt.  We use flat priors for sampling all nuisance parameters, as each of the parameters is well enough constrained that the choice of prior has no effect.

In this paper, where we include renormalisation of the input masses, we trade the Higgs pole mass for the $\MSBar$ mass $\msmh = \sqrt{-2\mu_{\sss H}^2}$, defined at the scale $\msms$.  We then compute the physical pole mass from the input parameters (see discussion in Sec.~\ref{sec:singletdm:spectrum}).  This relationship is affected by radiative corrections from the scalar singlet mass, so the relationship between $\msmh$ and the pole mass is not constant throughout the parameter space, and we must therefore scan a large range for $\msmh$.  The resultant value for the pole mass $m_h$ is constrained by the likelihood function described in Sec.~\ref{sec:singletdm:additional}.

We scan over a range of $\pm3\sigma$ around the best estimates of the strong coupling, top pole mass, nuclear matrix elements, the most probable DM speed in the Milky Way halo $v_\textrm{mean}$, and the Galactic escape velocity at the solar position $v_\textrm{esc}$.  We use the same parameter to control $v_\textrm{mean}$ and the rotation speed, $v_{\text{rot}}$ of the galactic disk, as these can be taken as approximately equal under the assumption of a smooth, spherical DM halo.  We apply a log-normal likelihood to the local DM density $\rho_0$, so we scan an asymmetric range about the central value for this parameter.  Details of the likelihoods that we apply to these parameters, along with references for their central values and measured uncertainties, can be found in Sec.~\ref{sec:singletdm:additional}.

We scan over the nuclear matrix elements and local DM density because they each have a significant impact on direct detection.  The strong coupling and Higgs mass enter into the cross-sections for annihilation and nuclear scattering of $S$ \cite{Cline:2013gha}.  In Ref.~\cite{SSDM}, we included 13 nuisance parameters.  In that study, we determined that varying the masses of the bottom, charm, strange, up and down quarks, the Fermi coupling and the electromagnetic couplings within their experimentally-allowed ranges did not have any significant effect on the results.  Here we therefore fix those parameters (Table \ref{tab:fixed_params}).  On the other hand, including the uncertainties of the local DM velocity profile would have a more important effect.  We therefore also include the most probable DM speed $v_\textrm{mean}$ and the local Galactic escape speed $v_\mathrm{esc}$ as nuisance parameters in our fits here.  This results in a total of 8 nuisance parameters, or 11 and 12 parameters in total for our respective scans of the $\mathbb{Z}_2$ and $\mathbb{Z}_3$ models.

The reduction in the total number of nuisance parameters here compared to Ref.~\cite{SSDM} is also intended to counter-act the increased computational requirements for this global fit.  The likelihood is significantly more demanding of computing resources due to the need to solve the RGEs and compute pole masses, and as a result takes longer to compute.  We have also replaced the relatively small prior on the Higgs pole mass in Ref.\ \cite{SSDM} with a much less constrained $\MSBar$ mass, in order to be able to effectively sample Higgs masses around the observed value across the whole scalar singlet parameter space.  Therefore, although we have fewer nuisance parameters in these global fits, they actually require more computational resources than those of Ref.~\cite{SSDM}.

\begin{table}[tp]
\centering
\caption{Names and values of parameters that we hold fixed in our fits.} \label{tab:fixed_params}
\begin{tabular}{l@{\ }c@{\ }r}
\hline
Parameter & & Fixed value \\
\hline
Electromagnetic coupling & $1/\alpha^{\MSBar}(m_Z)$        & $127.950$       \\
Fermi coupling & \phantom{$^{\MSBar}$}$G_\mathrm{F}$\phantom{$^{\MSBar}$} & $1.1663787\times10^{-5}$ \\
$Z$ boson pole mass & $m_Z$ & 91.1876\,GeV \\
$\tau$ lepton pole mass & $m_\tau$ & 1.77686\,GeV \\
Bottom quark mass & $m_b^{\MSBar}(m_b)$    & $4.18$\,GeV \\
Charm quark mass & $m_c^{\MSBar}(m_c)$ & $1.280$\,GeV \\
Strange quark mass & $m_s^{\MSBar}(2\,\text{GeV})$  & $96$\,MeV\\
Down quark mass & $m_d^{\MSBar}(2\,\text{GeV})$  &   $4.70$\,MeV  \\
Up quark mass & $m_u^{\MSBar}(2\,\text{GeV})$     & $2.20$\,MeV \\
\hline\end{tabular}
\end{table}

Our adopted masses for the $Z$ boson, $\tau$ lepton and the other quarks, as well as our chosen Fermi and electromagnetic couplings, come from the 2017 compilation of the Particle Data Group \cite{PDG17} (Table \ref{tab:fixed_params}).


\subsection{Scanning procedure}\label{sec:singletdm:scans}

Although many directions in parameter space are well constrained, efficient sampling of both the \ztwo and \zthree scalar singlet models still requires sophisticated sampling algorithms.  We scan the parameter space with a differential evolution sampler \diver \cite{ScannerBit}, and an ensemble Markov Chain Monte Carlo (MCMC) \twalk \cite{ScannerBit}.  Both algorithms are particularly well-suited to multi-modal problems in many dimensions.  \diver is an optimiser best suited to mapping the profile likelihood, whereas \twalk is better suited to obtaining the Bayesian posterior.  Using \diver, we first obtain well-sampled profile likelihoods, and make sure to identify all modes of the likelihood surface.  This provides information about the locations able to potentially contribute to the posterior.  We then obtain posterior distributions using \twalk, making sure that it does not fail to identify any of the modes found by \diver.

Sampling the resonance region $\ms \approx \mh/2$ can be challenging when scanning over a large mass range.  We run an additional \diver scan in this region to ensure sufficient sampling, employing a flat prior between $\ms=45$ and 70\,GeV.  The ``neck'' part of the resonance is even more difficult; here we perform a third, even more focussed scan, excluding any points with $S$ pole masses not in the range $\ms\in [61.8,63.1]$\,GeV.

We repeat all scans of the \zthree model with both flat and log priors on $\mu_3$, as the choice of prior on this parameter can have a significant impact on the completeness with which the profile likelihood of this model is sampled.

We perform identical scans with and without the requirement that the singlet fully stabilises the electroweak vacuum.  With this requirement imposed, models with a metastable electroweak vacuum (such as the SM) are excluded.  Although such models are not physically invalid, the ability to make the electroweak vacuum absolutely stable is theoretically appealing. We do not perform a scan over the low-mass range with this additional constraint, as it is ruled out except for the very top of the neck region ($\lhs \gtrsim 0.2$).  We also carry out additional scans with \diver using the older 2017 XENON1T constraint \cite{Aprile:2017iyp} instead of the recent 2018 result \cite{Aprile:2018dbl}, for comparison.

The population and convergence settings that we use for each sampler are given in Table~\ref{table:scanner_params}.  These settings are based on a series of extensive tests and optimisations \cite{ScannerBit}.  The \diver scans that we present here each used 3400 Intel Xeon Phi 7250 (Knights Landing) cores, for approximately 86\,hr in total across all scans.  For the 6 \twalk scans, instead of using a fixed tolerance associated with the \cpp{sqrtR} parameter, we found that more reliable sampling could be obtained in the current study by simply running on 1360 cores and halting scans after 23\,hr, using the \cpp{timeout_mins} parameter newly implemented in \twalk \textsf{1.0.1}.

The posteriors that we show come from the \twalk scans only.  Our profile likelihood plots are based on the final merged set of samples from all scans with common physical requirements and likelihoods.  This includes both \diver scans and any relevant \twalk scans, any targeted low-mass or neck scans, and scans with different priors on $\mu_3$.  Without the requirement of absolute vacuum stability, our final profile likelihoods (i.e.\ including XENON1T 2018 results) are based on a total of $4.9\times10^7$ and $1.9\times10^8$ samples for the $\mathbb{Z}_2$ and $\mathbb{Z}_3$ models, respectively. With the requirement of absolute vacuum stability, the profile likelihoods are based on $3.3\times10^7$ and $6.4\times10^7$ samples for the $\mathbb{Z}_2$ and $\mathbb{Z}_3$ models, respectively.

We produce posteriors and profile likelihoods with \pippi \cite{pippi}, basing our posteriors on the maximum posterior density requirement.

\begin{table}[tp]
\caption{Sampling parameters for global fits of the $\mathbb{Z}_2$- and $\mathbb{Z}_3$-symmetric scalar singlet models in this paper.}\label{table:scanner_params}
\centering
\begin{tabular}{l l l l}
\hline
Scanner & Parameter & Full range & Low mass\\
\hline
\diver & \cpp{NP} & 50,000 & 50,000 \\
  & \cpp{convthresh} & $10^{-4}$  & $10^{-5}$\\
  \hline
\twalk & \cpp{chain_number} &$ 3405$ &  \\
  & \cpp{timeout_mins} & 1380 \\
  & \cpp{sqrtR} & $<1.01$ &  \\
\hline\end{tabular}
\end{table}


\section{Physics framework \& likelihood details}\label{sec:singletdm:physics}

\subsection{Pole masses and $\MSBar$ parameters}\label{sec:singletdm:spectrum}

To investigate vacuum stability and perturbativity, and to
calculate observables contributing to the likelihood, we require pole masses and
running parameters consistent with known SM data.  We obtain these using the two-loop RGEs of \flexiblesusy \textsf{2.0.1} \cite{Athron:2014yba,Athron:2017fvs}, via
the \specbit \cite{SDPBit} interface within \gambit \cite{gambit}.  \flexiblesusy
uses \sarah \cite{Staub:2009bi,Staub:2010jh,Staub:2012pb,Staub:2013tta}, along with parts of \softsusy \cite{Allanach2002,Allanach:2013kza} and other higher-order corrections \cite{Athron:2016fuq, Chetyrkin:1999qi, Melnikov:2000qh, Chetyrkin:2000yt, Degrassi:2012ry, Martin:2014cxa, Bednyakov:2013eba,
Buttazzo2013, arXiv:1504.05200}.

As we vary the scalar singlet mass over two orders of
magnitude, we use the \textsf{EFTHiggs} mode of \FlexibleSUSY, which
uses the algorithm developed in Ref.~\cite{Athron:2016fuq} and
refined in Ref.~\cite{Athron:2017fvs}.  This implements a matching and running procedure for effective field
theories, which is appropriate when
$\msms \gg m_t$, while not compromising the precision of the Higgs
pole mass calculation (due to normal EFT uncertainties from missing
$\mathcal{O}(p^2 /\msms^2)$ terms) when $\msms$ is close to $m_t$.

We use two-loop SM RGEs between the electroweak scale and
the scale of new physics.  We take the scale of new physics to be the scalar singlet running mass $\msms$. At
$\msms$ we perform a matching between the SM and the
scalar singlet DM model\footnote{When $\msms < m_t$, we instead set the
matching scale to $m_t$, to avoid matching to the UV
theory below the scale where we extract SM parameters.} using the \FlexibleEFTHiggs matching conditions given in
Ref.~\cite{Athron:2017fvs}.  At scales larger than $\msms$, we use two-loop
RGEs for the scalar singlet model.

For the $\mathbb{Z}_2$ model, the inputs to the \FlexibleSUSY spectrum generator are the $\MSBar$
Lagrangian parameters $\lhs$, $\ls$, $\mu_{\sss H}$ and $\mu_{\sss S}$, defined
at the renormalisation scale $Q=\msms$.  For the $\mathbb{Z}_3$-invariant version, this parameter set is
extended to include $\mu_3(\msms)$.  The parameters $\lhs(\msms)$, $\ls(\msms)$ and
$\mu_3(\msms)$ are obtained directly from the model parameters sampled
by \scannerbit, while $\mu_{\sss H}(\msms)$ and $\mu_{\sss S}(\msms)$ are
obtained by inverting $\msmh^2 = -2\mu_{\sss H}^2$ and Eq.~\ref{Eq:m_S_MSBar}
respectively.\footnote{In the inversion of Eq.~\ref{Eq:m_S_MSBar} we
approximate the $\MSBar$ VEV as $v_0^2 = \frac{1}{\sqrt{2}
G_\mathrm{F}}$, where $G_\mathrm{F}$ is the Fermi constant from
Table \ref{tab:fixed_params}. This means we are effectively making a
very mild approximation in the prior for $\msms$, but no such
approximation is made in the spectrum calculation, and the impact on
the result is negligible.} \FlexibleSUSY fixes the remaining $\MSBar$ parameter
$\lh$ to ensure correct electroweak symmetry
breaking.  It also accepts additional input data in the form of an \texttt{SMINPUTS} block, as defined in the second SUSY Les
Houches Accord (SLHA2) \cite{Allanach:2008qq}. These
are given in Tables \ref{tab:SMparams} and \ref{tab:fixed_params}.

Once \FlexibleSUSY has determined a consistent set of \MSbar
parameters, it computes the singlet and Higgs pole masses, using
\begin{equation}
\mh^2 = -2\mu_{\sss H}^2 + \mathrm{\Sigma_H}
\end{equation}
and Eq.~\ref{Eq:m_S_pole}.  Here $\mathrm{\Sigma_S}$ and $\mathrm{\Sigma_H}$ incorporate the full one-loop self-energies generated with the help of
\sarah \textsf{4.12.2} \cite{Staub:2009bi,Staub:2010jh,Staub:2012pb,Staub:2013tta}.  $\mathrm{\Sigma_H}$ additionally includes all two-loop contributions from dominant orders,
$\mathcal{O}(\alpha_t^2 + \alpha_t \alpha_s)$. Note that in this approach,
we obtain the Higgs pole mass as an output rather than an input
parameter, and scan the parameter space by varying the input $\MSBar$
mass $\msmh$.  As the value of the scalar singlet mass can
have a significant impact on the relationship between
$\msmh(\msms)$ and $m_h$, we allow $\msmh$ to vary from
80--180\,GeV.  This is sufficient to permit a value of
$\msmh$ that can give a 125\,GeV pole mass
throughout the scalar singlet parameter space. We penalise all other points
using a Gaussian likelihood centred on the experimentally-measured mass ($\mh = 125.09 \pm 0.24$\,GeV; see Sec.~\ref{sec:singletdm:additional}).

Note that in this setup neither the Higgs pole mass $m_h$, nor the
quartic coupling $\lh$ are inputs.  Nonetheless, $m_h$ is well constrained by the likelihood, so it is important
that we can calculate both $\mh$ and $\lh$ consistently using the procedure described above.

\subsection{Vacuum stability and perturbativity}\label{sec:singletdm:vs}

With the running \MSbar parameters of the scalar singlet model
obtained as described in the previous section, it is now possible to
run the couplings from the electroweak scale to the Planck scale,
$M_{\text{Pl}} = 1.22\times 10^{19}$ GeV and test for vacuum
stability.

We classify the stability of the electroweak vacuum in three possible ways:
\begin{itemize}
\item[]\hspace{-5mm}\textbf{Stable.} If $\lh(Q)>0$ for all $Q<M_{\text{Pl}}$, the electroweak vacuum is the global minimum of the Higgs potential for all $Q$ up to the Planck scale, and is therefore absolutely stable with respect to quantum fluctuations.
\item[]\hspace{-5mm}\textbf{Metastable.} If $\lambda(Q_0)<0$ for any $Q_0<M_{\text{Pl}}$, but the electroweak vacuum has an expected lifetime that exceeds the age of the Universe.
\item[]\hspace{-5mm}\textbf{Unstable.} If $\lambda(Q_0)<0$ for any $Q_0<M_{\text{Pl}}$ and the electroweak vacuum has an expected lifetime that is less than the age of the Universe.
\end{itemize}

To distinguish between the latter two cases, and incorporate a
likelihood penalty associated with vacuum decay, one should calculate
the decay rate of the electroweak vacuum. A detailed description of
how to obtain the decay rate and estimate the probability that the
vacuum would decay within the age of the Universe can be found in
section 2.5 of Ref.~\cite{SDPBit} and references therein.

At large field values, the potential can be approximated as $V \approx \frac14 \lh|H|^4$.
This can be used to find the so-called ``bounce'' solution to the
Euclidean equation of motion, and obtain the bounce
action \cite{Coleman1977}
\begin{align}
B=\frac{8\pi^2}{3\left|\lh\right|}.
\end{align}
The rate of bubble nucleation per unit volume per unit time can be estimated from the bounce action using
\begin{align}
\Gamma \approx \Lambda_B^4 e^{-B}\label{Eqn:BubNucRate},
\end{align}
where $\Lambda_B$, the scale at which $\lh$ is minimised, has
been introduced following Ref.~\cite{Sher1989}.  As we are
interested in the probability that the Universe would have decayed in our
past light cone, we introduce the lifetime of the Universe,
$T_U \approx e^{140} / M_{\text{Pl}}$, and use it to define the volume of the past lightcone, $T_U^4$.  The predicted number of decays in our past lightcone is therefore $T_U^4\Lambda_B^4e^{-B}$.  We can hence define a likelihood contribution for no decay having occurred in our past light cone as
\begin{align}
\mathcal{L}=\exp\left[-\left(e^{140}\frac{\Lambda_B}{M_{\textrm{Pl}}}\right)^4\exp\left({-\frac{8\pi^2}{3\left|\lh(\Lambda_B)\right|}}\right)\right],	\label{eqn:prob2}
\end{align}
based on the Poisson probability of the Universe having
decayed out of the electroweak vacuum by the present day.

The actual predicted lifetime in years is
\begin{equation}
\frac{\tau}{\text{yr}}  = 2.09\times 10^{-32} \left(\frac{\text{GeV}}{r}\right), \label{eqn:lifetime}
\end{equation}
where the rate $r$ is given by,
\begin{align}
r \approx T_U^3\Lambda_B^4 e^{-B}.
\end{align}
For the SM, this gives a predicted lifetime of $\sim$$1.1\times 10^{99}$ years.

Because the dominant contribution from a scalar singlet to the running of $\lh$ is always positive, the electroweak vacuum can only become more stable in the models we consider in this paper than it is in the SM.  As the probability of vacuum decay is already very small even in the metastable SM, the effect of going from a metastable vacuum to an absolutely stable one has a negligible impact on the composite likelihood.\footnote{For this reason, in the summary above, we make a number of simple approximations that are standard in the literature. For a discussion of these and references to more precise calculations see section 2.5 of Ref.\ \cite{SDPBit}.} However, because the scenario of absolute stability is theoretically appealing, we repeat our global fits with the strict condition that all models must be absolutely stable, invalidating all parameter combinations that give a metastable vacuum.

In the $\mathbb{Z}_3$ model, low-scale vacuum stability gives an additional constraint on the $\mu_3$ parameter.  If $\mu_3$ is large, the scalar potential can posess $\mathbb{Z}_3$-breaking minima already at the weak scale, which would be degenerate with or deeper than the SM vacuum.  This can be avoided by placing an upper bound on the $\mu_3$ parameter.  We adopt the condition given in Ref.~\cite{Belanger2013a} for an absolutely stable SM vacuum, as an upper limit on $\mu_3$:
\begin{align}
\mu_3 \leq 2\sqrt{\ls}\ms \, . \label{eqn:ew_stability_condition}
\end{align}
This constraint can be relaxed slightly by allowing for the possibility of a $\mathbb{Z}_3$-breaking minimum with a lower potential energy than the SM vacuum, but an SM vacuum with a decay half-life longer than the age of the Universe (see Ref.~\cite{Belanger2013a}).  We do not consider this possibility, as part of our interest in studying scalar singlet DM, particularly in this global fit, is the appeal of removing metastability from the SM altogether.

We will also require that the scalar singlet couplings remain positive, such that the scalar singlet potential is bounded from below.  This means that we can isolate our study of the electroweak vacuum to the Higgs dimension only.  This analysis neglects the possibility of a second minimum forming in the $S$ direction of the potential, which is possible when $\mu_S^2<0$ and $\lhs$ is sufficiently large \cite{Profumo2010a}.  Due to the nature of the RGEs for the dimensionless scalar couplings, $\lhs$ and $\ls$, these couplings only grow with scale.

We let $\Lambda_P$ denote the scale where the dimensionless couplings become larger than our upper bound for perturbativity $\sqrt{4\pi}\approx 3.54$.  If $\Lambda_P<\ms$, we invalidate the point; otherwise, we record the scale $\Lambda_P$ for later analysis.

There is an important caveat to our definition of vacuum stability and how we apply this as a constraint on the parameter space.
In many cases, increasing the values of the dimensionless couplings in the scalar singlet sector ($\lhs$ and $\ls$) results in the theory becoming non-perturbative at energy scales as low as the electroweak scale.  Because perturbation theory is no longer applicable in this case, we cannot compute the running of the quartic Higgs coupling to the typical scales of instability, so our analysis does not encounter a minimum and thus renders the electroweak vacuum ``stable''.  Such parameter combinations therefore pass the test for stability.  This caveat is acceptable, because such models can still be filtered out (if desired) based on the extremely low scale at which perturbativity is broken, as given by $\Lambda_P$.  Nevertheless, it is important to consider the order in which we apply these constraints when interpreting the results in Sections \ref{sec:singletdm:z2uv} and \ref{sec:singletdm:z3uv}.

\begin{figure}[t!]
\includegraphics[width=1\columnwidth]{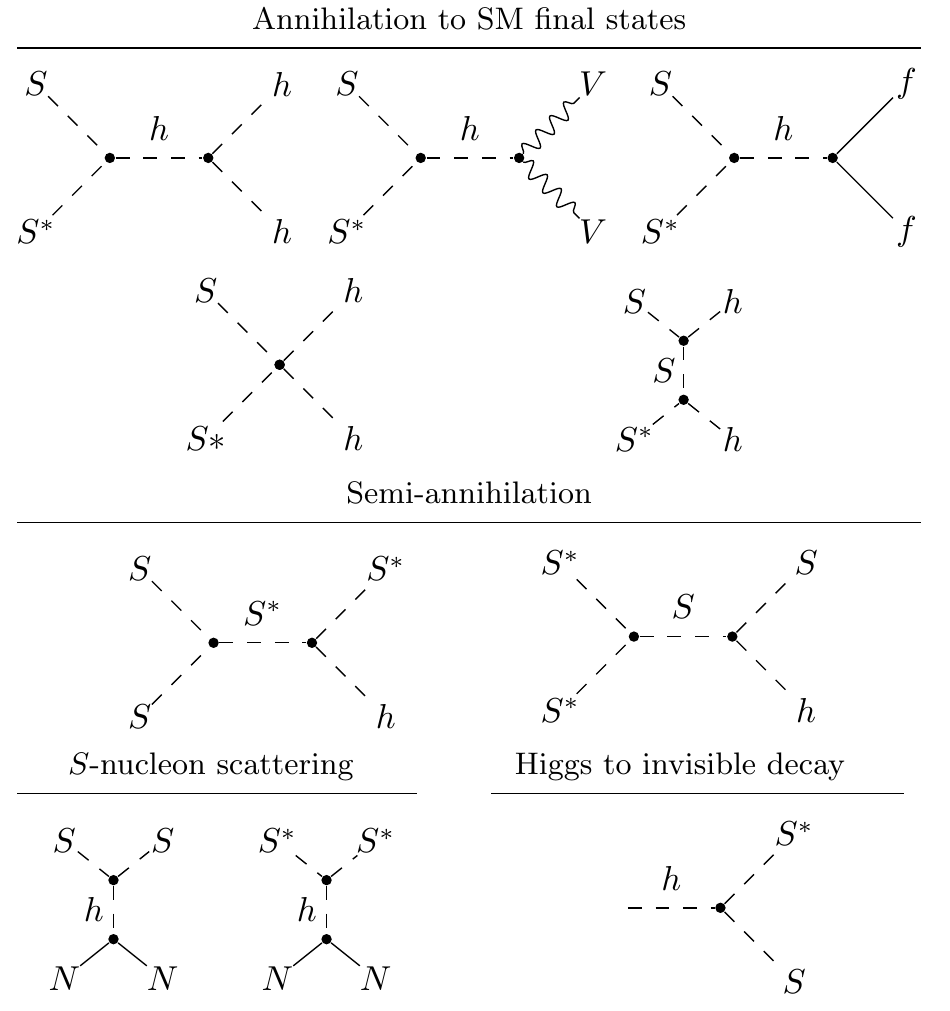}
\caption[The diagrams for annihilation, semi-annihilation, scalar-nucleon scattering and Higgs invisible decays in the $\mathbb{Z}_3$ scalar singlet model.]{The diagrams for annihilation, semi-annihilation, scalar-nucleon scattering and Higgs invisible decays in the $\mathbb{Z}_3$ scalar singlet model.  Here $N$ denotes nucleons, $f$ are SM fermions and $V$ SM gauge bosons.  Except for the semi-annihilation processes, the equivalent diagrams apply in the $\mathbb{Z}_2$ scenario but with $S^*$ replaced with $S$.} \label{fig:diagrams}
\end{figure}

\subsection{Relic density}\label{sec:singletdm:rd}

In the early Universe, scalar singlet DM would have been in thermal equilibrium with SM particles.  The annihilation processes in the top two rows of Fig.~\ref{fig:diagrams} would have occurred frequently compared to the Hubble expansion rate. As the Universe expanded and cooled, the density of the scalar fields would have decreased, making forward annihilation reactions extremely rare and causing the DM abundance to freeze out.  To find this abundance, we solve the Boltzmann equation \cite{Gondolo:1990dk}
\begin{equation}
\frac{dn_{\sss S}}{dt}+3Hn_{\sss S}=-\langle\sigma v_\mathrm{rel}\rangle\left(n_{\sss S}^2-n_{{\sss S},\mathrm{eq}}^2  \right)\,. \label{eqn:boltzmann1}
\end{equation}
Here $H$ is the Hubble rate, $\langle\sigma v_\mathrm{rel}\rangle$ is the thermal average of the relative velocity of DM particles times their self-annihilation cross-section, the number density of DM is given by $n_{\sss S}$, and its equilibrium number density by $n_{{\sss S},\mathrm{eq}}$.

When semi-annihilation processes are possible, as in the $\mathbb{Z}_3$ scalar singlet model, then Eq.~(\ref{eqn:boltzmann1}) must be modified.  The tree-level semi-annihilation processes, where two DM particles can annihilate to a DM particle and an SM particle, are shown in Fig.~\ref{fig:diagrams}.  In the $\mathbb{Z}_3$ model, the relic abundance consists of equal parts $S^*$ and $S$, as each annihilation processes requires both an $S$ and $S^*$, and the semi-annihilation process can occur equally rapidly via $SS\rightarrow S^*h$ and $S^*S^*\rightarrow Sh$.  We can therefore treat $S$ and $S^*$ as the same particle in the Boltzmann calculation, by including a factor of $1/2$ \cite{Belanger2013a}.
\begin{align}
\begin{split}
\frac{dn_{\sss S}}{dt}+3Hn_{\sss S}=&-\langle\sigma v_\mathrm{rel}\rangle\left(n_{\sss S}^2-n_{{\sss S},\mathrm{eq}}^2  \right)\\&-\frac{1}{2}\langle\sigma v_\mathrm{rel}\rangle_{SS\rightarrow hS}\left(n_{\sss S}^2-n_{\sss S}n_{{\sss S},\mathrm{eq}}  \right)\,, \label{eqn:boltzmann2}
\end{split}
\end{align}
where  $\langle\sigma v_\mathrm{rel}\rangle$ is the thermally averaged self-annihilation cross-section without semi-annihilations, and $\langle\sigma v_\mathrm{rel}\rangle_{SS\rightarrow hS}$ is the equivalent for the semi-annihilation channel.  We define a semi-annihilation fraction
\begin{align}
\alpha=\frac{1}{2}\frac{\langle\sigma v_\mathrm{rel}\rangle_{SS\rightarrow hS}}{\langle\sigma v_\mathrm{rel}\rangle+\frac{1}{2}\langle\sigma v_\mathrm{rel}\rangle_{SS\rightarrow hS}},\label{eqn:sa_fraction}
\end{align}
which we record for each sampled point in the \zthree parameter space.  To deal with semi-annihilations in the $\mathbb{Z}_3$ model, we compute $\Omega_{\sss S} h^2$ using \micromegas \textsf{3.6.9.2} \cite{Belanger:2013oya}, with the setting \yaml{fast = true}.

As in Ref.~\cite{SSDM}, we employ the measured relic density $\Omega_\text{DM} h^2 = 0.1188\pm 0.0010$~\cite{Planck15cosmo} as an upper limit, allowing models where the $S$ relic abundance indicates that it is only a fraction of the observed DM.  We use a marginalised Gaussian upper limit likelihood \cite{gambit} for this purpose, adopting the default 5\% theoretical uncertainty offered by \darkbit and combining it in quadrature with the uncertainty on the measured value.  We self-consistently rescale all direct and indirect signals for the thermal $S$ relic density at each point in the parameter space.

\subsection{Direct detection}\label{sec:singletdm:dd}

Scalar singlet DM is strongly constrained by limits on the DM-nucleon scattering cross-section from direct detection.  The corresponding tree-level processes are represented in the bottom left diagrams of Fig.~\ref{fig:diagrams}.

We apply direct detection constraints using \darkbit, drawing on the \ddcalc \cite{DarkBit} implementations of the experimental results of LUX \cite{LUXrun2}, PandaX \cite{Tan:2016zwf,Cui:2017nnn} and XENON1T \cite{Aprile:2017iyp,Aprile:2018dbl}. We emphasise that since all three experiments have similar sensitivity, a consistent combination of the respective likelihoods is essential to infer accurate constraints on the parameter space.

For a given experiment, the likelihood of observing $N$ direct detection events, given a predicted number of signal events $N_{\mathrm{p}}$, follows a Poisson distribution
\begin{equation} \label{eqn:Poisson}
  \mathcal{L}(N|N_{\mathrm{p}})
  = \frac{(b+N_{\mathrm{p}})^{N} \, e^{-(b+N_{\mathrm{p}})}}{N!} \; ,
\end{equation}
where $b$ denotes the expected number of background events within the analysis region.  We interpolate between values in pre-calculated tables contained in \ddcalc in order to determine the detector efficiencies and acceptance effects.  The likelihood in Eq.~(\ref{eqn:Poisson}) is then obtained by recasting the experimental results contained in \ddcalc \cite{DarkBit} for each experiment.

In particular, for the recent XENON1T results \cite{Aprile:2018dbl} we employ the data collected within the core mass of 0.65\,t, which in comparison to the full 1.3\,t dataset has a substantially smaller level of surface and neutron background events. We take the total detection efficiency as a function of the recoil energy from Ref.\ \cite{Aprile:2018dbl}, weighting it by an additional factor $0.65 /1.3$. Moreover, we only consider events within the reference region defined as the area between the median of the nuclear recoil band and the $2\sigma$ quantile, leading to an additional factor 0.475 in the detection efficiency. In our analysis, we then divide the events into two energy bins, based on a separation of the S1 signal into the intervals $[3\,\text{PE},\,35\,\text{PE}]$ and  $[35\,\text{PE},\,70\,\text{PE}]$. To this end, we convert a given nuclear recoil energy into an expected S1 signal using Fig.~3 of Ref.\ \cite{Aprile:2018dbl}, and determine its probability to fall in either of the S1 bins by assuming S1 to be Poisson distributed. Furthermore, we assume that the electron recoil background is constant in S1, while we take the energy dependence of the neutron background from Ref.\ \cite{Aprile:2015uzo}. Assuming for simplicity that the remaining (sub-dominant) background contributions fall into the first energy bin, this gives 0.46  and 0.34 expected background events for the lower and upper energy bins of our analysis, respectively, compared to 0 and 2 observed events. With these assumptions, we obtain a 90\% CL upper bound on the spin-independent scattering cross-section of DM in good agreement with the published bound, and also reproduce the slight preference (less than $2 \sigma$) for a non-zero cross-section at large DM masses.

\subsection{Indirect detection}\label{sec:singletdm:id}

Searches for anomalous gamma-ray emission in dwarf spheroidal galaxies constrain the DM annihilation cross-section.  The expected flux of gamma rays is
\begin{equation}
  \Phi_i = \sum_j\frac{\langle\sigma v\rangle_{0,j}}{8\pi \ms^2}\int_{E_\text{\text{min},i}}^{E_{\text{max},i}} dE \,
  \frac{dN_{\gamma,j}}{dE} \;,
  \label{eqn:gamLikePhi}
\end{equation}
for an energy bin of width $\Delta E_i \equiv E_{\text{max},i} - E_{\text{min},i}$, where $\langle\sigma v\rangle_{0,j}\equiv \sigma v_j|_{v\to0} \equiv \sigma v_j|_{s\to4\ms^2}$ is the partial annihilation cross-section into final state $j$ in the zero-velocity limit, and $dN_{\gamma,j}/dE$ is the differential photon multiplicity for annihilations into the $j$th final state.

We use a combination of analytic expressions from Ref.~\cite{Cline:2013gha} and \micromegas to compute the annihilation and semi-annihilation cross-sections for indirect detection. At tree level, the zero temperature annihilation cross-section for a pair of scalar singlet particles to SM states, $\langle\sigma v\rangle_{0}$, is given by the processes in the top two rows of Fig.~\ref{fig:diagrams} for the $\mathbb{Z}_3$ model, and equivalent processes in the $\mathbb{Z}_2$ model with $S=S^*$. The \textit{effective} cross-sections for annihilation to SM final states in the $\mathbb{Z}_3$ model are a factor of two \textit{smaller} than in the $\mathbb{Z}_2$ model, accounting for the fact that only particle-antiparticle pairs can annihilate. We compute these by scaling the $\mathbb{Z}_2$ cross-sections down by a factor of two. We obtain the semi-annihilation cross-section directly from \micromegas, with there being no equivalent in the $\mathbb{Z}_2$ model.

With the necessary cross-sections computed we then obtain the predicted spectrum ${dN_{\gamma}}/{dE}$ for each model point by using a Monte-Carlo showering simulation, detailed in Ref.~\cite{DarkBit}.  We then use this to compute a combined likelihood for all the dwarf spheroidals in the \textit{Fermi}-LAT six-year \texttt{Pass 8} dataset \cite{LATdwarfP8}.  The details of this likelihood are given in Ref.~\cite{SSDM}.

\subsection{Higgs invisible width}\label{sec:singletdm:inv}

When $\ms<\mh/2$, the Higgs may decay to two $S$ bosons (Fig.~\ref{fig:diagrams}).  The resulting $S$ bosons would be invisible at the LHC, so they would be identified as a missing contribution to the total decay width. For a model with a $\mathbb{Z}_3$-charged scalar, the decay width of the Higgs to $S$ bosons is
\begin{equation}
  \Gamma^{\mathbb{Z}_3}_{h\rightarrow{\sss SS^*}} = \frac{\lhs^2 v_0^2}{16\pi \mh}\left(1 -4 \ms^2/\mh^2\right)^{1/2},
  \label{Gamma_SS}
\end{equation}
where $v_0$ is the Higgs VEV.  In the $\mathbb{Z}_2$ model the final states are identical, so we must include a symmetry factor of $1/2$ to avoid double counting,
\begin{align}
\Gamma^{\mathbb{Z}_2}_{h\rightarrow{\sss SS}} =   \frac{1}{2}\Gamma^{\mathbb{Z}_3}_{h\rightarrow{\sss SS^*}}\, .
  \label{Gamma_SS_2}
\end{align}
Eqs.~(\ref{Gamma_SS}) and (\ref{Gamma_SS_2}) show that constraints on the Higgs invisible width exclude large $\lhs$ for small singlet masses.

\begin{figure*}[tbp]
\centering
\includegraphics[width=1\columnwidth]{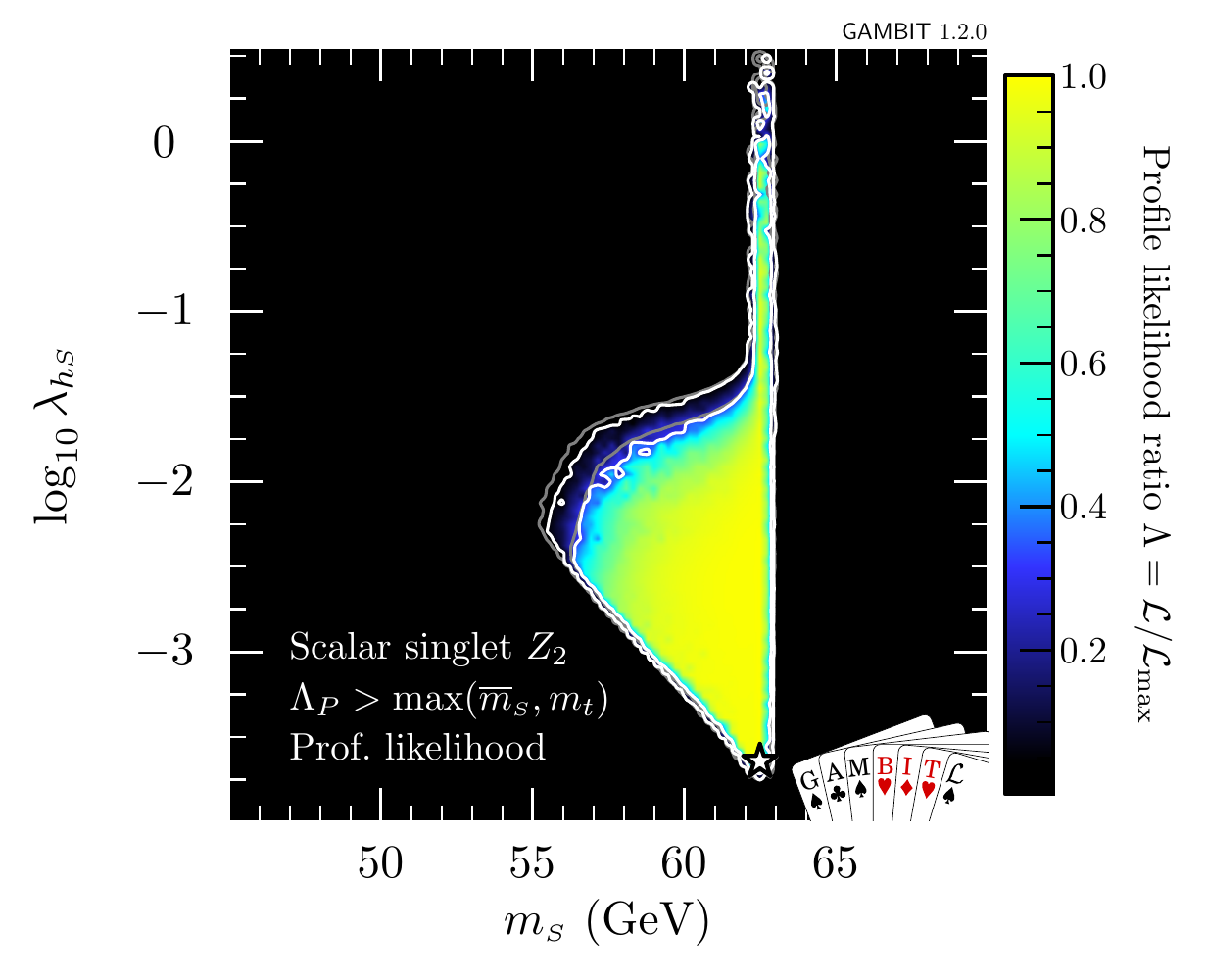}
\includegraphics[width=1\columnwidth]{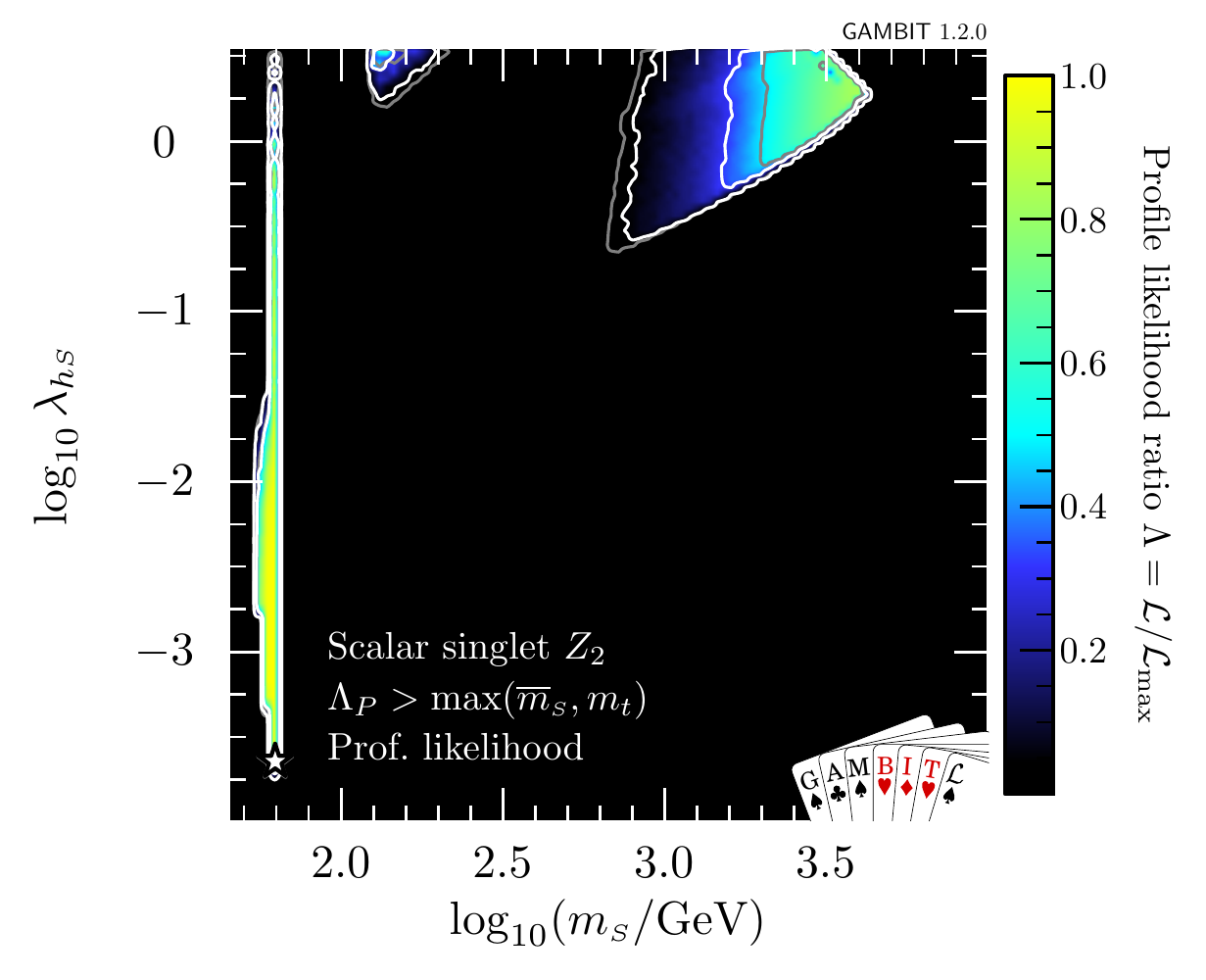}\\
\includegraphics[width=1\columnwidth]{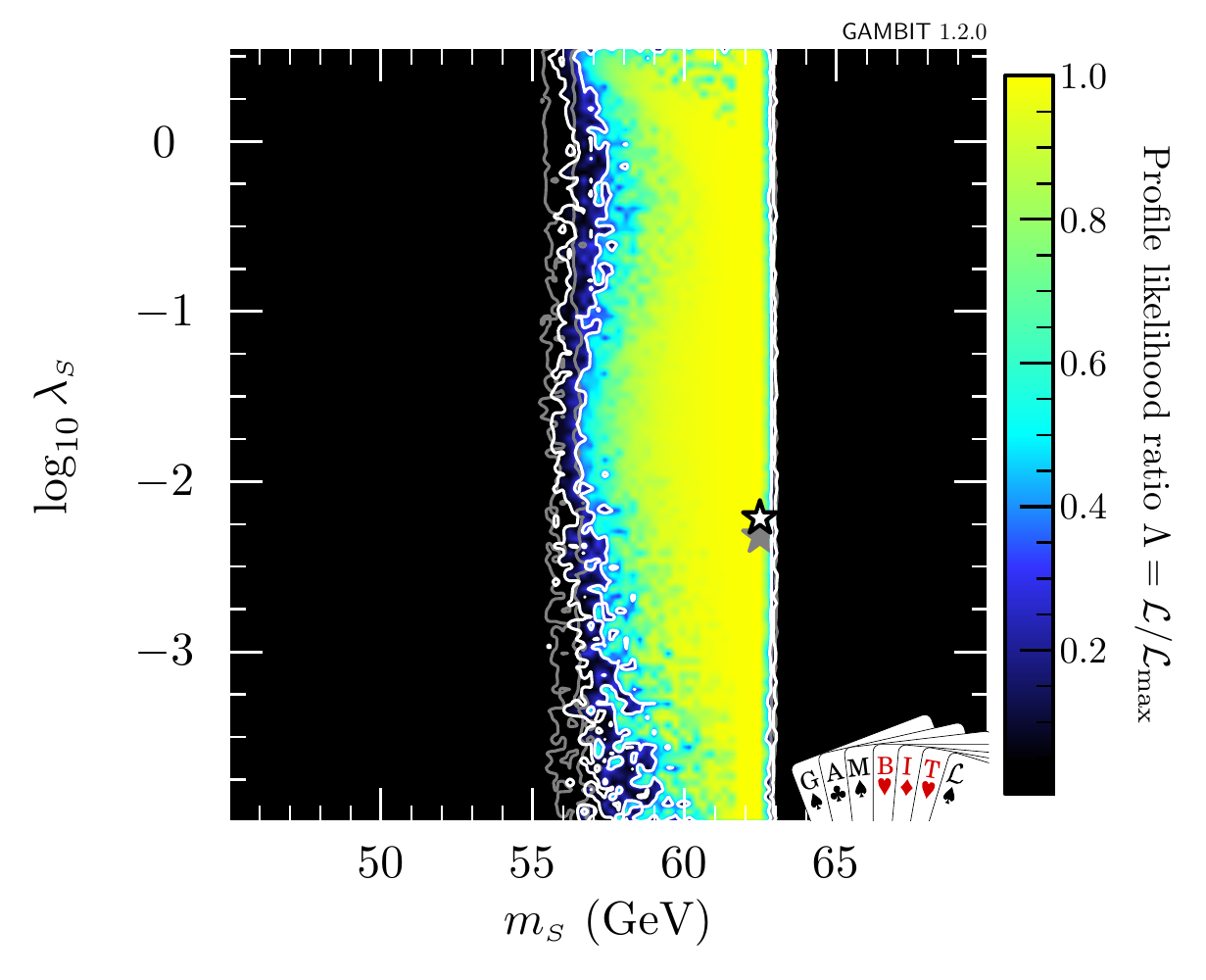}
\includegraphics[width=1\columnwidth]{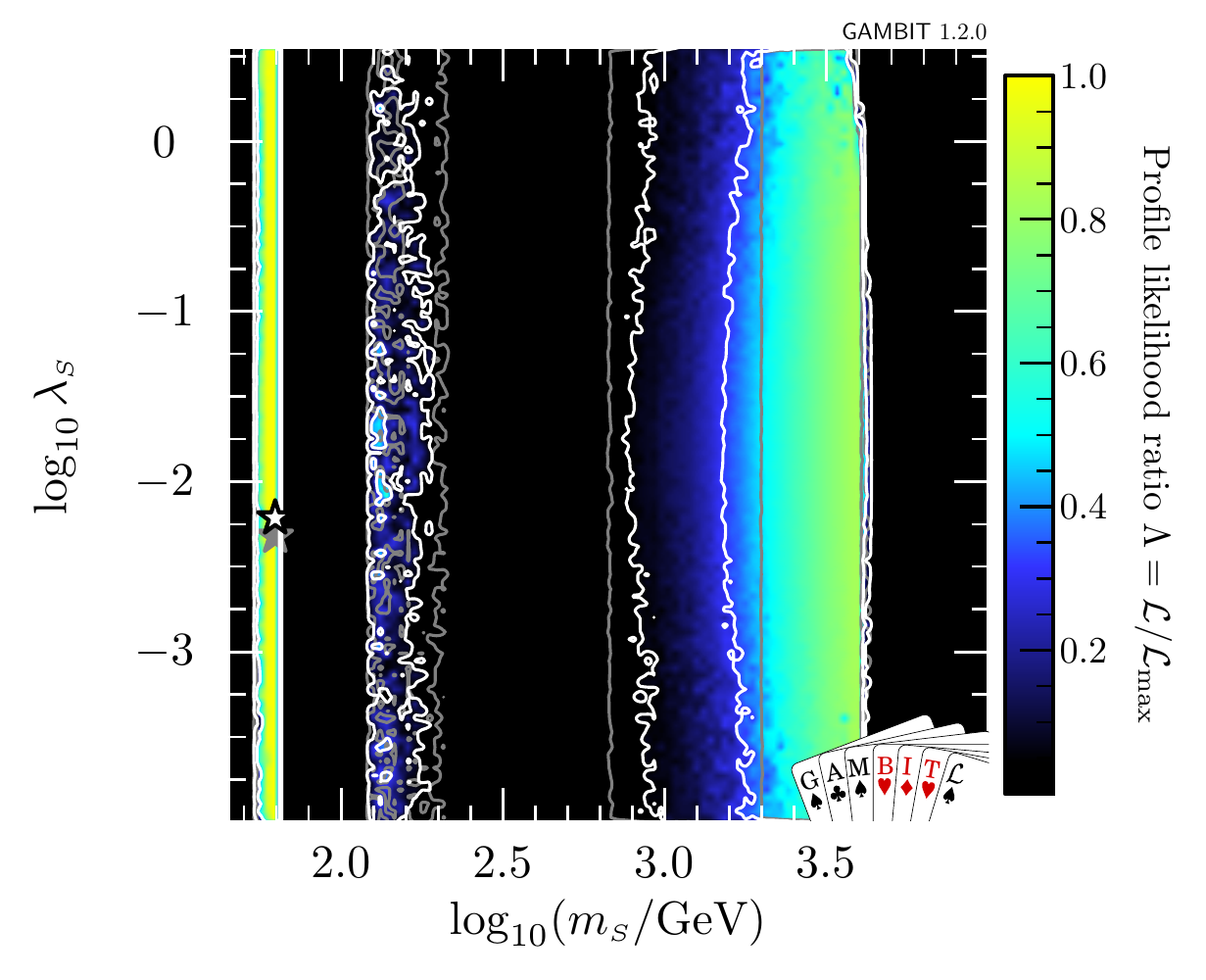}
\caption{Profile likelihoods for the \ztwo scalar singlet model, with the requirement that $\Lambda_P>\max(\msms,m_t)$ only.  Results are shown in the $\ms$--$\lhs$ (\textit{top}) and $\ms$--$\ls$ (\textit{bottom}) planes.  Left panels show a zoomed-in view of the resonance region; right panels show the full mass range.  Contour lines indicate $1\sigma$ and $2\sigma$ confidence regions, and best fit points are indicated with stars.  Shading and white contours show the result of including the 2018 XENON1T analysis \cite{Aprile:2018dbl}, whereas grey annotations illustrate the impact of using the 2017 analysis~\cite{Aprile:2017iyp} instead.}
\label{fig:Z2_full}
\end{figure*}

\begin{figure}[tbp]
\centering
\includegraphics[width=1\columnwidth]{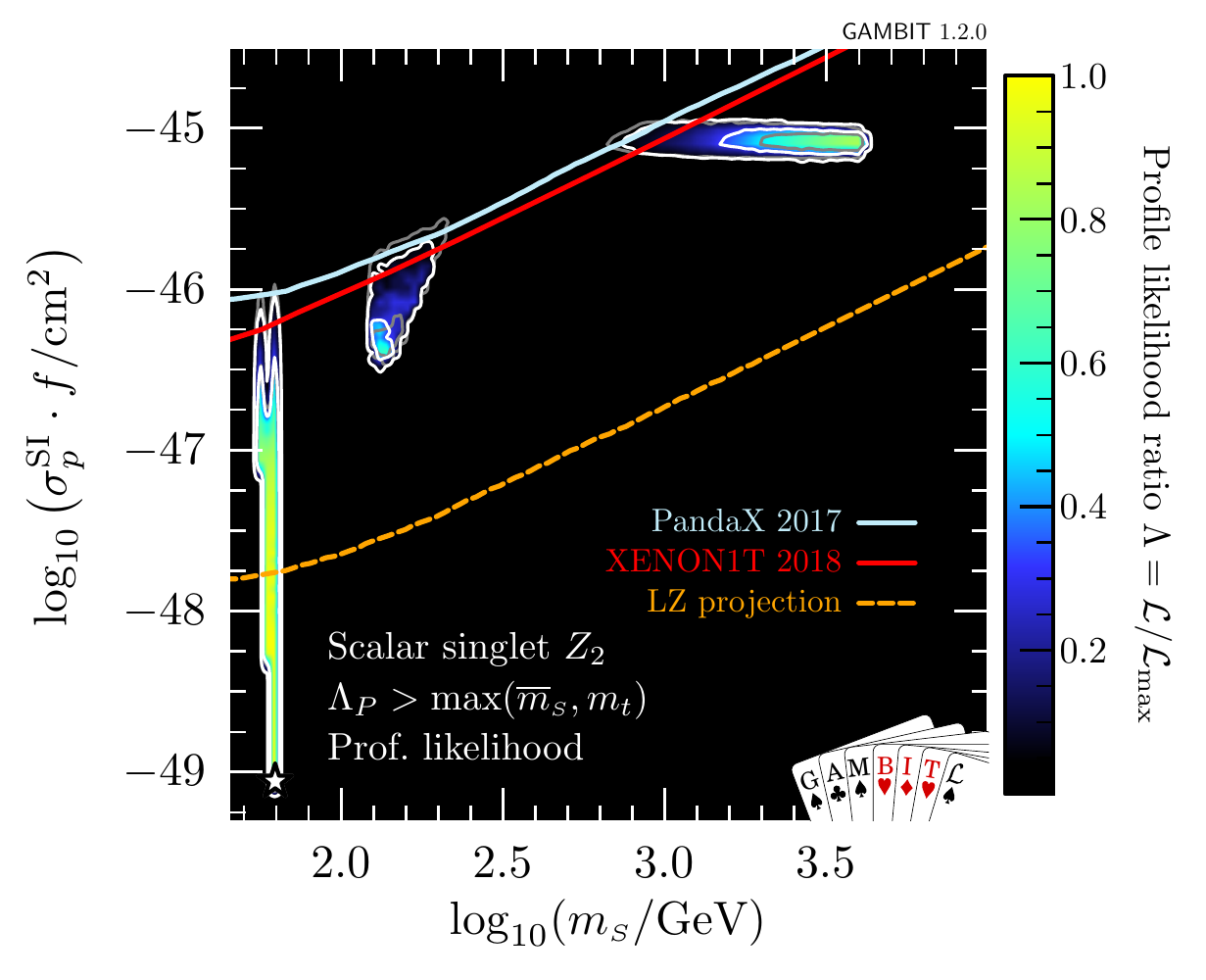}
\caption{Profile likelihood for the $\mathbb{Z}_2$ scalar singlet model with the requirement $\Lambda_P>\max(\msms,m_t)$.  The preferred regions are expressed as a function of $\ms$ and the spin-independent direct detection cross-section for scattering with protons rescaled to the predicted relic abundance $\sigma_p^\text{SI}\cdot f$, and compared to the exclusion bounds from various direct detection experiments. Contour lines indicate $1\sigma$ and $2\sigma$ confidence regions, and best fit points are indicated with stars.  Shading and white contours show the result of including the 2018 XENON1T analysis \cite{Aprile:2018dbl}, whereas grey annotations illustrate the impact of using the 2017 analysis~\cite{Aprile:2017iyp} instead. Coloured solid lines indicate published limits from PandaX \cite{Cui:2017nnn} and XENON1T \cite{Aprile:2018dbl}, and the dashed line is the projected sensitivity of LZ \cite{LZ}.}
\label{fig:Z2_DD}
\end{figure}

With SM-like couplings (which the Higgs possesses in the \ztwo and \zthree models), the upper limit on the invisible branching fraction of the Higgs is 19\% at 95\% confidence level \cite{Belanger:2013xza}. We employ the implementation of the full likelihood associated with this result in \decaybit \cite{SDPBit}.

\subsection{Additional likelihoods}\label{sec:singletdm:additional}

We also include simple likelihoods for the nuisance parameters varied in our fits (Table \ref{tab:SMparams}), via \darkbit \cite{DarkBit} and \precisionbit \cite{SDPBit}.  These quantities are well constrained by existing data.

We implement a log-normal likelihood for the local DM density, with a central value of $\bar{\rho_0}=0.4$ GeV/cm$^3$ (e.g.~\cite{Catena:2009mf}) and an uncertainty of $\sigma_{\rho_0} = 0.15$\,GeV\,cm$^{-3}$,
\begin{equation}
\mathcal{L}_{\rho_0} =  \frac{1}{\sqrt{2\pi} \sigma'_{\rho_0} \rho_0} \exp \left(- \frac{\ln(\rho_0 / \bar\rho_0)^2}{2 {\sigma^{\prime 2}_{\rho_0}}} \right) ,
\end{equation}
where $\sigma'_{\rho_0} = \ln(1 + \sigma_{\rho_0}/\rho_0)$. More details can be found in Ref.~\cite{DarkBit}.

We model the speed distribution of DM in the Milky Way as Maxwell-Boltzmann, truncated at the local Galactic escape velocity $v_\mathrm{esc}$. We apply a Gaussian likelihood to the mean of this distribution, characterised by a central value of 240\,km\,s$^{-1}$ and a standard deviation of 8\,km\,s$^{-1}$.  This is based on a calculation of the circular rotation speed of the Sun, $v_{\text{rot}}$ \cite{Reid:2014boa}.  We also constrain the escape velocity using a Gaussian likelihood based on $v_\mathrm{esc} = 550 \pm 35$\,km\,s$^{-1}$, derived from measurements of stellar velocities in the RAVE survey \cite{Smith:2006ym}.

We apply Gaussian likelihoods to the nuclear parameters as well, based on the estimates $\sigma_s = 43 \pm 8$\,MeV \cite{Lin:2011ab} and $\sigma_l = 50 \pm 15$\,MeV~\cite{Bishara:2017pfq}.  More detailed discussion of our adopted nuclear and velocity likelihoods can be found in Refs.~\cite{DarkBit, HiggsPortal}.

For the Higgs mass, the top quark mass and the strong coupling, we use Gaussian likelihoods based on $\mh = 125.09 \pm 0.24$\,GeV \cite{Aad:2015zhl,PDG17}, $m_t = 173.34\pm 0.76$\,GeV \cite{ATLAS:2014wva,PDG17} and $\alpha_{s}(m_{Z}) = 0.1181 \pm 0.0011$ \cite{PDG17}.

\section{The status of the $\mathbb{Z}_2$ model}\label{sec:singletdm:z2uv}

\subsection{No vacuum constraint}

In this section, we present global fits of the $\mathbb{Z}_2$ scalar singlet model, with a full spectrum calculation and RGE running up to the Planck scale. In the most general case, we allow either a metastable or an absolutely stable electroweak vacuum.  We furthermore require that the dimensionless couplings remain in the perturbative regime (which we define to be less than $\sqrt{4\pi}$), up to the greater of $\msms$ and $m_t$.  That is, we demand that $\Lambda_P>\max(\msms,m_t)$.  The profile likelihoods for this scenario are presented in Fig.~\ref{fig:Z2_full} in the $\ms$--$\lhs$ (top) and $\ms$--$\ls$ (bottom) planes.  All three of the regions mentioned in Sec.\ \ref{sec:singletdm:model} (resonance, neck and high-mass) are clearly visible in the upper panels. Due to the strength of the latest direct detection constraints, and the fact that we rescale the expected signals by the thermal relic density at each point in the parameter space, the high-mass region is split into a TeV-scale mode and an intermediate-mass mode situated just above $\ms=\mh$.\footnote{A detailed discussion of the shape of rescaled direct detection constraints, and therefore the appearance of the intermediate-mass mode, can be found in Ref.\ \cite{Cline:2013gha} in the context of Fig.\ 6 of that paper.}

The restriction to $\Lambda_P>\max(\msms,m_t)$ results in a reduction of the volume of the allowed region compared to our results in Ref.~\cite{SSDM}. Any model with values of $\lhs$ or $\ls$ greater than $\sqrt{4\pi}$ at the input scale $\msms$ violates the perturbativity condition even before RGE running, so our profile likelihoods extend only to $\sqrt{4\pi}$ in $\lhs$ and $\ls$. In contrast, Ref.~\cite{SSDM} allowed up to $\lhs = 10$.

At very large $\ms$ and $\lhs$, the Higgs quartic coupling is driven up by large loop corrections in the scalar sector, causing it to become non-perturbative.  This excludes a region at $\log_{10}(\lhs) \approx 0.2$, $\log_{10}(\ms/\mathrm{GeV}) \approx 3.6$ that was previously allowed in Ref.\ \cite{SSDM}.  This consequence of the perturbativity requirement at high $\ms$ combines with the relic density constraint, which pushes up at the allowed parameter region from below, to provide a robust upper limit on the singlet mass of $m_s < 4.5\,\mathrm{TeV}$.

The profile likelihood of the scalar quartic coupling $\ls$ is reasonably uniform over the prior range. This is unsurprising, given that $\ls$ has little phenomenological impact in this model; indeed, this is why it was not included in Ref.\ \cite{SSDM}. However, as we will show, it can be important for stabilising the electroweak vacuum and/or influencing the range of scales over which the model can remain perturbative.

In Fig.\ \ref{fig:Z2_DD}, we show the spin-independent nuclear scattering cross-section $\sigma_p^\text{SI}$ for the \ztwo scalar singlet, rescaled by the fraction $f\equiv \Omega_{\sss S} / \Omega_{\sss \text{DM}}$ of the relic density explained by each point in parameter space.  Because we scale the expected signals of each model by $f$ when computing their likelihoods, this rescaling is necessary when visually comparing predicted cross-sections to published exclusion curves (which assume $f=1$).  Compared to our earlier results \cite{SSDM}, significant amounts of parameter space are now excluded from the high-mass modes.  The new perturbativity constraint removes parameter space at low $\sigma_p^\text{SI}$, whereas the advent of XENON1T cuts into the allowed region from above.  LZ \cite{LZ} will probe a large fraction of the remaining parameter space, even including a substantial part of the resonance region.

In Figs.\ \ref{fig:Z2_full} and \ref{fig:Z2_DD}, we also illustrate the impact of the latest XENON1T data \cite{Aprile:2018dbl} by comparing the 1 and 2$\sigma$ CL regions with the inclusion of the 2018 (white contours) and 2017 constraints (grey contours).  We see that the majority of the impact of XENON1T compared to Ref.\ \cite{SSDM} was provided already in 2017, with a comparatively modest additional constraint imposed by the 2018 data.  Indeed, the small excess above background expectation at high recoil energies in the 2018 data leads to a small increase in the size of the 1$\sigma$ preferred region at large $\ms$, where the predictions of the model are consistent with the observed excess.

\begin{figure*}[tbp]
\centering
\includegraphics[width=1\columnwidth]{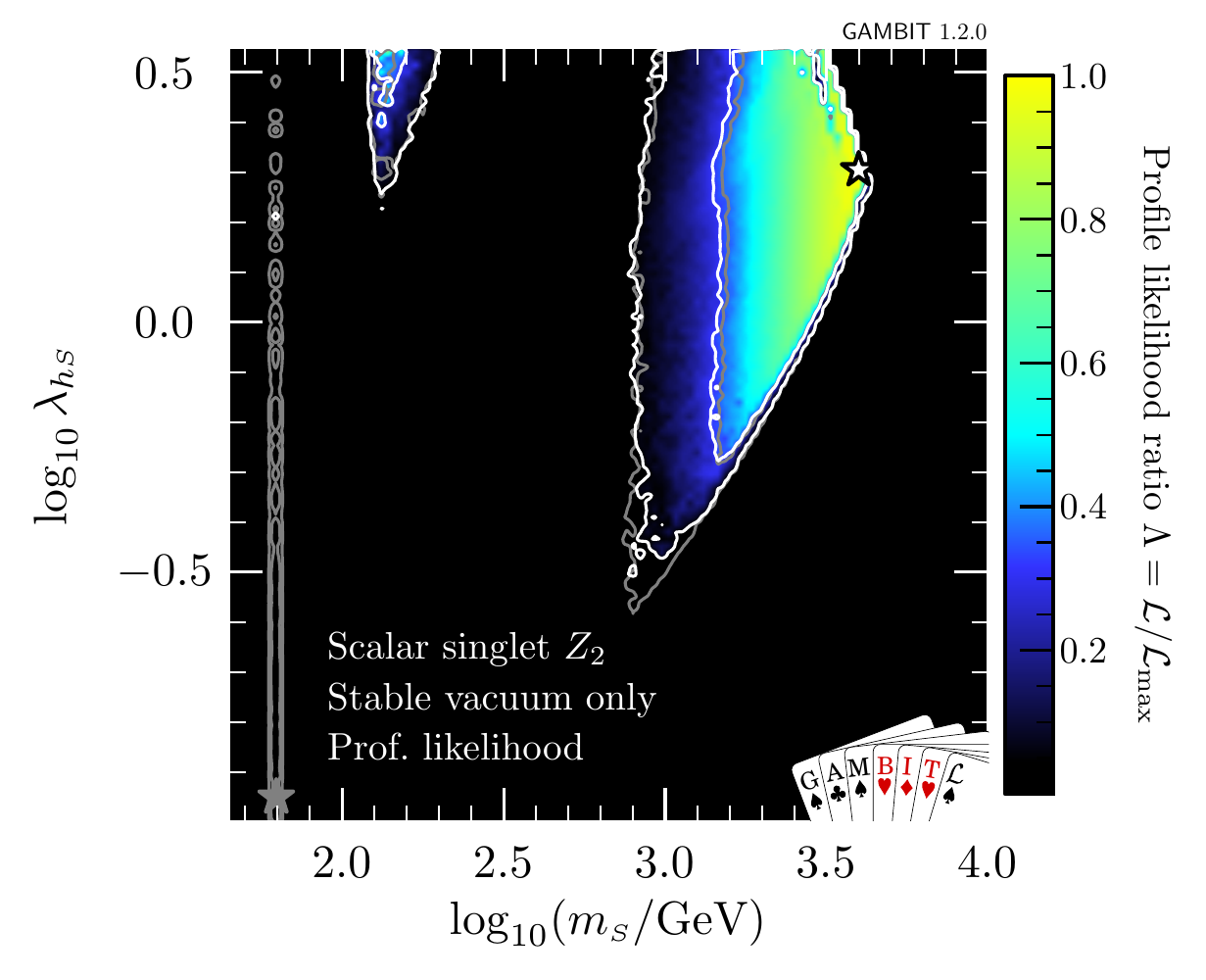}
\includegraphics[width=1\columnwidth]{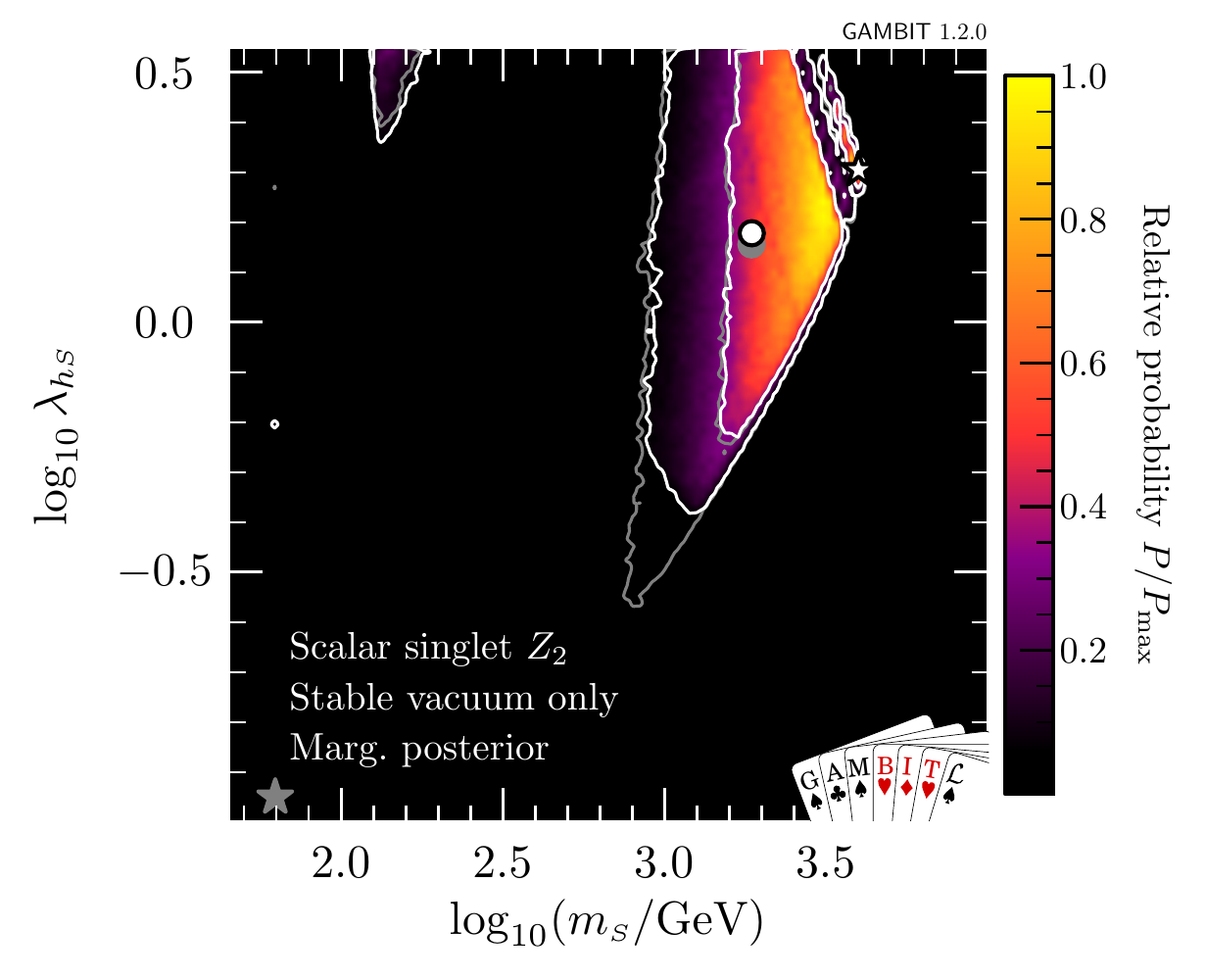}
\caption{Impact of the requirement of vacuum stability on the \ztwo scalar singlet model, expressed in terms of profile likelihoods (\textit{left}) and posterior probability densities (\textit{right}).  Bullets indicate posterior means and stars indicate best fit points.  Shading and white annotations correspond to scans where the singlet is required to absolutely stabilise the electroweak vacuum.  For comparison, we also show the preferred regions without this requirement in grey.}
\label{fig:Z2_vs}
\end{figure*}

\subsection{Absolutely stable vacuum}\label{sec:singletdm:z2uv:vs}

Next, we restrict the model further by imposing the additional constraint of absolute vacuum stability (Fig. \ref{fig:Z2_vs}). We find that values of $\lhs\gtrsim 0.2$ are required to prevent the Higgs quartic coupling becoming negative and thereby stabilise the electroweak vacuum.  As a result, the low-mass resonance mode around $\ms \sim m_h/2$ is almost entirely ruled out except for the very top of the neck region, where a few points are found with $\lhs\gtrsim 0.2$ and a stable vacuum.  This essentially leaves just the high mass-modes, centred on approximately $100\,$GeV and $1$\,TeV, where $\lhs$ is large enough to stabilise the vacuum.

In Fig. \ref{fig:Z2_vs} we also show the marginalised posterior for the $\mathbb{Z}_2$ model. As in Ref.~\cite{SSDM}, we see that even without the requirement of absolute vacuum stability, there is a clear preference for the high-mass region over the resonance region, due to the need to fine-tune nuisance parameters in order to fit all existing data at any given point in the resonance region.  With the inclusion of vacuum stability, the same effect can be seen to disfavour the medium-mass mode, where $\ms$ is $\mathcal{O}(100)$\,GeV.

Both the profile likelihoods and the marginalised posterior of Figs.\ \ref{fig:Z2_full} and \ref{fig:Z2_vs} show a small diagonal strip where valid solutions are difficult to come by at large $\ms$ and $\lhs$, just below (and running parallel to) the border of the allowed region where the Higgs quartic coupling becomes non-perturbative.  In this region, the pole mass calculation for the Higgs runs into numerical instabilities, and fails to converge.  This is a numerical artefact; large $\lhs$-dependent radiative corrections cause the Higgs pole mass calculation at $\msms$ to fail in the UV (singlet) theory, but technically this particular iteration could be avoided, seeing as the Higgs pole mass that we actually adopt comes instead from the SM EFT.  The true results in this region would therefore smoothly interpolate those from the surrounding region.  This effect confirms that predictions are becoming less stable, due to large one-loop corrections, as we approach the perturbativity limit, and that indeed we should not adopt any larger perturbative cutoff on the dimensionless couplings than $\sqrt{4\pi}$.  This problem can also be partially compensated for by varying other nuisance parameters (in particular, the top mass) within their allowed ranges, as can be seen by the fact that this effect has a much larger impact on the posterior than the profile likelihood.

\begin{figure*}[tbp]
\centering
\includegraphics[width=1\columnwidth]{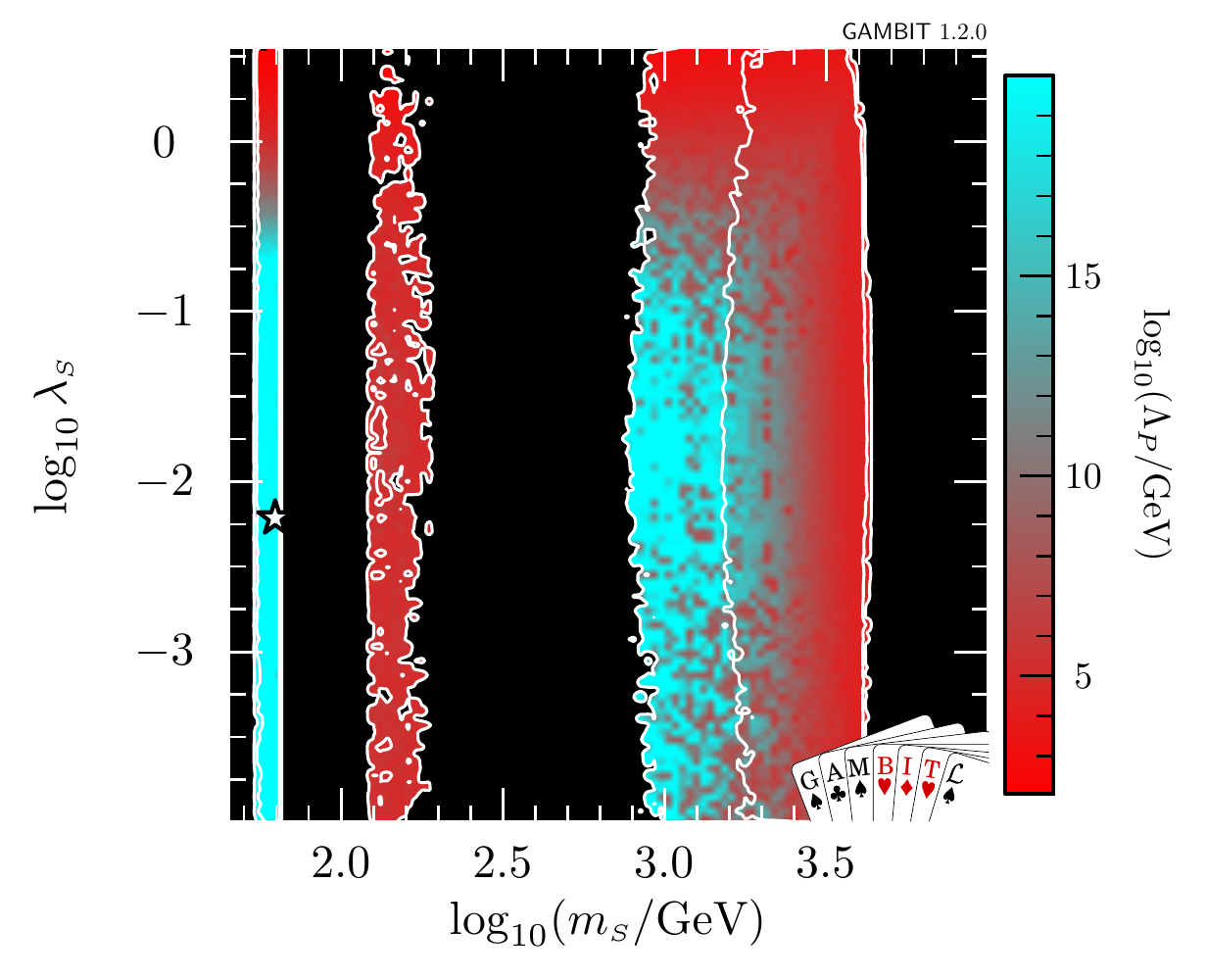}\includegraphics[width=1\columnwidth]{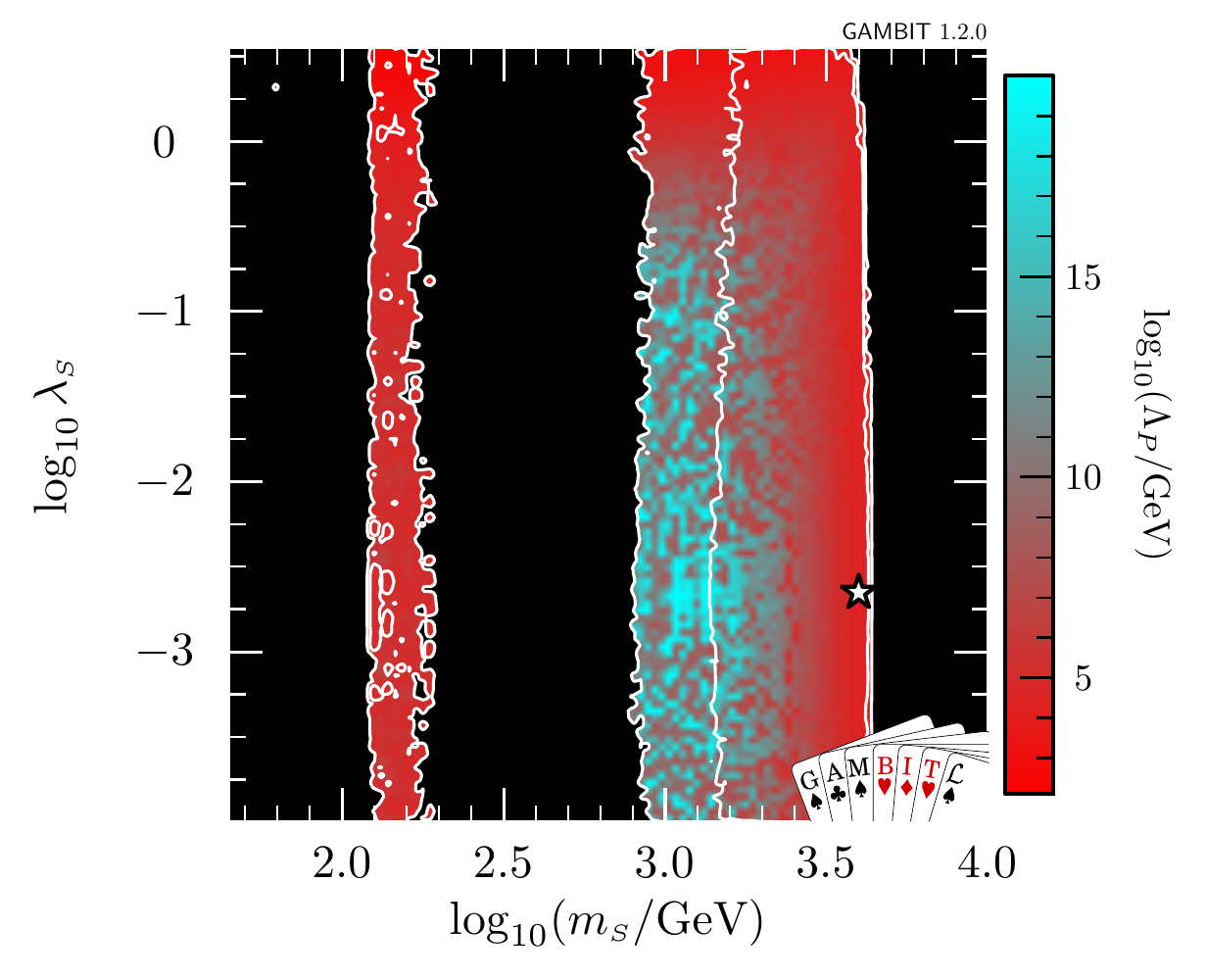}
\caption{Scale of perturbativity violation with respect to $\ms$ and $\ls$ for the $\mathbb{Z}_2$ scalar singlet model with the requirement that $\Lambda_P>\max(\msms,m_t)$ only (\textit{left}) and with the additional requirement of absolute vacuum stability (\textit{right}).  The $1\sigma$ and $2\sigma$ confidence regions are delineated by white contours, and the best-fit by a white star.}
\label{fig:Z2_perturb_ls}
\end{figure*}

Let us now take a closer look at just \textit{how} the $\mathbb{Z}_2$
scalar singlet model can satisfy the vacuum stability constraint. As
discussed in Sec.\ \ref{sec:singletdm:vs}, we apply the vacuum
stability condition by excluding points where we can show perturbatively
that the electroweak vacuum is metastable, because $\lh$ becomes negative before the Planck scale $M_{\text{Pl}}=1.22\times10^{19}$\,GeV.
However, this means that in Fig.\ \ref{fig:Z2_vs}, we do not distinguish between two quite different
cases: \begin{itemize}
\item[i)] at high scales all couplings remain perturbative and
$\lh \geq 0$,
\item[ii)] some couplings simply run to non-perturbative values
before $M_{\text{Pl}}$.
\end{itemize}
In the case of i), we have explicitly shown that the scalar singlet model can help to stabilise the electroweak vacuum.
In ii), the stability of the electroweak vacuum may be restored by non-perturbative effects, but we are
unable to determine whether this is the case or not from our perturbative calculations. It is therefore
important to discriminate between these two cases.

In Fig.~\ref{fig:Z2_perturb_ls}, we plot the scale at which perturbativity is violated in the $\ms$--$\ls$ parameter plane, choosing the scale by profiling the likelihood over the other parameters (i.e.\ plotting $\Lambda_P$ for the best-fit points found in the scan at each combination of $\ms$ and $\lhs$).  We plot the value of $\Lambda_P$ only within the $2\sigma$ contours, as determined by the profile likelihood.  Note that there can exist parameters with a larger value of $\Lambda_P$ that only have a slightly worse $\mathcal{L}$, and are still within $2\sigma$ of the best-fit point.  As we run the couplings to a maximum scale of $1\times10^{20}\,$GeV (well above the Planck scale, where quantum gravitational effects become important), points with $\Lambda_P$ equal to this value should be interpreted as valid to \textit{at least} $1\times10^{20}\,$GeV.

\begin{figure*}[tbp]
\centering
\includegraphics[width=1\columnwidth]{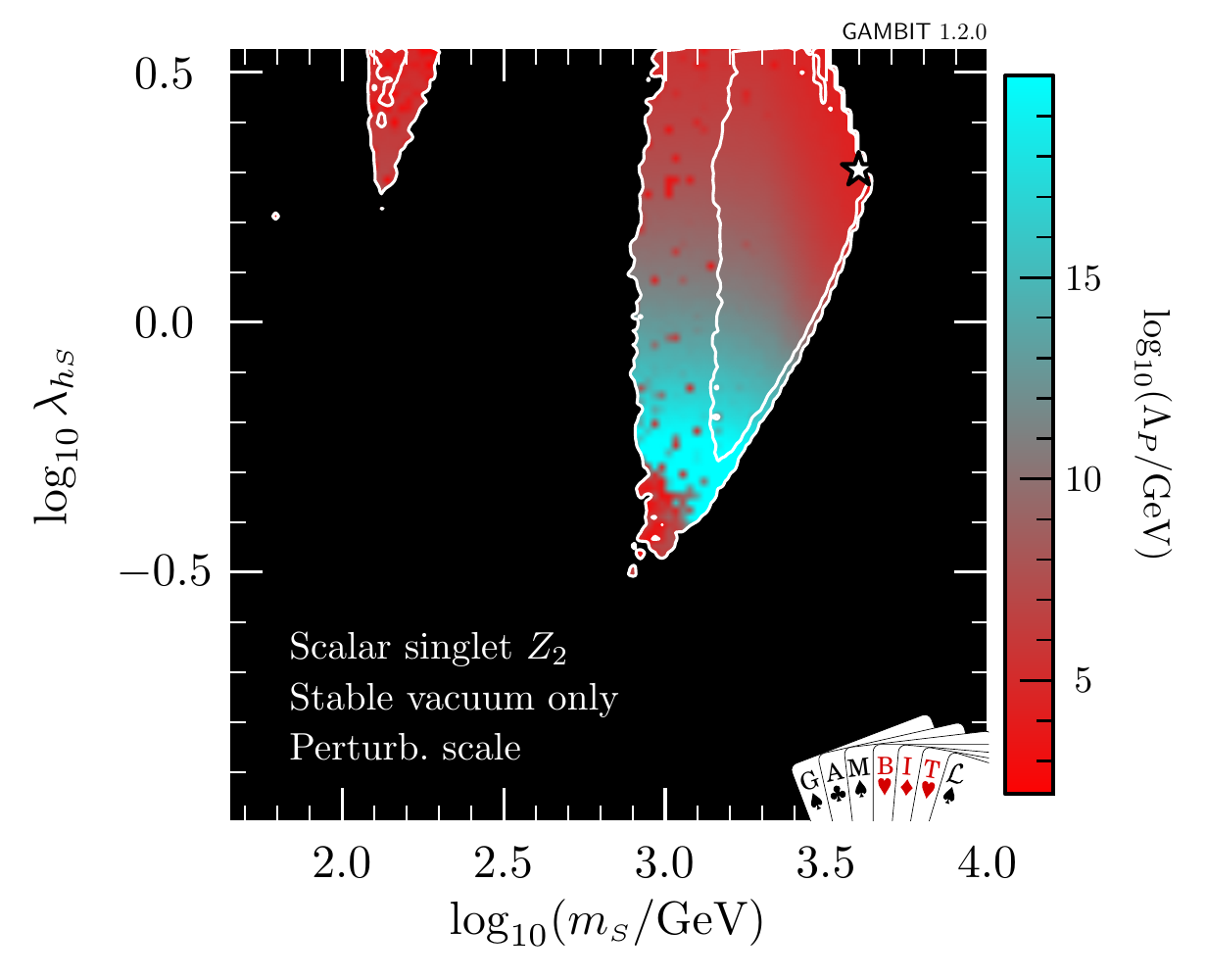}\includegraphics[width=1\columnwidth]{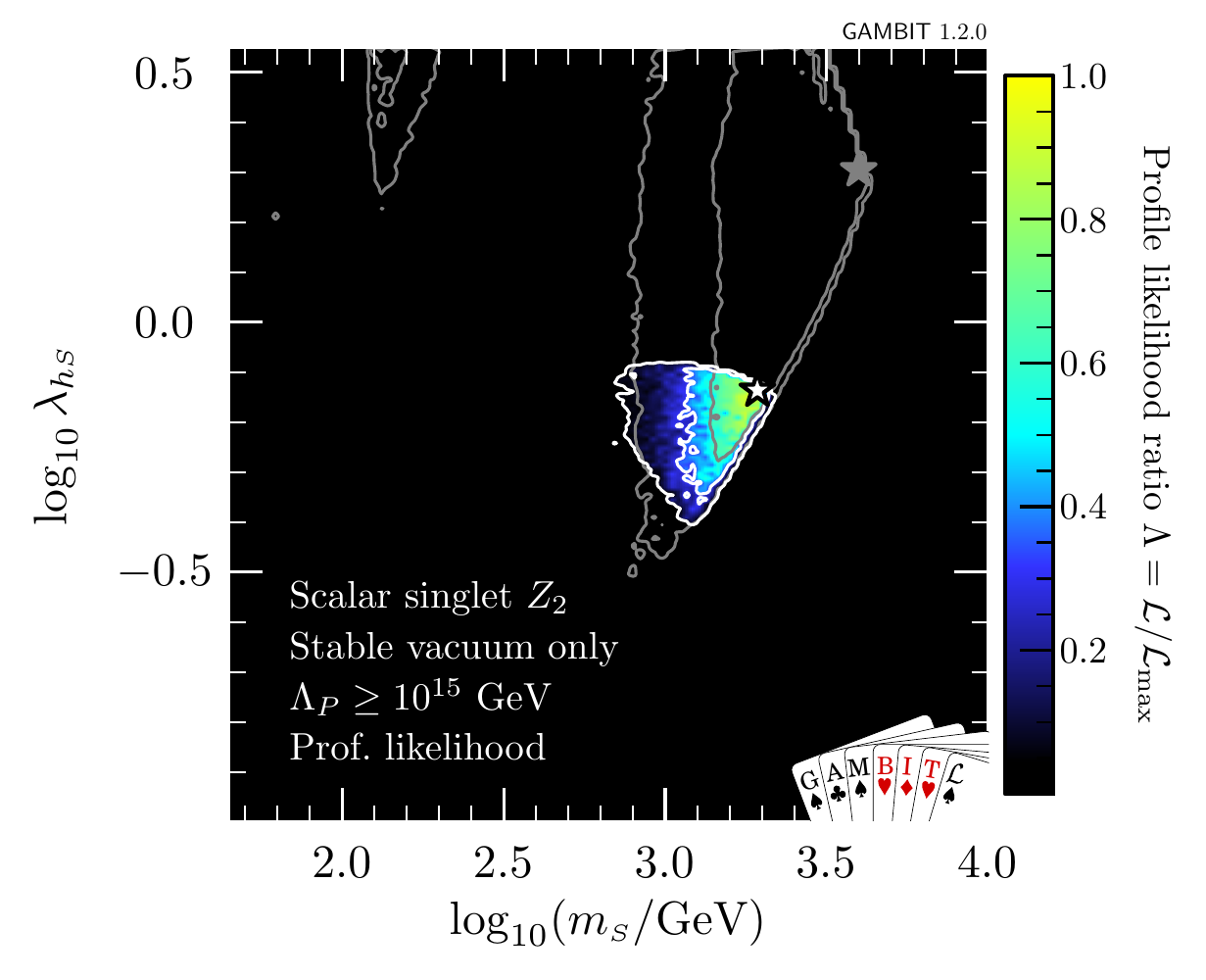}
\caption{\textit{Left:} scale of perturbativity violation for the $\mathbb{Z}_2$ scalar singlet model with the requirement of a stable electroweak vacuum. \textit{Right:} profile likelihood when furthermore imposing the requirement $\Lambda_P>10^{15}\,$GeV.  The $1\sigma$ and $2\sigma$ confidence regions are delineated by white contours, and the best-fit by a white star.  Grey contours on the right panel correspond to the $1\sigma$ and $2\sigma$ confidence regions of the left panel.}
\label{fig:Z2_cut}
\end{figure*}

A number of observations can be made from Fig.~\ref{fig:Z2_perturb_ls}. First of all, we note that for $\ls \gtrsim 0.7$ the scale of perturbativity is always low, as the singlet quartic coupling quickly runs to non-perturbative values. The dependence on $\ms$ is more complicated and can be best understood by comparing to Fig.~\ref{fig:Z2_full}. In the low-mass resonance mode ($\ms\sim m_h/2$), $\lhs$ is typically very small and the scale of perturbativity violation can be very high as long as $\ls$ is sufficiently small. The mode at $\ms\sim 100\,$GeV, on the other hand, requires $\lhs > 1$, which renders the spectrum invalid at scales well below $10^{10}\,$GeV irrespective of $\ls$. In the high-mass region ($\ms\sim 1\,\mathrm{TeV}$) it is possible to find points with a scale of perturbativity near or beyond the Planck scale, in particular towards smaller masses (corresponding to smaller $\lhs$).

In Fig.~\ref{fig:Z2_cut} (\textit{left}), we show $\Lambda_P$ as function of $\lhs$ and $\ms$ in the high-mass region, imposing absolute vacuum stability. There is a rough correlation between $\lhs$ and $\Lambda_P$, which is only broken for the small $\lhs$ tip of the high-mass mode, where $\Lambda_P$ decreases rapidly. This is because such small values of $\lhs$ are insufficient to stabilise the electroweak vacuum, so our requirement that $\lh$ not run negative only finds solutions where $\ls$ contributes to the running of $\lh$. However, since the impact of $\ls$ on the running of $\lh$ is indirect and only mild, large values of $\ls$ are required, rendering the model non-perturbative below the scale of vacuum instability (thus rendering it ``stable'' according to our definition).  This can also be seen as the cause for the difference in the profile likelihood and the posterior in the tip of this region in Fig.\ \ref{fig:Z2_vs}, reflecting the fact that $\ls$ must be tuned in order to find permitted models in this area.

The competing interests of vacuum stability and perturbativity become more problematic when we ask what values of $\Lambda_P$ are acceptable.  The metastability of the electroweak vacuum in the SM is the result of the Higgs quartic coupling becoming negative near the grand unified theory (GUT) scale, at $\sim\,10^{15}\,$GeV.  If we are concerned about vacuum stability, then we should generally also demand that our theory is perturbative to at least this scale. The electroweak vacuum is stable in the \ztwo theory in some parts of the otherwise allowed parameter space, but in others the model simply becomes non-perturbative at scales well below the GUT scale. It therefore makes sense to impose another selection requirement on our samples, in order to identify only those points that remain perturbative to high scales.

In the right panel of Fig.~\ref{fig:Z2_cut}, we show the profile likelihood after excluding all models where $\Lambda_P<10^{15}\,$GeV (in addition to requiring a stable electroweak vacuum). Due to the competing influences of direct detection, the relic density, vacuum stability and perturbativity, the allowed parameter space of the model is reduced considerably. Nevertheless, even when limiting the parameter space to points with $\Lambda_P>10^{15}\,$GeV, the model is certainly not ruled out. In the next section, we will explicitly identify parameter points for which couplings remain perturbative up to the typical instability scales, and the electroweak vacuum is stabilised, showing that these points can still give a good fit to the data -- and can even explain the entirety of DM.

\begin{figure}[tbp]
\centering
\includegraphics[width=1\columnwidth]{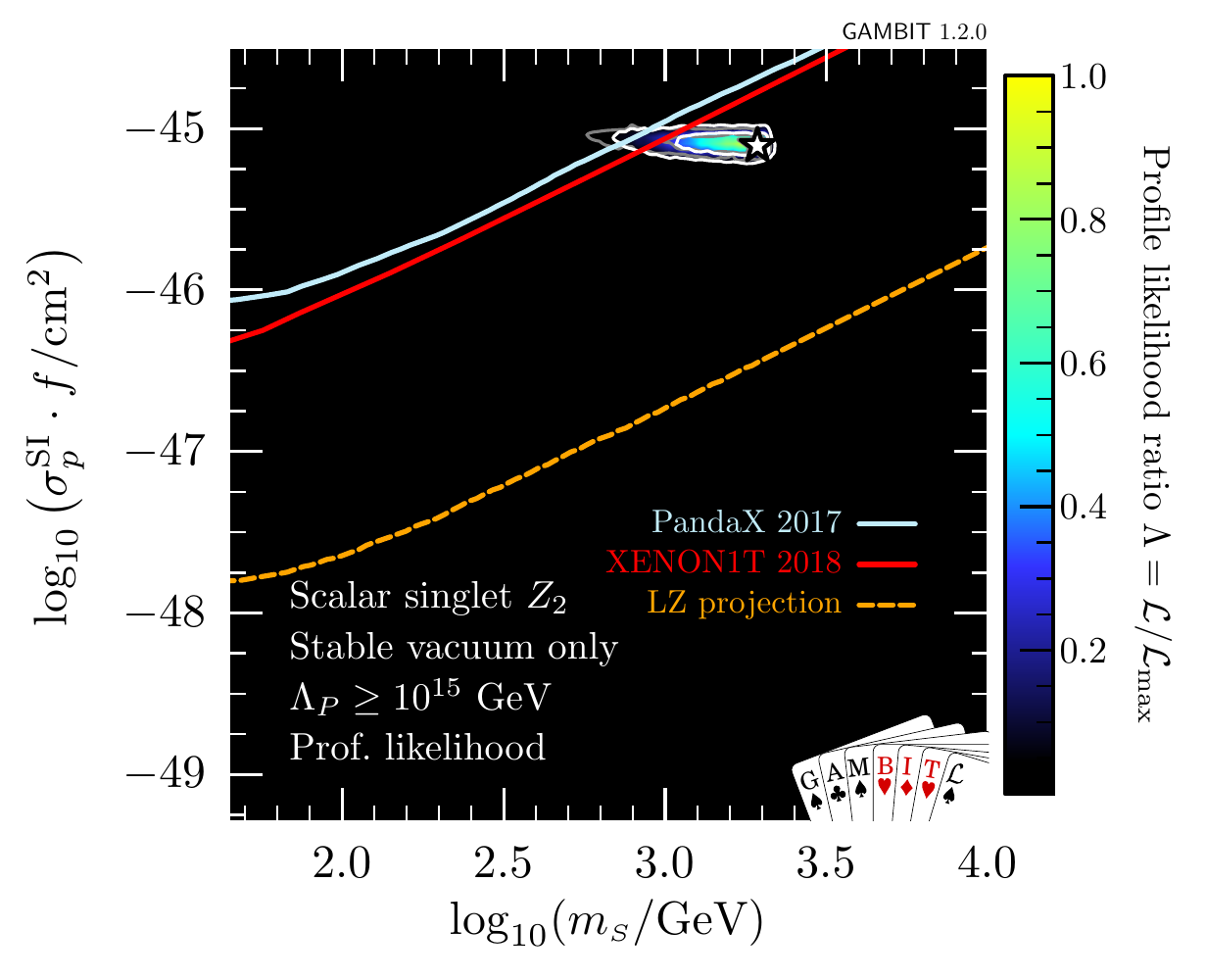}
\caption{As in Fig.\ \ref{fig:Z2_DD}, but with the added requirements of vacuum stability and perturbativity to large scales, $\Lambda_P>10^{15}\,$GeV.}
\label{fig:Z2_DD_cut}
\end{figure}

In Fig.~\ref{fig:Z2_DD_cut}, we show the impacts of demanding absolute vacuum stability and perturbativity to high scales on signals at direct detection experiments.  If the \ztwo model is to stabilise the electroweak vacuum, it must lie in a narrow region with effective nuclear scattering cross-section $5\times 10^{-46} < \sigma_p^\text{SI}\cdot f < 10^{-45}$\,cm$^2$ and mass $600$\,GeV $<$ $\ms < 2$\,TeV.  Here we again show both the result with the 2018 (shading and white contours) and 2017 XENON1T likelihoods (grey contours).  Even with vacuum stability imposed, the new XENON1T data remains consistent with the allowed region.  Future multi-ton experiments such as LZ \cite{LZ} will definitively detect or exclude this scenario.

\subsection{Best-fit point}\label{sec:singletdm:z2uv:best}

\begin{table*}[tbp]
\center{{\footnotesize
\setlength\tabcolsep{1.4pt}
\begin{tabular}{l@{\hspace{4mm}}l@{\hspace{4mm}}l@{\hspace{4mm}}l@{\hspace{4mm}}l@{\hspace{4mm}}l @{\hspace{4mm}}l @{\hspace{4mm}}l @{\hspace{4mm}}l@{\hspace{4mm}}l}
Stable &  &Relic &  &  &&   &  & & \\
vac.  & $\Lambda_P$ (GeV) & density & $\ls$ & $\lhs$ & $\ms$ (GeV) &  $\Omega_{\sss S} h^2$  & $\log(\mathcal{L})$ & $\Delta\ln\mathcal{L}$ & $\sigma^{\text{SI}}_p$ (cm$^2$) \\
  \toprule
$\sim$ & $\geq 10^{20}$ & $\leq$ & $\num{6.240e-03}$ & $\num{2.316e-04}$ & $\num{6.248e+01}$ & $\num{9.2472e-02}$ & 43.11 & 0.32 & $\num{1.15e-49}$\\
\checkmark & $\num{2.59e+04}$ & $\leq$ & $\num{2.278e-03}$ & $\num{2.031e+00}$ & $\num{3.969e+03}$ & $\num{1.0407e-01}$ & 43.30 & 0.50 & $\num{9.33e-46}$\\
\checkmark & $\num{1.00e+15}$ & $\leq$ & $\num{6.590e-04}$ & $\num{7.357e-01}$ & $\num{1.928e+03}$ & $\num{9.7236e-02}$ & 43.92 & 1.12 & $\num{9.77e-46}$\\
\checkmark & $\num{9.12e+15}$ & \checkmark & $\num{2.589e-03}$ & $\num{6.804e-01}$ & $\num{1.938e+03}$ & $\num{1.1316e-01}$ & 44.25 & 1.45 & $\num{8.36e-46}$\\
 \bottomrule
\end{tabular}}
\caption{Details of the best-fit points for the \ztwo scalar singlet model when different physical restrictions are imposed on the model. Points that have an absolutely stable electroweak vacuum are indicated by a tick in the first column. Points with a singlet relic density within 1$\sigma$ of the \textit{Planck} observed value ($\Omega_{\sss S} h^2\sim\Omega_{\sss \text{DM}} h^2$) are indicated with a tick in the third column. We omit the values of the nuisance parameters, as they are not significantly different to the central values of their respective likelihood functions.}\label{tab:best_fit_z2}}
\end{table*}

\begin{table*}[tbp]
\center{
\begin{tabular}{l r r r r r r r r r}
 &  & \multicolumn{8}{c}{ $\Delta\ln\mathcal{L}$}\\
 \cmidrule(lr){3-10}
Likelihood contribution & Ideal &  $\mathbb{Z}_2$A&  $\mathbb{Z}_2$B&  $\mathbb{Z}_2$C &  $\mathbb{Z}_2$D &$\mathbb{Z}_3$A&  $\mathbb{Z}_3$B&  $\mathbb{Z}_3$C &  $\mathbb{Z}_3$D\\
\hline
Relic density & $5.989$ & 0 & $0.001$ & 0 & $0.120$ & 0 & 0 & $0.034$ & $0.142$ \\
LUX Run II 2016 & $-1.467$ & $0.001$ & $0.112$ & $0.221$ & $0.207$ & $0.001$ & $0.095$ & $0.528$ & $0.592$ \\
PandaX 2016 & $-1.886$ & 0 & $0.071$ & $0.140$ & $0.131$ & $0.001$ & $0.059$ & $0.339$ & $0.380$ \\
PandaX 2017 & $-1.550$ & $0.001$ & $0.156$ & $0.298$ & $0.280$ & $0.002$ & $0.130$ & $0.678$ & $0.752$ \\
XENON1T 2018 & $-3.440$ & $0.210$ & $0.003$ & $0.218$ & $0.179$ & $0.209$ & $0.074$ & $1.465$ & $1.770$ \\
$\gamma$ rays (\textit{Fermi}-LAT dwarfs) & $-33.244$ & $0.105$ & $0.148$ & $0.165$ & $0.170$ & $0.105$ & $0.112$ & $0.196$ & $0.207$ \\
Higgs invisible width & $0.000$ & 0 & 0 & 0 & 0 & 0 & 0 & 0 & 0 \\
Hadronic elements $\sigma_s$, $\sigma_l$ & $-6.625$ & 0 & $0.001$ & $0.016$ & $0.019$ & 0 & 0 & $0.099$ & $0.043$ \\
Local DM density $\rho_0$ & $1.142$ & 0 & $0.010$ & $0.039$ & $0.101$ & 0 & $0.001$ & $0.547$ & $0.499$ \\
DM velocity $v_0$ & $-2.998$ & 0 & 0 & 0 & $0.001$ & 0 & 0 & $0.001$ & $0.013$ \\
DM escape velocity $v_{esc}$ & $-4.474$ & 0 & 0 & 0 & $0.005$ & 0 & 0 & $0.002$ & 0 \\
$\alpha_s$ & $5.894$ & 0 & 0 & 0 & $0.002$ & 0 & $0.001$ & $0.004$ & $0.001$ \\
Higgs mass & $0.508$ & 0 & 0 & 0 & $0.043$ & 0 & 0 & $0.082$ & $0.004$ \\
Top quark mass & $-0.645$ & 0 & 0 & $0.022$ & $0.196$ & 0 & 0 & 0 & $0.041$ \\
Vacuum stability & $0.000$ & 0 & 0 & 0 & 0 & 0 & 0 & 0 & 0 \\
\hline
Total &  & 0.317 & 0.503 & 1.121 & 1.455 & 0.318 & 0.473 & 3.975 & 4.443 \\
\end{tabular}
\caption{\label{tab:maxlike} Individual contributions to the $\Delta$ log-likelihood for the various best-fit points (see Tabs.~\ref{tab:best_fit_z2} and~\ref{tab:best_fit_z3}) compared to an `ideal' case.  We take this to be the background-only likelihood for exclusions, and the central observed value for detections.  Note that because each likelihood is dimensionful, the absolute values are less meaningful than the offset with respect to another point (see section 8.3 of Ref.~\cite{gambit} for more details on the normalisation).  The best-fit points are labelled as follows: A represents a fit with the only constraint that $\Lambda_P>\max(\msms,m_t)$, B is a fit with the additional constraint of absolute vacuum stability, C includes the constraint of $\Lambda_P>10^{15}$\, GeV and D also includes the requirement that $\Omega_Sh^2$ be within 1$\sigma$ of the observed relic density.}}
\end{table*}

The best-fit point for our global fit of the $\mathbb{Z}_2$ model is located at $\ls = \num{6.24e-03}$, $\lhs = \num{2.32e-04}$ and $\ms = \num{62.48}\,$GeV. This point is located in the low-mass resonance region, the electroweak vacuum is metastable with a lifetime of $\sim$\,$\num{1.1e+99}$ years, the minimum of $\lh$ occurs at $\sim$$3\times10^{13}\,$GeV, and the model is perturbative up to at least $10^{20}\,$GeV.  Details of this point can be found in Table~\ref{tab:best_fit_z2}. The mass at this point is within $0.03\,$GeV of the best-fit found in Ref.~\cite{SSDM}, and the portal coupling is approximately a factor of three smaller.  Given that the profile likelihood is quite flat with respect to $\lhs$ around the best-fit, the difference in $\lhs$ is not significant.

Indeed, we expect this best-fit point to be similar to what we found in Ref.~\cite{SSDM}, as the constraint $\Lambda_P>\max(\msms,m_t)$ and the variation of $\ls$ do not have a significant impact on the phenomenology at small couplings. Nevertheless, we find $\Delta\ln \mathcal{L} = 0.317$ relative to the ideal likelihood (where each individual likelihood takes its maximum value), compared to $\Delta\ln \mathcal{L} = 0.107$ in Ref.~\cite{SSDM}. This difference is due to the contribution from the new 2018 XENON1T likelihood, which exhibits a slight preference for a non-zero DM signal and hence slightly disfavours the low mass region (see Table~\ref{tab:maxlike}).

When the constraint of absolute vacuum stability is imposed, the location of the best fit necessarily moves away from the resonance region, where $\lhs$ is too small to stabilise the vacuum.  In this case we find a best-fit point at $\ls =\num{2.28e-03}$, $\lhs =\num{2.03e+00}$ and $\ms =\num{3.97}\,$TeV.  In this case we find $\Delta\ln \mathcal{L} = 0.503$, which corresponds to a slight penalty over the metastable case. Note in particular that this second point gives a better fit to the data from XENON1T, but the combined likelihood from all direct detection experiments is worse due to the contribution from LUX and PandaX. Although the vacuum is classified as stable at this point, it is an example of a point where the couplings are so large that they cannot be run all the way to the typical scale of vacuum instability, leading to $\Lambda_P\sim26\,$TeV.  This reduces the theoretical appeal of this point.

\begin{figure*}[tbp]
\centering
\includegraphics[width=1\columnwidth]{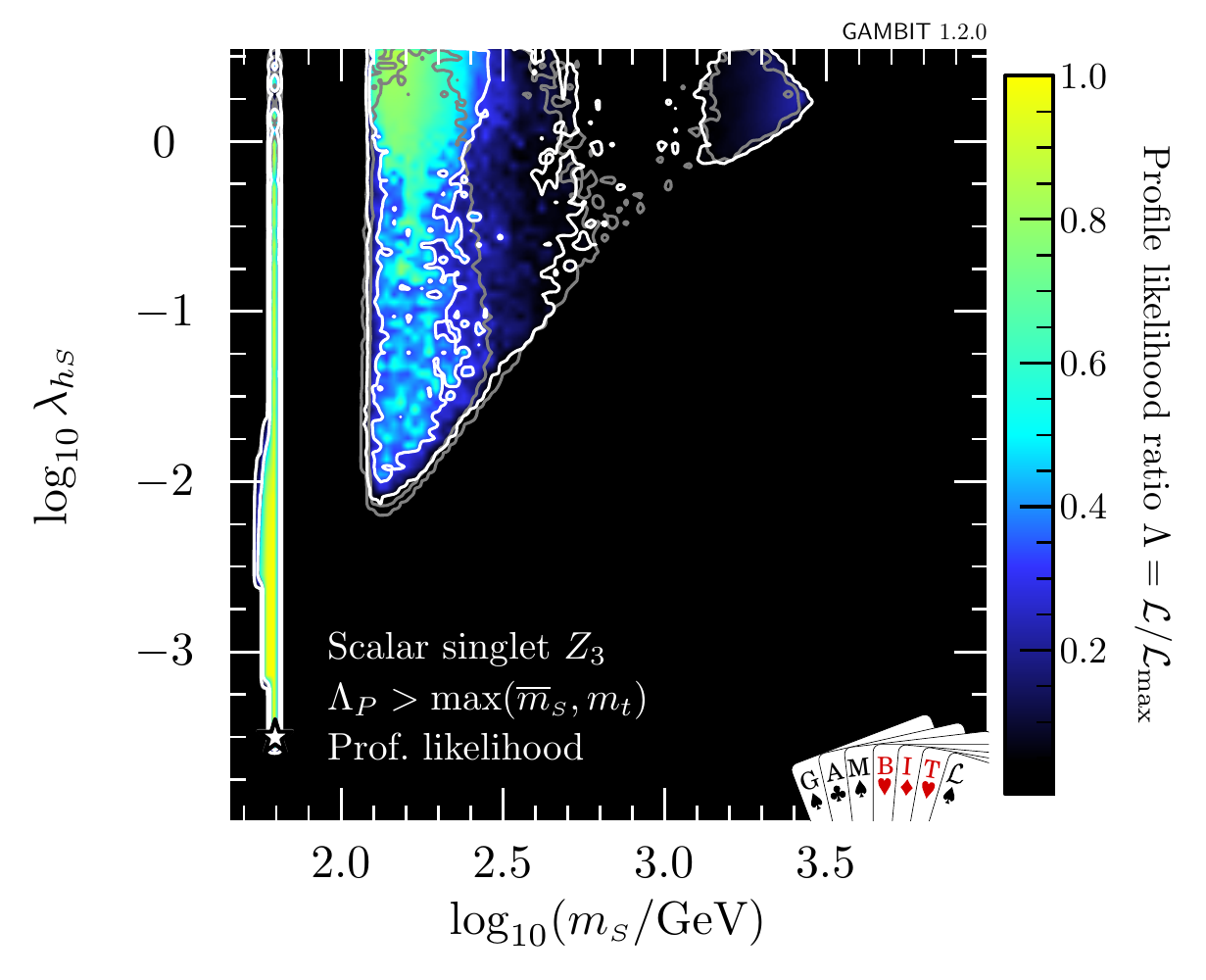}
\includegraphics[width=1\columnwidth]{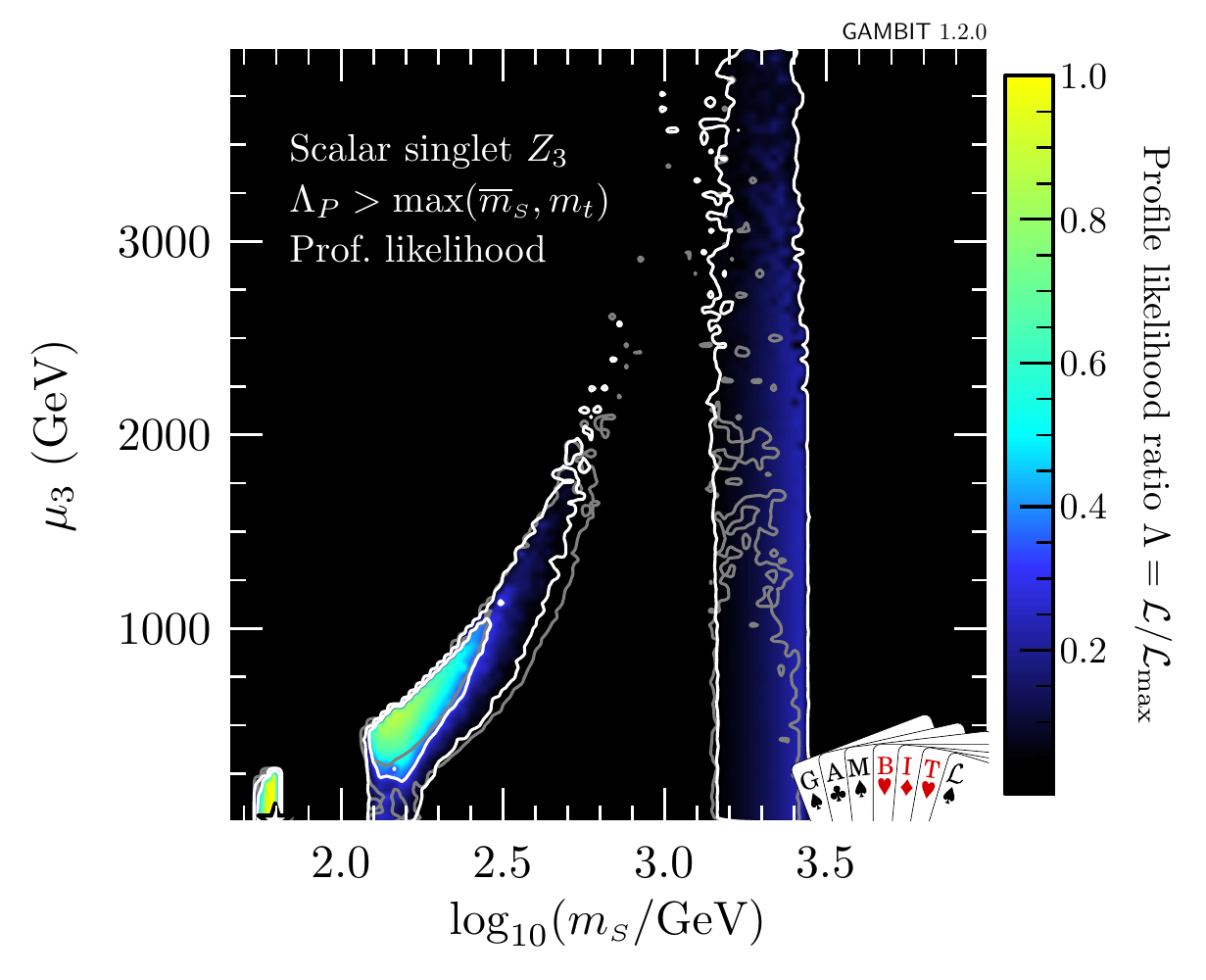}\\
\caption{Profile likelihoods for the $\mathbb{Z}_3$ scalar singlet model, with the requirement that $\Lambda_P>\max(\msms,m_t)$ only. Results are shown in the $\lhs$--$\ms$ (\textit{left}) and in the $\mu_3$--$\ms$ (\textit{right}) planes. Contour lines indicate $1\sigma$ and $2\sigma$ confidence regions, and best fit points are indicated with stars.  Shading and white contours show the result of including the 2018 XENON1T analysis \cite{Aprile:2018dbl}, whereas grey annotations illustrate the impact of using the 2017 analysis~\cite{Aprile:2017iyp} instead.}
\label{fig:Z3_full}
\end{figure*}

By excluding all samples with $\Lambda_P<10^{15}\,$GeV, we can find points that have a stable vacuum and are more theoretically interesting (see Fig.~\ref{fig:Z2_cut}). We find a best-fit point that is absolutely stable and has $\Lambda_P = 1.0\times 10^{15}\,$GeV, at $\ls =\num{6.59e-04}$, $\lhs = \num{7.36e-01}$, $\ms = \num{1.93}\,$TeV.  This point has $\Delta\ln \mathcal{L} = 1.121$, with the largest contributions coming from direct detection likelihoods. This corresponds to a likelihood ratio $\Lambda= 0.448$ relative to the overall best-fit point, which places it inside the generally-preferred $1\sigma$ parameter region.

Although the requirement of an absolutely stable vacuum and perturbative couplings up to at least the GUT scale leads to some mild tension with direct detection limits, it is intriguing to observe that this tension has not grown with the latest XENON1T results, even though the expected sensitivity of XENON1T would have been sufficient to comprehensively test these solutions. The reason is that in precisely this parameter region, the model accommodates the slight preference for a non-zero DM signal in XENON1T. It will therefore be extremely interesting to include the results from the next generation of direct detection experiments in a similar analysis.

Finally, we consider points with relic densities within $1\sigma$ of the \textit{Planck} measured value, with stable vacua, and that remain perturbative to at least $10^{15}\,$GeV. The best-fitting of these points is located at $\ls = \num{2.59e-03}$, $\lhs = \num{6.80e-01}$ and $\ms = \num{1.94}\,$TeV, and has $\Delta\ln \mathcal{L} = 1.455$, still within 1$\sigma$ of the global best-fit point. The four best-fit points, the corresponding relic densities and the scale of perturbativity violation are presented in Table~\ref{tab:best_fit_z2}. The individual likelihood contributions for each point are given in Table~\ref{tab:maxlike}.

By interpreting $\Delta \ln\mathcal{L}$ as half the ``likelihood $\chi^2$'' of Baker \& Cousins \cite{Baker:1983tu} and assuming either one or two degrees of freedom, we can obtain an approximate $p$-value for each of our best-fit points.  For the best fit with metastability allowed, we find $p\approx 0.4\text{--}0.7$.  With vacuum stability required, $p$ drops to 0.3--0.6. For the case where the couplings are perturbative up to $10^{15}\,$GeV and the electroweak vacuum is absolutely stable, we find $p \approx 0.15\text{--}0.3$.  This decreases to $p \approx 0.1\text{--}0.25$ when also requiring that the $S$ relic density is within $1\sigma$ of the \textit{Planck} value. Each of these $p$-values is acceptable, although requiring the UV properties of perturbativity and vacuum stability does have a notable impact.

\section{The status of the $\mathbb{Z}_3$ model}\label{sec:singletdm:z3uv}

\subsection{No vacuum constraint}

We now turn to the \zthree scalar singlet model. The main difference compared to the \ztwo model is the presence of an additional parameter $\mu_3$, which has a significant impact on phenomenology because it can lead to semi-annihilations. Fig.~\ref{fig:Z3_full} presents the profile likelihoods in the $\ms$--$\lhs$ parameter plane (left) and in the $\ms$--$\mu_3$ parameter plane (right), based on scans over the full range of $\ms$. Note that the allowed region for $\mu_3$ is constrained by the vacuum stability condition given in Eq.~(\ref{eqn:ew_stability_condition}), particularly at small singlet masses.

\begin{figure*}[tbp]
\centering
\includegraphics[width=1\columnwidth]{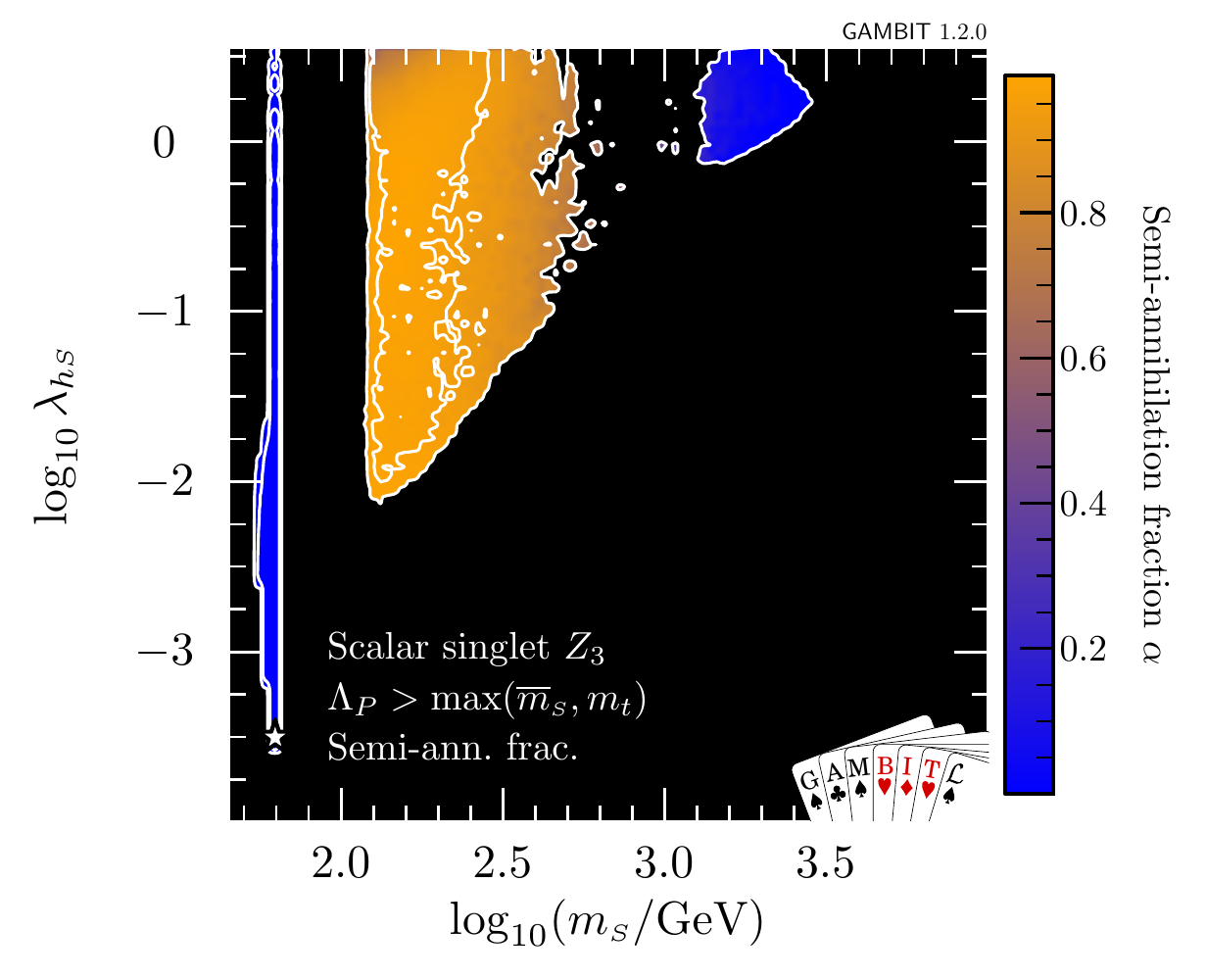}
\includegraphics[width=1\columnwidth]{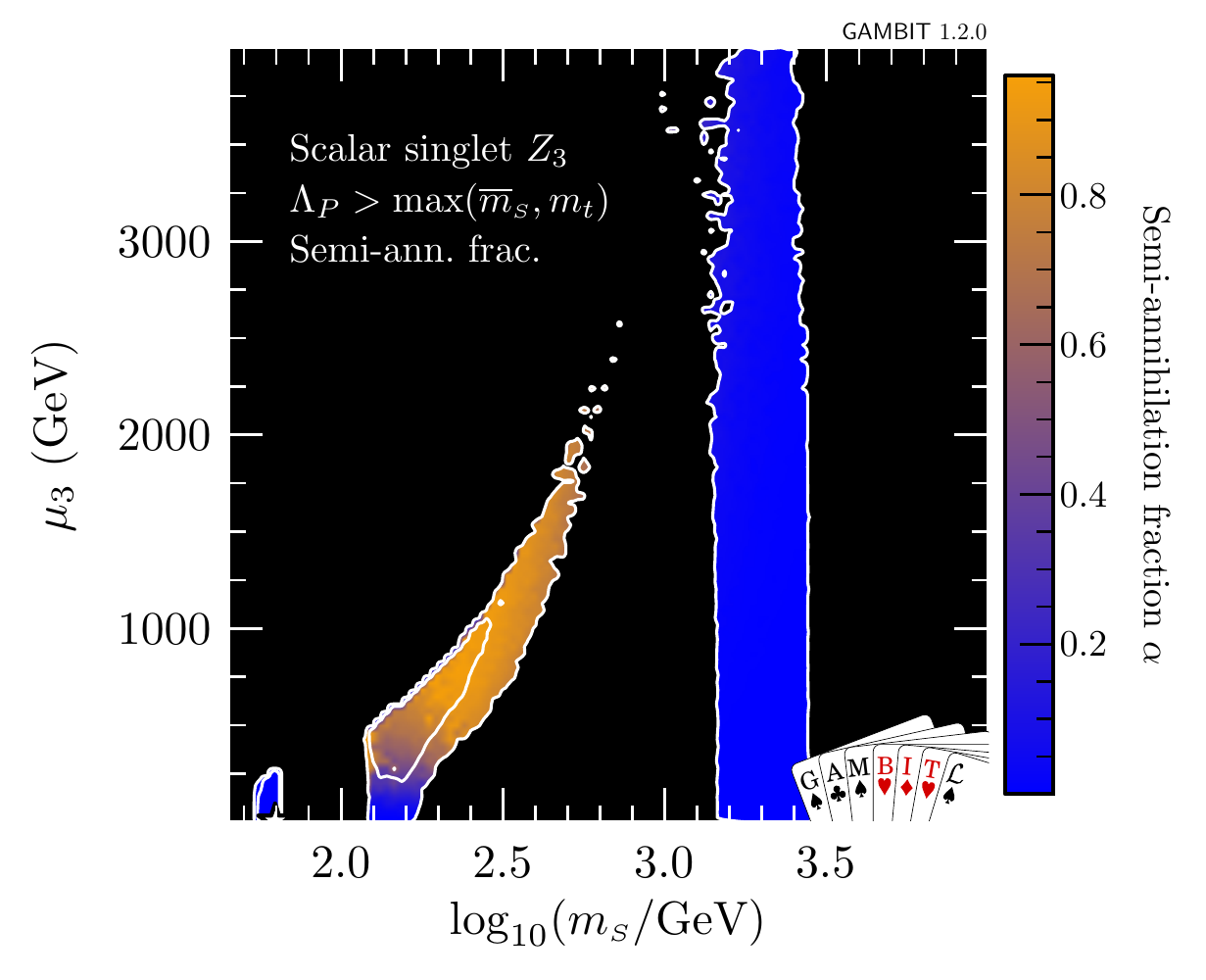}
\caption{The semi-annihilation fraction $\alpha$ with respect to $\lhs$ and $\ms$ (\textit{left}) and with respect to $\mu_3$ and $\ms$ (\textit{right}) for the $\mathbb{Z}_3$ scalar singlet model without the requirement of an absolutely stable electroweak vacuum, but imposing $\Lambda_P>\max(\msms,m_t)$. The $1\sigma$ and $2\sigma$ confidence regions are delineated by white contours, and the best-fit by a white star.}
\label{fig:Z3_semi}
\end{figure*}

\begin{figure}[tbp]
\centering
\includegraphics[width=1\columnwidth]{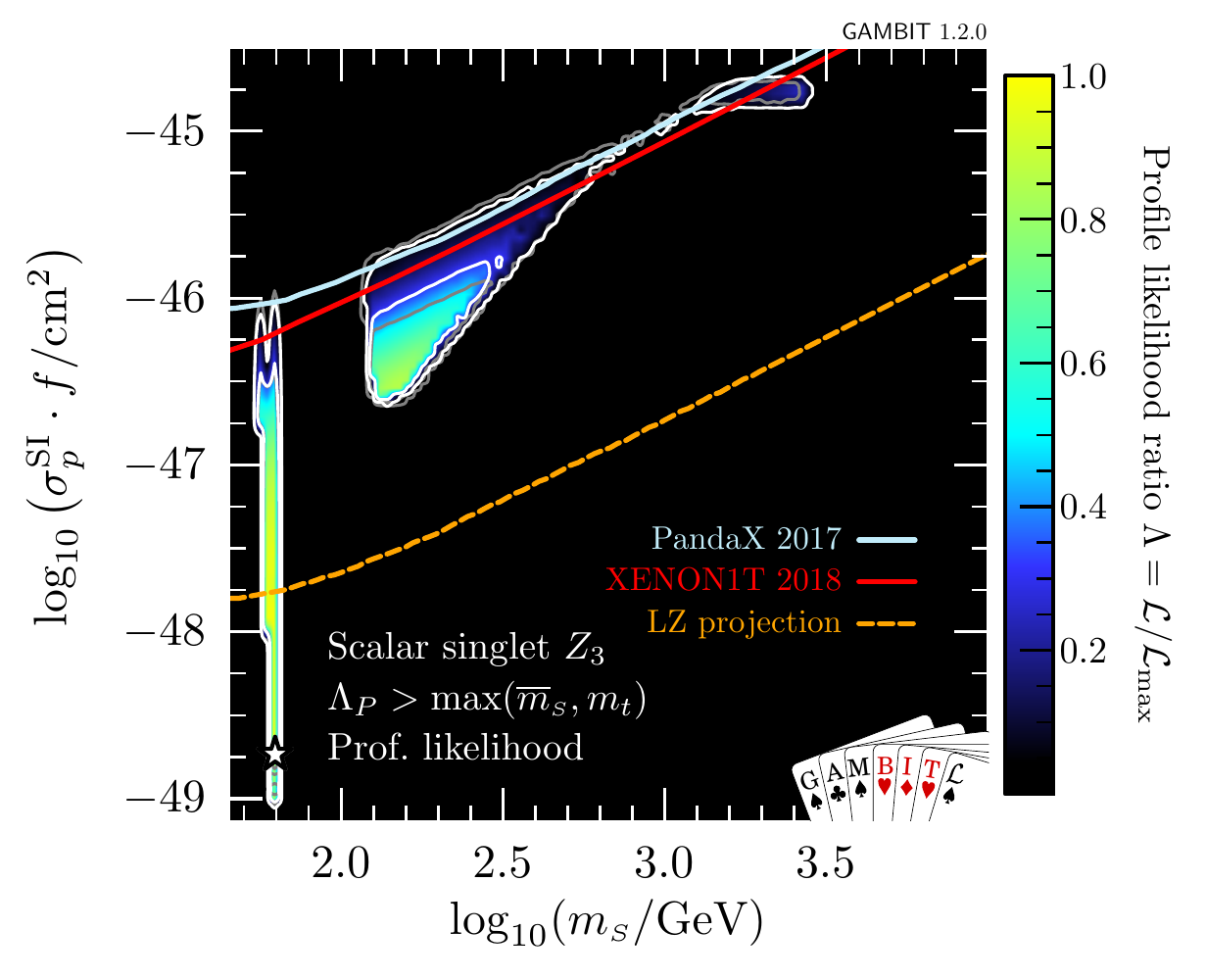}
\caption{Profile likelihood for the $\mathbb{Z}_3$ scalar singlet model with the requirement $\Lambda_P>\max(\msms,m_t)$.  Regions are shown as a function of $\ms$ and the spin-independent direct detection cross-section for scattering with protons, rescaled by the predicted relic abundance $\sigma_p^\text{SI}\cdot f$, and compared to the exclusion bounds from various direct detection experiments. Contour lines indicate $1\sigma$ and $2\sigma$ confidence regions, and best fit points are indicated with stars.  Shading and white contours show the result of including the 2018 XENON1T analysis \cite{Aprile:2018dbl}, whereas grey annotations illustrate the impact of using the 2017 analysis~\cite{Aprile:2017iyp} instead. Other lines indicate limits from PandaX \cite{Cui:2017nnn} and XENON1T \cite{Aprile:2018dbl}, and the projected sensitivity of LZ \cite{LZ}.}
\label{fig:Z3_DD}
\end{figure}

We find that the allowed resonance region in the $\mathbb{Z}_3$ model is practically identical to the corresponding region in the $\mathbb{Z}_2$ model.  We therefore do not include a version of Fig.~\ref{fig:Z3_full} zoomed in to low masses. At larger masses, however, there are notable differences between the \ztwo and \zthree models.  The allowed parameter region with $\ms \sim 200\,$GeV is substantially larger in the \zthree model, and extends to much smaller values of $\lhs$. This difference in shape can be understood by considering the fraction of semi-annihilation. In Fig.~\ref{fig:Z3_semi} we plot the semi-annihilation fraction $\alpha$, defined in eq.~(\ref{eqn:sa_fraction}), within the $2\sigma$ confidence regions. As expected, the extended allowed parameter region in the intermediate mass range corresponds to $\alpha \approx 1$, meaning that the semi-annihilation channel dominates.  As a result, the same relic abundance can be achieved with smaller values of the portal coupling $\lhs$.  In other words, the bound $\Omega_{\sss S}h^2\le\Omega_{\sss \text{DM}}h^2$ can be evaded at much lower values of $\lhs$, by invoking a large contribution from semi-annihilation. Similarly, at large values of $\lhs$ the contribution from semi-annihilation brings the relic density lower than in the $\mathbb{Z}_2$ model, and direct detection constraints are more easily avoided (given that we consistently rescale signals for the local density of singlet particles).

\begin{figure}[tbp]
\centering
\includegraphics[width=1\columnwidth]{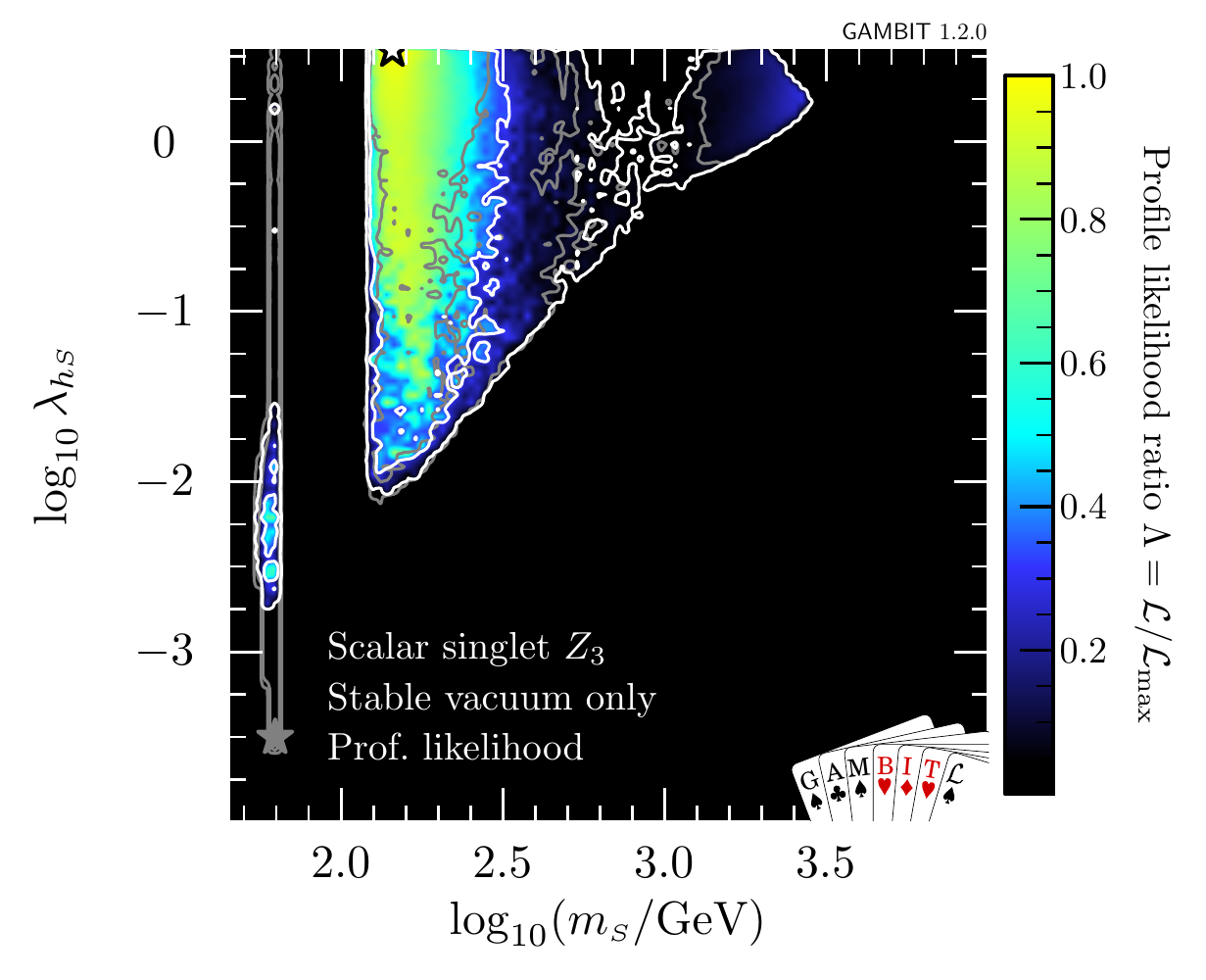}\\
\includegraphics[width=1\columnwidth]{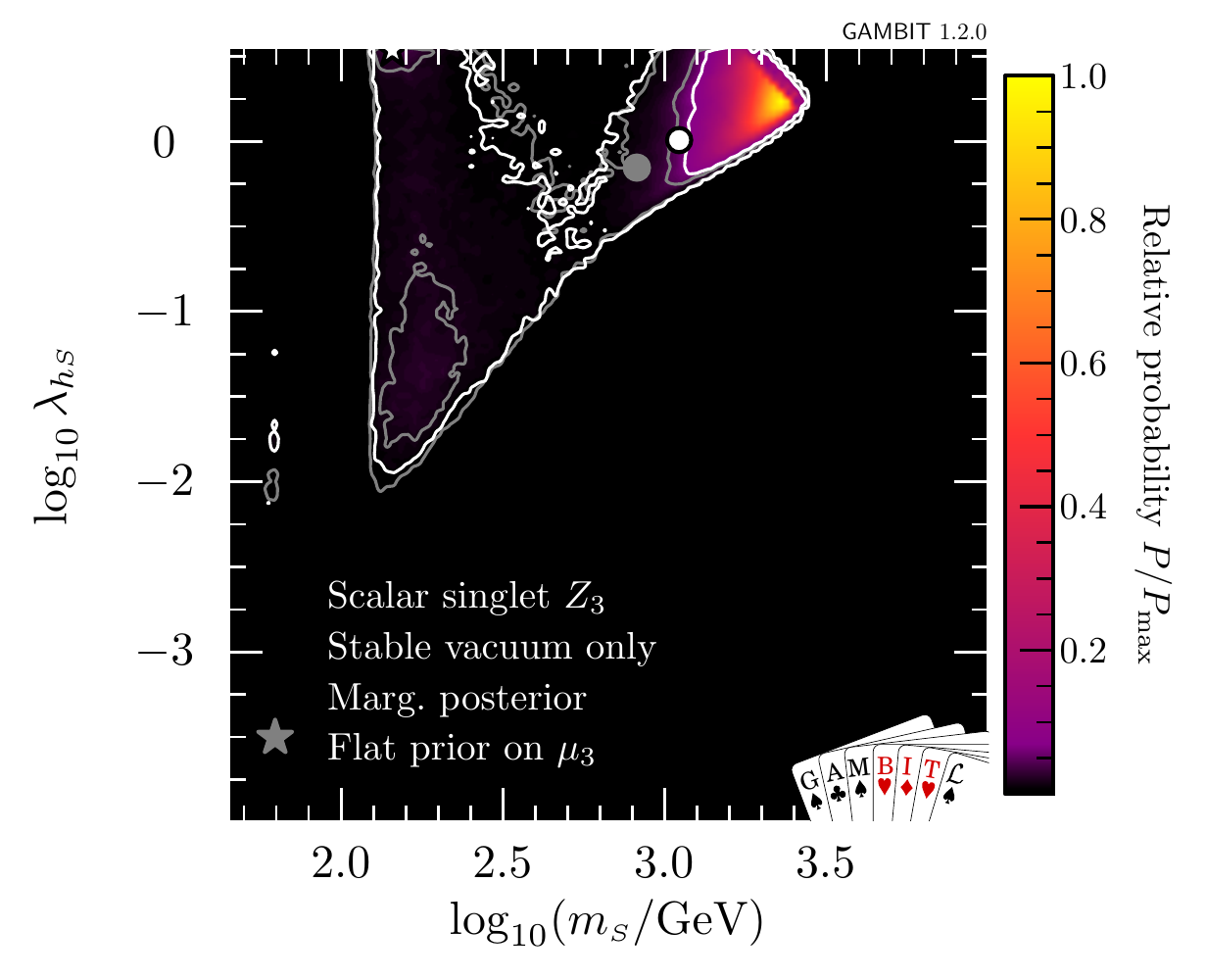}\\
\includegraphics[width=1\columnwidth]{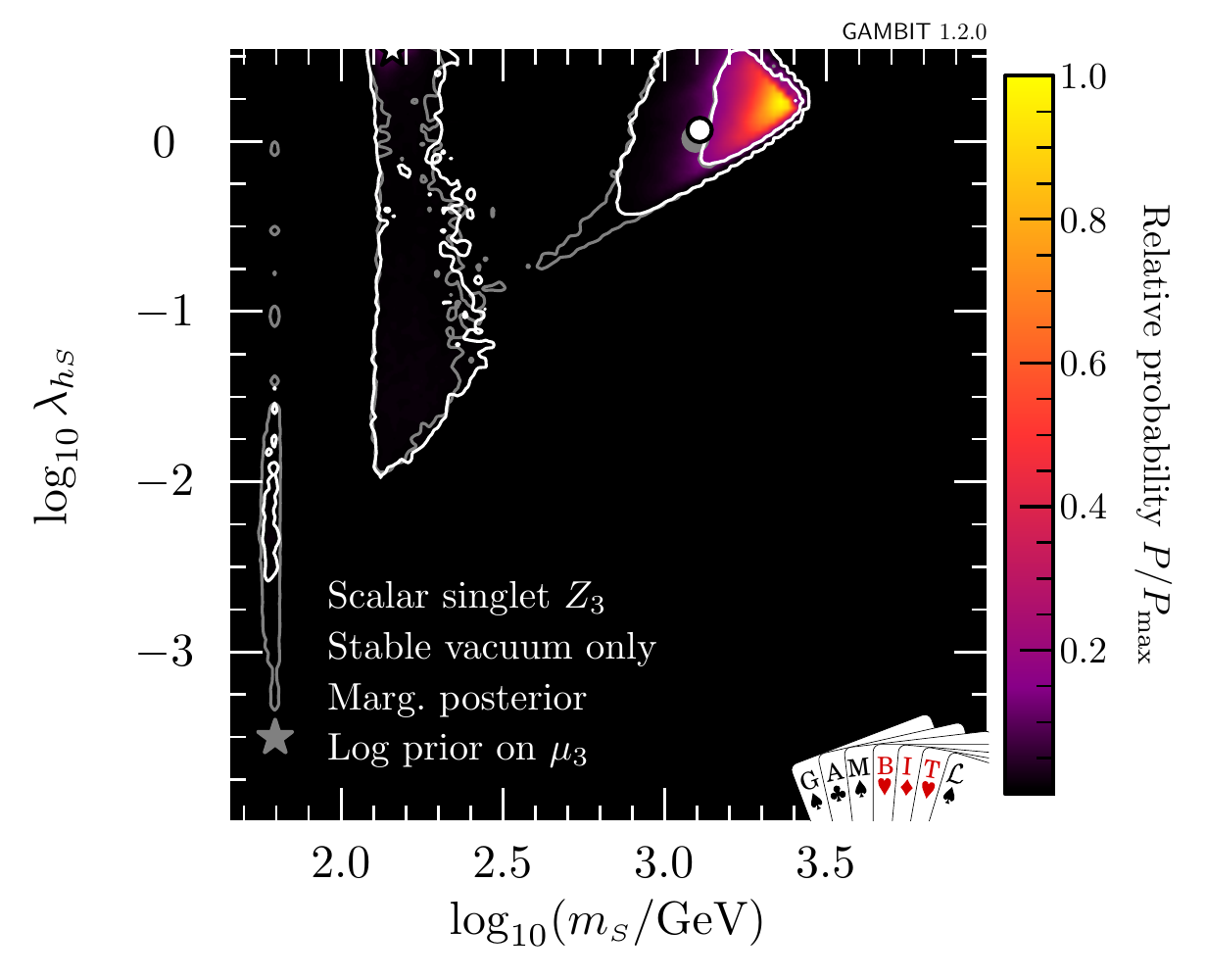}
\caption{Impact of the requirement of vacuum stability on the $\mathbb{Z}_3$ scalar singlet model, expressed in terms of profile likelihoods (\textit{top}) and posterior probability densities with flat prior on $\mu_3$ (\textit{centre}) and with logarithmic prior on $\mu_3$ (\textit{bottom}). Bullets indicate posterior means and stars indicate best fit points.  Shading and white annotations correspond to scans where the singlet is required to absolutely stabilise the electroweak vacuum.  For comparison, we also show the preferred regions without this requirement in grey.}
\label{fig:Z3_post}
\end{figure}

For larger masses, semi-annihilations become less efficient, as the semi-annihilation fraction is proportional to $\mu_3\lhs^2/\ms^6$ at leading order~\cite{Belanger2013a}. As a result, the shape of the allowed parameter region is similar to the $\mathbb{Z}_2$ model for $\ms \gtrsim 1 \, \mathrm{TeV}$. The likelihood of this region is however much smaller, and is in fact outside the global $1\sigma$ confidence region. The reason is that the $\mathbb{Z}_3$ model requires a complex scalar, whereas we have considered a real scalar for the $\mathbb{Z}_2$ model. The coupling $\lhs$ must therefore be a factor of two larger in the \zthree model to achieve the necessary effective annihilation cross-section required to avoid DM overproduction, increasing the degree of tension between the relic density constraint and bounds from direct detection experiments.

Fig.~\ref{fig:Z3_DD} illustrates the impact of current and future direct detection experiments on the parameter space of the $\mathbb{Z}_3$ model.  Here we have again rescaled $\sigma_p^\text{SI}$ by the relic density fraction $f = \Omega_s/\Omega_\text{DM}$. Similar to the case of the $\mathbb{Z}_2$ model, the resonance region extends three orders of magnitude below current direct detection limits, and even a next-generation experiment such as LZ will not be able to probe the full parameter space at $\ms \simeq \mh/2$. However, the allowed parameter region at $\ms \sim 200\,$GeV, even though it is  substantially larger than its counterpart in the $\mathbb{Z}_2$ model, will eventually be fully explored by direct detection searches.

\subsection{Absolutely stable vacuum}\label{sec:singletdm:z3uv:vs}

As with the \ztwo model, we now investigate the preferred parameter regions more closely by imposing additional physical requirements. In the top panel of Fig.~\ref{fig:Z3_post}, we show how the profile likelihoods change when requiring absolute vacuum stability. The corresponding marginalised posteriors are shown in the central panel of Fig.~\ref{fig:Z3_post} for a scan with flat prior on $\mu_3$, and in the bottom panel for a scan with logarithmic prior on $\mu_3$. Although the likelihood is maximised for $\ms\sim100\,\mathrm{GeV}$, the majority of the posterior mass is found at higher masses, $\ms > 1\,\mathrm{TeV}$. The reason is that at larger masses, the relic abundance becomes independent of $\mu_3$ and therefore benefits strongly from marginalisation (rather than profiling) over $\mu_3$.  Comparing the middle and lower panels, this conclusion is independent of the choice of prior on $\mu_3$.  The choice of prior on $\mu_3$ has a relatively small impact overall, essentially just translating into a stronger preference for large $\ms$ when taken flat rather than logarithmic, due to the restriction to lower values of $\mu_3$ at lower $\ms$ coming from Eq.\ \ref{eqn:ew_stability_condition}.

\begin{figure*}[tbp]
\centering
\includegraphics[width=1\columnwidth]{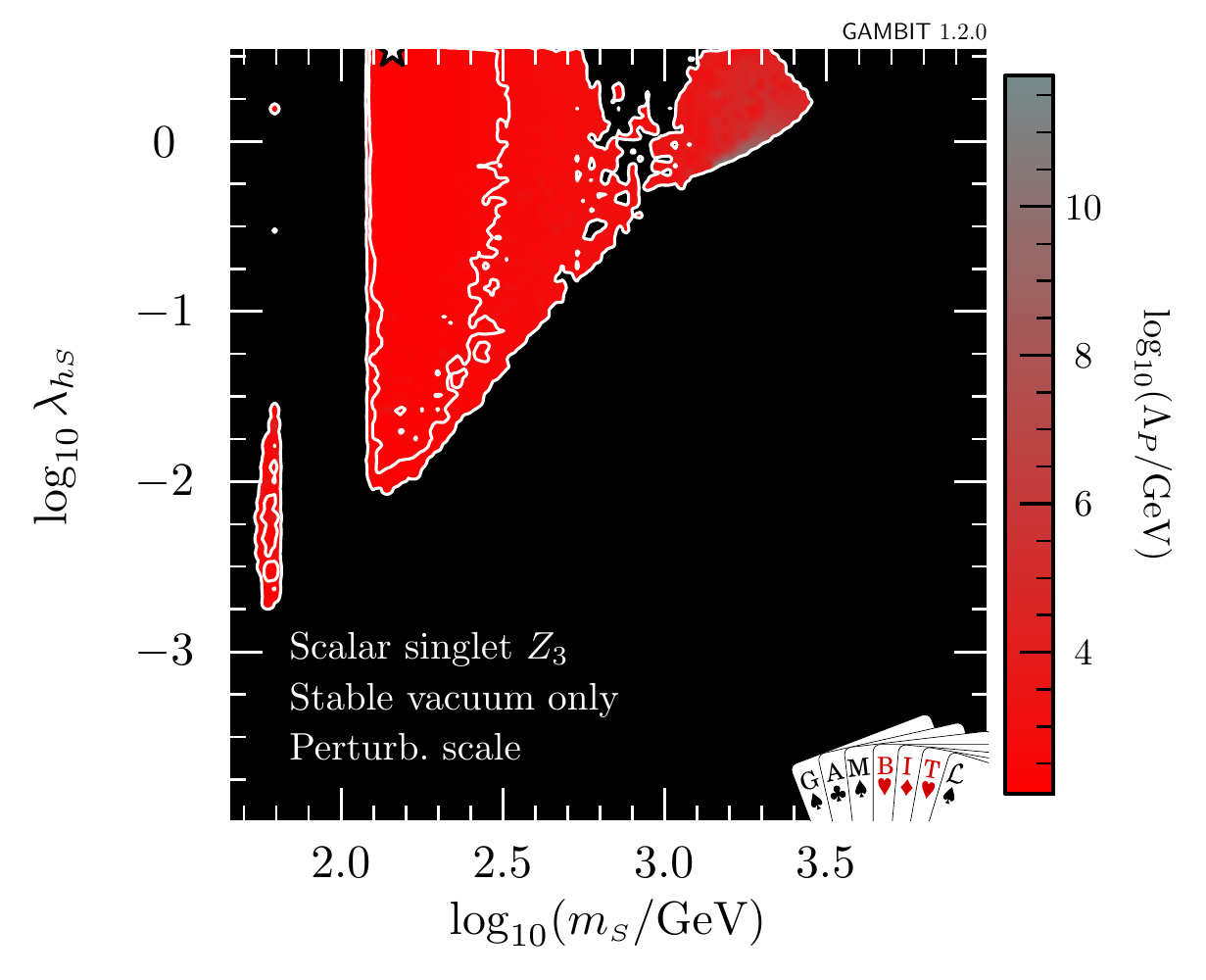}
\includegraphics[width=1\columnwidth]{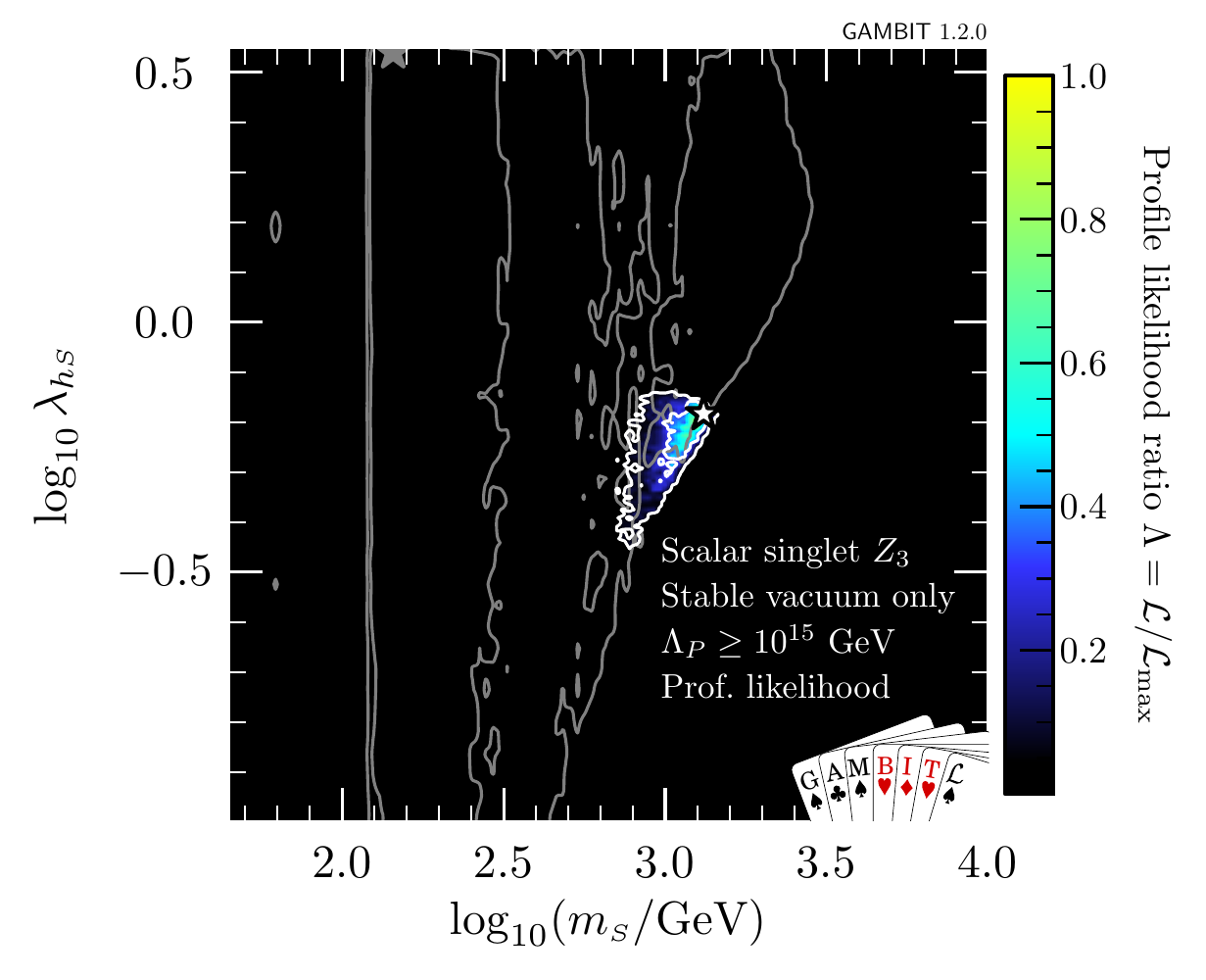}
\caption{\textit{Left:} scale of perturbativity violation for the $\mathbb{Z}_3$ scalar singlet model with the requirement of a stable electroweak vacuum. \textit{Right:} profile likelihood when also imposing the requirement $\Lambda_P>10^{15}\,$GeV.  The $1\sigma$ and $2\sigma$ confidence regions are delineated by white contours, and the best-fit by a white star.  Grey contours on the right panel correspond to the $1\sigma$ and $2\sigma$ confidence regions of the left panel.}
\label{fig:Z3_cut}
\end{figure*}

Fig.~\ref{fig:Z3_post} demonstrates that most of the parameter space opened up by semi-annihilations in the intermediate mass range remains viable when imposing absolute vacuum stability.   This observation raises the question whether the stabilisation is due to the influence of $\mu_3$ on the running of $\lh$ or whether the new parameter simply leads to a breakdown of perturbativity. We therefore show the scale of perturbativity violation in the left panel of Fig.~\ref{fig:Z3_cut}. Indeed, we find that $\Lambda_P$ is extremely low (less than $10^{10}\,$GeV) throughout the $2\sigma$ preferred region of the new parameter space, such the points with a `stable' vacuum are not actually as theoretically appealing as might have naively been expected on the basis of Fig.\ \ref{fig:Z3_post}.  This is also the reason for the persistence of part of the resonance reason in Fig.~\ref{fig:Z3_post} after absolute vacuum stability is required, unlike in the corresponding \ztwo plot (Fig.\ \ref{fig:Z2_vs}). \zthree models remaining in this region after absolute vacuum stability is required are just those with the highest values of $\ls$, allowing them to combine with non-zero values of $\mu_3$ to send $\ls$ non-perturbative at relatively low scales, and thereby avoid having to actually stabilise the electroweak vacuum.

The reason that large couplings are required in the intermediate mass range is related to the need for semi-annihilations in this part of parameter space. A large semi-annihilation fraction requires that the coupling $\mu_3$ is larger than about $300\,$GeV (see Fig.~\ref{fig:Z3_semi}). This in turn forces $\ls$ to be large in order to satisfy Eq.~(\ref{eqn:ew_stability_condition}), which leads to the couplings becoming non-perturbative at a low scale.\footnote{Note, in particular, that $\mu_3$ does not directly impact the running of $\ls$; as a dimension-1 parameter, it cannot enter the RGE of $\ls$ (which is dimensionless).} Thus, the new regions of parameter space opened up by semi-annihilations in the $\mathbb{Z}_3$ model are of limited theoretical appeal from the point of view of stabilising the electroweak vacuum.

\begin{figure}[tbp]
\centering
\includegraphics[width=1\columnwidth]{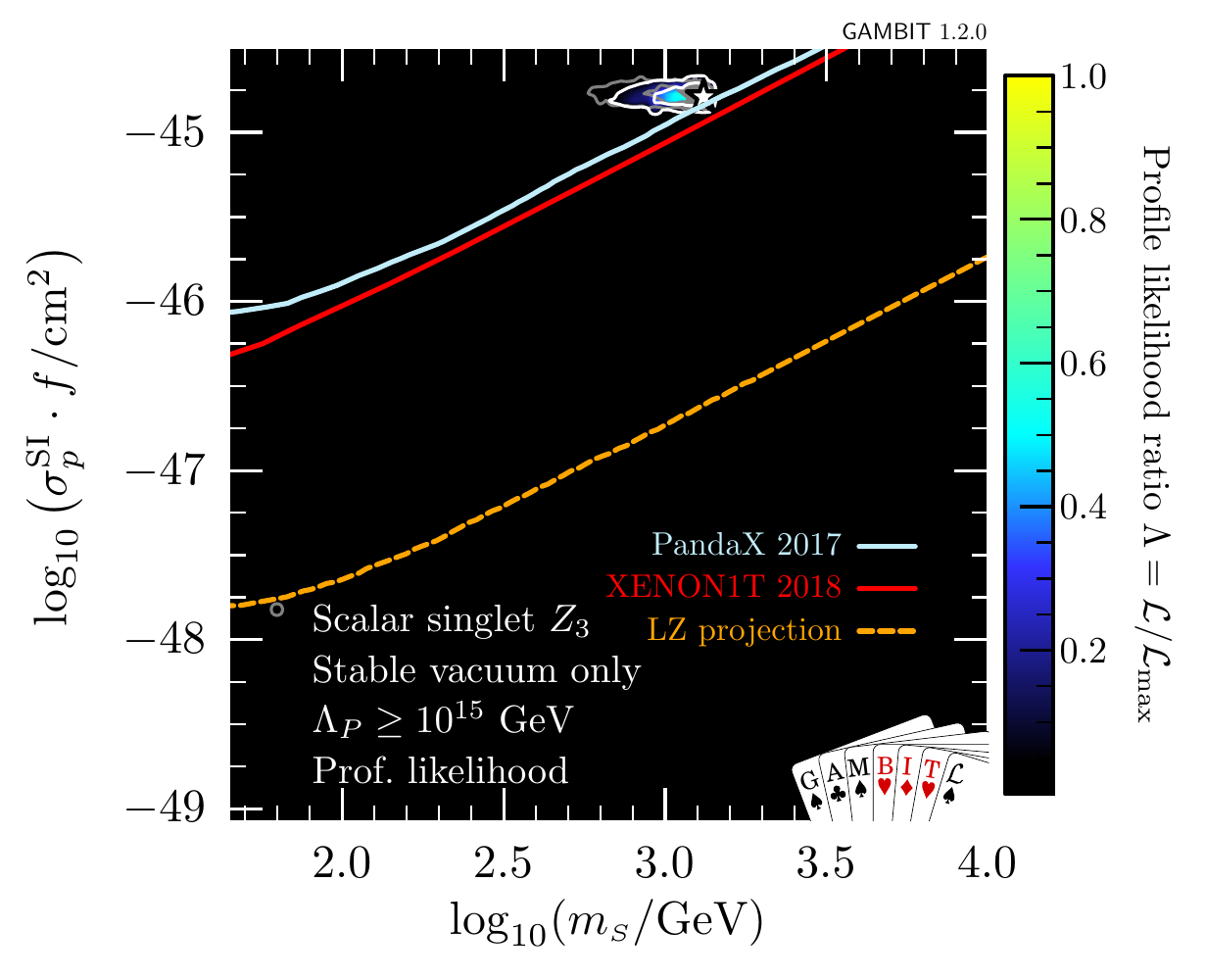}
\caption{As in Fig.\ \ref{fig:Z3_DD}, but with the added requirements of vacuum stability and perturbativity to large scales, $\Lambda_P>10^{15}\,$GeV.}
\label{fig:Z3_DD_cut}
\end{figure}

In the right panel of Fig.~\ref{fig:Z3_cut}, we present the profile likelihood after also imposing the requirement $\Lambda>10^{15}\,$GeV. We find that this not only removes the resonance region,\footnote{That is, except for the very tip of the neck at large $\lhs$, as noted earlier in the context of the \ztwo model and visible in one luckily-sampled bin in the bottom-left of Fig.\ \ref{fig:Z3_DD_cut}.} but also the intermediate mass range. In fact, only a tiny region around $\ms \sim 1\,\mathrm{TeV}$ remains. Judging from Fig.~\ref{fig:Z3_full}, we expect this remaining parameter space to have a much lower likelihood than the resonance region. In Fig.~\ref{fig:Z3_DD_cut} we show the nuclear scattering cross-section (rescaled as usual by the fraction $f$ of the DM relic density constituted by singlet scalars) as a function of $\ms$.  This plot qualitatively confirms our expectation that this region should lie outside the preferred region in the global scan, as the 90\% C.L. upper bound from XENON1T \cite{Aprile:2018dbl} already excludes the $2\sigma$ confidence regions. In the following, we will make this point more explicit by studying the best-fit point in this region, and showing that it is in considerable tension with data.

\subsection{Best-fit point}\label{sec:singletdm:z3uv:best}

The best-fit point in the $\mathbb{Z}_3$ model with metastability allowed is at $\ls =\num{4.83e-01}$, $\lhs =\num{3.21e-04}$, $\mu_3 = 11.8\,$GeV and $\ms = \num{62.48}\,$GeV.  This point has a lifetime of $\sim$$\num{1.4e+99}$ years, and a minimum in its Higgs quartic coupling at $\sim$$3\times10^{13}\,$GeV. For this point we find $\Delta\ln \mathcal{L} = 0.318$, essentially the same as for the equivalent best-fit point in the $\mathbb{Z}_2$ model. This result is expected, as the semi-annihilation fraction at this point is $\alpha = 0$.

\begin{table*}[t!]
\center{{\footnotesize
\setlength\tabcolsep{1.4pt}
\begin{tabular}{l@{\hspace{2mm}}l@{\hspace{2mm}}l@{\hspace{2mm}}l@{\hspace{2mm}}l@{\hspace{2mm}}l @{\hspace{2mm}}l @{\hspace{2mm}}l @{\hspace{2mm}}l @{\hspace{2mm}}l@{\hspace{2mm}}l}
Stable &  &Relic &  &  &&   &  & & \\
vac.  & $\Lambda_P$ (GeV) & density & $\ls$ & $\lhs$ & $\ms$ (GeV) & $\mu_3$ (GeV) & $\Omega_{\sss S} h^2$  & $\log(\mathcal{L})$ & $\Delta\ln\mathcal{L}$ & $\sigma^{\text{SI}}_n$ (cm$^2$)\\
  \toprule
$\sim$ & $\num{6.33e+08}$ & $\leq$ & $\num{4.827e-01}$ & $\num{3.207e-04}$ & $\num{6.248e+01}$ & $\num{1.180e+01}$ & $\num{9.7001e-02}$ & 43.12 & 0.32 & $\num{2.21e-49}$\\
\checkmark & $\num{1.68e+02}$ & $\leq$ & $\num{3.523e+00}$ & $\num{3.498e+00}$ & $\num{1.438e+02}$ & $\num{5.367e+02}$ & $\num{6.2920e-07}$ & 43.27 & 0.47 & $\num{5.02e-42}$\\
\checkmark & $\num{1.29e+15}$ & $\leq$ & $\num{3.083e-02}$ & $\num{6.604e-01}$ & $\num{1.314e+03}$ & $\num{3.419e+02}$ & $\num{1.0990e-01}$ & 46.77 & 3.98 & $\num{1.72e-45}$\\
\checkmark & $\num{1.89e+15}$ & \checkmark & $\num{1.194e-01}$ & $\num{5.917e-01}$ & $\num{1.206e+03}$ & $\num{5.399e+02}$ & $\num{1.1367e-01}$ & 47.24 & 4.44 & $\num{1.67e-45}$\\
 \bottomrule
\end{tabular}}
\caption{Details of the best-fit points for the \zthree scalar singlet model when different physical restrictions are imposed on the model. Points that have an absolutely stable electroweak vacuum are indicated by a tick in the first column. Points with a singlet relic density within 1$\sigma$ of the \textit{Planck} observed value ($\Omega_{\sss S} h^2\sim\Omega_{\sss \text{DM}} h^2$) are indicated with a tick in the third column. We omit the values of the nuisance parameters, as they are not significantly different to the central values of their respective likelihood functions.}\label{tab:best_fit_z3}}
\end{table*}

With the additional constraint of absolute vacuum stability, the best-fit is located at $\ls =\num{3.52e+00}$, $\lhs =\num{3.50e+00}$, $\mu_3 = \num{537}\,$GeV and $\ms =\num{144}\,$GeV.  In this case the semi-annihilation fraction is $\alpha =0.72$ and we find $\Delta\ln \mathcal{L} = 0.473$. Compared to the equivalent point in the $\mathbb{Z}_2$ model, this represents a small improvement due to the contribution from semi-annihilations.  However, $\Lambda_P$ at this point is only $\num{168}\,$GeV, due to the large value of $\ls$, making it less than appealing in a theoretical sense.

Demanding that $\Lambda_P\ge 10^{15}\,$GeV, we find a best-fit point with an absolutely stable vacuum, $\Lambda_P = 1.29\times10^{15}\,$GeV and $\alpha =0.004$.  This point is located at $\ls = \num{3.08e-02}$, $\lhs = \num{6.60e-01}$, $\mu_3 = \num{342}\,$GeV and $\ms = \num{1.31}\,$TeV. This point has $\Delta\ln \mathcal{L} = 3.975$, with the dominant contributions coming from the most recent direct detection experiments. This corresponds to a likelihood ratio $\Lambda=0.026$, which places this point more than $2\sigma$ away from the overall best-fit point.  In other words, we find considerable tension in the $\mathbb{Z}_3$ model between direct detection limits and the requirement for the model to be absolutely stable and perturbative to at least the GUT scale.

Finally, we consider a point that is perturbative to at least $10^{15}\,$GeV, has a stable electroweak vacuum and has a singlet relic density within $1\sigma$ of the \textit{Planck} measured value.  The best-fit point under these requirements is located at $\ls =\num{1.19e-01}$, $\lhs = \num{5.92e-01}$, $\mu_3 = 540\,$GeV and $\ms = \num{1.21}\,$TeV. This point has $\Omega_Sh^2=0.1137$ and $\Lambda_P = 1.89\times10^{15}\,$GeV as well as $\Delta\ln \mathcal{L} = 4.443$, making it even more strongly disfavoured than the corresponding best-fit model with subdominant singlet DM.

We present the four best-fit points, their relic densities and the scales at which their couplings become non-perturbative in Table~\ref{tab:best_fit_z3}.

As with the \ztwo model, we can obtain approximate $p$-value ranges by assuming either one or two degrees of freedom. For the best-fit point with metastability allowed, we find $p\approx 0.4\text{--}0.7$, while the best fit with vacuum stability has $p\approx 0.3\text{--}0.6$. Both of these ranges are very similar to those for equivalent constraints on the $\mathbb{Z}_2$ model, despite semi-annihilations opening up a large region of parameter space. For the model with $\Lambda_P > 10^{15}\,$GeV, we find $p\approx 0.005\text{--}0.02$, reducing to $p\approx 0.003\text{--}0.01$ when also imposing the relic density requirement.  This illustrates once again that the $\mathbb{Z}_3$ model, which requires a complex scalar, is disfavoured by data as a joint mechanism to stabilise the electroweak vacuum and to provide a DM candidate.

\section{Conclusions}\label{sec:singletdm:conc}

In this work we have investigated two realisations of scalar singlet DM: a real scalar stabilised by a $\mathbb{Z}_2$ symmetry and a complex scalar stabilised by a $\mathbb{Z}_3$ symmetry. In addition to potentially accounting for the observed DM relic abundance via the freeze-out mechanism, these models have the attractive feature that in large regions of parameter space they remain valid up to very large scales. This makes it possible to study the RGE evolution of the various parameters and determine the impact of the scalar singlet on the running of the Higgs quartic self-coupling. Indeed, we find that this additional contribution may stabilise the electroweak vacuum, thus resolving an apparent deficiency of the SM.

Nevertheless, models of scalar singlets face a large number of experimental and theoretical constraints. The most important experiments are those aimed at direct detection of DM, most notably the very recent results from XENON1T \cite{Aprile:2018dbl}.  In spite of observing a small upward fluctuation,  XENON1T places strong constraints on the DM-nucleon scattering cross-section. The most important theoretical requirement that we have considered is for couplings to remain perturbative up to the scales where the stability of the electroweak vacuum may become an issue.

By performing a global fit to all available data, we have shown that it is still possible to explain DM and stabilise the electroweak vacuum through the addition of a scalar singlet field charged under a $\mathbb{Z}_2$ symmetry, while at the same time satisfying all experimental constraints. Although in much of the allowed parameter space we find the scale at which couplings become non-perturbative to be quite low, there is an allowed parameter region with scalar masses of about $2\,$TeV, where the theory remains perturbative up to at least $10^{15}\,$GeV. Moreover, it is possible in this parameter region for scalar singlets to constitute all of the DM. This parameter region is in slight tension with direct detection, but is consistent with the small preference for a non-zero signal contribution at high DM mass in recent XENON1T results.  The next generation of experiments will therefore fully explore the viability of this scenario.

The alternative possibility of a complex scalar singlet with a $\mathbb{Z}_3$ symmetry opens up large regions of the parameter space, because the semi-annihilation channel allows the same relic density to be achieved for much smaller singlet-Higgs couplings than in equivalent parts of the \ztwo parameter space. However, the presence of a large trilinear coupling drives the couplings non-perturbative at a relatively low scale, making it impossible to calculate the running of Higgs self-couplings to high scales. When requiring a stable electroweak vacuum as well as perturbative couplings up to at least $10^{15}\,$GeV, the semi-annihilation channel ceases to be relevant and the remaining parameter space resembles the one of the $\mathbb{Z}_2$ model. However, the relic density constraint is more severe for a complex scalar than for a real scalar, so that the $\mathbb{Z}_3$ model is in fact more tightly constrained. Indeed, the parameter region with $\ms \sim 1\,$TeV is disfavoured more than 95\% confidence, irrespective of whether or not the scalar singlets constitute all of DM.

Scalar singlets have frequently been advocated as one of the simplest realisations of the WIMP idea. The non-observation of a DM signal in any type of experiment designed to search for WIMPs therefore clearly increases pressure on these models. While the low-mass (resonance) region remains challenging to probe experimentally, we have focused on the more interesting high-mass region, which may help to address the issue of a metastable electroweak vacuum. While this solution is now essentially ruled out for the case of a complex scalar singlet (stabilised e.g.\ by a $\mathbb{Z}_3$ symmetry), it remains an interesting possibility for real scalars (with a $\mathbb{Z}_2$ stabilising symmetry). The next generation of direct detection experiments will be able to reach a definite verdict on these models, including the exciting possibility that they may confirm the slight excess seen in XENON1T.

\begin{acknowledgements}
We thank our colleagues within the GAMBIT community, in particular Sanjay Bloor, Jos\'e-Eliel Camargo-Molina, Jan Conrad and Martin White, for helpful discussions, and Fatih Ertas for beta-testing \ddcalc \textsf{2.0.0}. We acknowledge PRACE for awarding us access to Marconi at CINECA, Italy. We are also grateful to the UK Materials and Molecular Modelling Hub, which is partially funded by EPSRC (EP/P020194/1), for additional computational resources.  PA is supported by Australian Research Council Future Fellowship FT160100274, JMC by the Natural Sciences and Engineering Research Council (NSERC) of Canada, FK by DFG Emmy Noether Grant No.\ KA 4662/1-1, JM by the Imperial College London President's PhD Scholarship, PS by STFC (ST/K00414X/1, ST/N000838/1, ST/P000762/1) and SW by ERC Starting Grant `NewAve' (638528).
\end{acknowledgements}

\appendix

\section{Supplementary material}

All samples, input files and best-fit points from this paper are freely available for download from Zenodo \cite{the_gambit_collaboration_2018_1298566}.

\bibliography{R1.5}

\end{document}